\documentclass[prd,aps,showpacs,nofootinbib,nobibnotes]{revtex4}
\usepackage{epsfig}

\newcommand{\ETAL}{{\it et al.}}

\newcommand{\MAT}{{\rm m}}
\newcommand{\BAR}{{\rm b}}
\newcommand{\TOT}{{\rm tot}}
\newcommand{\LSS}{{\scriptscriptstyle \rm LSS}}
\newcommand{\OBS}{{\rm obs}}
\newcommand{\MAX}{{\rm max}}

\newcommand{\SW}{{\rm SW}}
\newcommand{\DOP}{{\rm Dop}}

\newcommand{\PS}{{\cal P}_\phi}
\newcommand{\ALM}[2]{a_{{#1}\,{#2}}}
\newcommand{\CLMLPMP}[4]{C_{{#1}\,{#2}}^{{#3}\,{#4}}}
\newcommand{\SK}[1]{\mbox{s}_{#1}}

\newcommand{\UUNIT}[2]{\,{\mbox{#1}^{#2}}}

\newcommand{\ee}{{\rm e}}
\newcommand{\YLM}[2]{Y_{#1}^{#2}}
\newcommand{\PLM}[2]{P_{#1}^{#2}}
\newcommand{\GNA}[2]{C_{#1}^{#2}}
\newcommand{\JAC}[3]{P_{#3}^{({#1}\,,\,{#2})}}

\newcommand{\YXKLM}[4]{{\cal Y}^{[{#1}]}_{{#2}\,{#3}\,{#4}}}
\newcommand{\RXKL}[3]{R^{[{#1}]}_{{#2}\,{#3}}}
\newcommand{\UPSTKS}[3]{\Upsilon^{[{#1}]}_{{#2}\,{#3}}}
\newcommand{\XITKSLM}[5]{\xi^{[{#1}]\,{#3}}_{{#2}\,{#4}\,{#5}}}

\newcommand{\GRP}[2]{\rm {#1}({#2})}
\newcommand{\MOD}[1]{\:\;{\rm mod}({#1})}
\newcommand{\GCD}{{\rm gcd}}
\newcommand{\CNP}[2]{\left(\begin{array}{c}{#1} \\ {#2} \end{array} \right)}

\newcommand{\DIR}[1]{\delta^{{\rm D}}({#1})}
\newcommand{\KRON}[2]{\delta_{{#1}\,{#2}}}
\newcommand{\ddd}{{\rm d}}

\newcommand{\bx}{{\bf x}}
\newcommand{\br}{{\bf r}}
\newcommand{\bk}{{\bf k}}
\newcommand{\bn}{{\bf n}}
\newcommand{\bhn}{{\bf\hat n}}

\newcommand{\gsim}{\mathrel{%
   \rlap{\raise 0.511ex \hbox{$>$}}{\lower 0.511ex \hbox{$\sim$}}}}
\newcommand{\lsim}{\mathrel{
   \rlap{\raise 0.511ex \hbox{$<$}}{\lower 0.511ex \hbox{$\sim$}}}}

\newcommand{\bb}{\bf}
\newcommand{\RTR}{{\bb R}^3}
\newcommand{\STR}{{\bb S}^3}
\newcommand{\HTR}{{\bb H}^3}

\newcommand{\QKLM}[3]{{\cal Q}_{{#1}\,{#2}\,{#3}}}
\newcommand{\BKLM}[3]{B_{{#1}\,{#2}\,{#3}}}
\newcommand{\CKLM}[3]{C_{{#1}\,{#2}\,{#3}}}
\newcommand{\ETAKSLM}[5]{\eta^{[{#1}]\,{#3}}_{{#2}\,{#4}\,{#5}}}
\newcommand{\ALKLMLTMT}[5]{\alpha_{{#1}\,{#2}\,{#3}\,{#4}\,{#5}}}
\newcommand{\ST}{{\scriptscriptstyle {\rm T}}}
\newcommand{\CHU}{\chi}
\newcommand{\CHN}{\bar \chi}

\newcommand{\SZFIG}{\normalsize}

\begin{document}
\title{Simulating Cosmic Microwave Background maps in multi-connected spaces}

\author{Alain Riazuelo}
\email{riazuelo@spht.saclay.cea.fr}
\affiliation{Service de Physique Th\'eorique,
             CEA/DSM/SPhT, Unit{\'e} de recherche associ\'ee au
             CNRS, CEA/Saclay F--91191 Gif-sur-Yvette c\'edex,
             France}

\author{Jean-Philippe Uzan}
\email{uzan@iap.fr}
\affiliation{Institut d'Astrophysique de Paris, GR$\varepsilon$CO, FRE
             2435-CNRS, 98bis boulevard Arago, 75014 Paris, France \\
             Laboratoire de Physique Th\'eorique, CNRS-UMR 8627,
             Universit\'e Paris Sud, B\^atiment 210, F--91405 Orsay
             c\'edex, France}

\author{Roland Lehoucq}
\email{lehoucq@cea.fr}
\affiliation{CE-Saclay, DSM/DAPNIA/Service d'Astrophysique,
             F--91191 Gif-sur-Yvette c\'edex, France,\\ Laboratoire
             Univers et Th\'eories, CNRS-FRE 2462, Observatoire de
             Paris, F--92195 Meudon c\'edex, France}

\author{Jeffrey Weeks}
\email{weeks@northnet.org}
\affiliation{15 Farmer St.,  Canton  NY  13617-1120, USA}

\date{17 February 2003}
\begin{abstract}
This article describes the computation of cosmic microwave background
anisotropies in a universe with multi-connected spatial sections and
focuses on the implementation of the topology in standard CMB computer
codes.  The key ingredient is the computation of the eigenmodes of the
Laplacian with boundary conditions compatible with multi-connected
space topology. The correlators of the coefficients of the
decomposition of the temperature fluctuation in spherical harmonics
are computed and examples are given for spatially flat spaces and one
family of spherical spaces, namely the lens spaces.  Under the
hypothesis of Gaussian initial conditions, these correlators encode
all the topological information of the CMB and suffice to simulate CMB
maps.

\vspace*{0.35cm}
\noindent
{\footnotesize Preprint numbers: SPhT-T02/182, LPT-02/123,
{\tt astro-ph/0212223}}
\end{abstract}
\maketitle

\section{Introduction}
\label{sec_intro}

Future cosmic microwave background (CMB) experiments such as the
MAP~\cite{map} and later the Planck satellites~\cite{planck} will
provide full sky maps of CMB anisotropies (up to the galactic
cut). These datasets offer the opportunity to probe the topological
properties of our universe. A series of tests to detect the topology,
including the use of the angular power
spectrum~\cite{sokolov93,staro93,stevens93,costa95}, the distribution
of matched patterns such as circles~\cite{cornish98}, correlation of
antipodal points~\cite{levin98} and non Gaussianity~\cite{inoue00}
have been proposed (see Refs.~\cite{lachieze95,uzan99,levin02} for
reviews of CMB methods to search for the topology). The study of the
detectability of the topology by any of these methods first requires
simulating maps with the topological signature for a large set of
topologies.  These maps will allow one to test the detection methods,
estimate their run time, and, once all sources of noise are added,
determine to what extent a given method detects the topological signal
(in the same spirit as the investigation of the ``crystallographic''
methods based on galaxy catalogs~\cite{us3}).

In a simply connected\footnote{Geometers and cosmologists often
refer to simple-connectedness as the ``trivial topology''.
However, trivial topology has a different meaning in the context
of point-set topology: in that formalism, the trivial topology is
the smallest topology on a set $X$, namely the one in which the
only open sets are the empty set and the entire set $X$.},
spatially homogeneous and isotropic universe, the angular
correlation function depends only on the angle between the two
directions and the coefficients $\ALM{\ell}{m}$ of the
decomposition of the temperature fluctuation in spherical
harmonics, which are uncorrelated for different sets of $\ell$ and
$m$. Multi-connectedness breaks the global isotropy and sometimes
the global homogeneity of the universe, except in projective space
(see, e.g., Ref.~\cite{lwugl}). Consequently, the CMB temperature
angular correlation function will depend on the two directions of
observation, not only on their relative angle, and possibly on the
position of the observer. This induces correlations between the
$\ALM{\ell}{m}$ of different $\ell$ and $m$. Such correlations are
hidden when one considers only the angular correlation function
and its coefficients, the so-called $C_\ell$, in a Legendre
polynomial decomposition, because they pick up only the isotropic
part of the information and are therefore a poor indicator of the
topology. This work aims to detail the whole computation of the
correlation matrix
\begin{equation}
\label{eq:I:1}
\CLMLPMP{\ell}{m}{\ell'}{m'}
 \equiv \left< \ALM{\ell}{m} \ALM{\ell'}{m'}^* \right> ,
\end{equation}
which encodes all the topological properties of the CMB, and from
which one can compute the usual $C_\ell$, simulate maps, and so
on.

The study of the detectability of a topological signal (if it exists)
in forthcoming CMB datasets requires simulating high quality maps
containing the topological signature for a wide class of
topologies. Up to now, most CMB studies considered only compact
Euclidean
spaces~\cite{sokolov93,staro93,stevens93,costa95,levin99,inoue01} and
some compact hyperbolic
spaces~\cite{aurich00,inoue00b,cornish00,bond1,bond2,bond3}, and
focused mainly on the $C_\ell$. The approach developed in this
article, and first introduced in Ref.~\cite{lwugl}, is well suited to
simulate the required CMB maps in any topology once the eigenmodes of
the Laplacian have been determined. It paves the way to the simulation
of maps for a wide range of topologies, particularly spherical ones.

Recent measurements of the density parameter $\Omega$ imply that
the observable universe is ``approximately flat''\footnote{The
popular expression ``flat universe'' is misleading, because in
General Relativity the ``universe'' is not three-, but
four-dimensional and the Friedmann-Lema\^{\i}tre solutions are
dynamical, so that the universe is not flat --- only its spatial
section may be flat or nearly so. In what follows, we will
implicitly assume we are talking about the {\em three-dimensional
spacelike sections} of the universe when talking about flat,
hyperbolic, or spherical spaces/universes.}, perhaps with a slight
curvature. The exact constraint on the total density parameter
obtained from CMB experiments depends on the priors used during
the data analysis. For example with a prior on the nature of the
initial conditions, the Hubble parameter and the age of the
universe, recent analysis of the DASI, BOOMERanG, MAXIMA and DMR
data~\cite{sievers,netterfield,archeops2} lead to $\Omega = 0.99
\pm 0.12$ at $1\sigma$ level, and to $\Omega = 1.04 \pm 0.05$ at
$1\sigma$ level if one takes into account only the DASI, BOOMERanG
and CBI data.  Including stronger priors can indeed sharpen the
bound. For instance, including information respectively on large
scale structure and on supernovae data leads to $\Omega = 1.01_{-
0.06}^{+ 0.09}$ and $\Omega = 1.02_{- 0.08}^{+ 0.09}$ at $1\sigma$
level while including both finally leads to $\Omega = 1.00_{-
0.06}^{+ 0.10}$. This has been recently improved by the Archeops
balloon experiments~\cite{archeops2,archeops1} which get, with a
prior on the Hubble constant, $\Omega = 1.00_{- 0.02}^{+ 0.03}$.
In conclusion, it is fair to assert that current cosmological
observations set a reliable bound $0.9 < \Omega < 1.1$. These
results are consistent with Friedmann-Lema\^{\i}tre universe
models with spherical, flat or hyperbolic spatial sections.  In
the spherical and hyperbolic cases, $\Omega \simeq 1$ implies that
the curvature radius must be larger than the horizon radius. In
all three cases --- spherical, flat, and hyperbolic --- the
universe may be simply connected or multiply
connected.\\

The possibility of detecting the topology of a nearly flat universe
was discussed in Ref.~\cite{wlu02}. It was noted that the chances of
detecting a multiply connected topology are worst in a large
hyperbolic universe. The reason is that the typical translation
distance between a cosmic source and its nearest topological image
seems to be on the order of the curvature radius, and that when
$\Omega \simeq 1$ the distance to the last scattering surface is less
than the half of that distance.  See
Refs.~\cite{gomero1,aurichsteiner,inoue3,gomero3,weeks} for some
studies on detectability of nearly flat hyperbolic universes. In a
multiply connected flat universe the topology scale is completely
independent of the horizon radius, because Euclidean geometry ---
unlike spherical and hyperbolic geometry --- has no preferred scale
and admits similarities.  Note that in the Euclidean case, there are
only ten compact topologies, which reduces and simplifies the analysis
(in particular regarding the computation of the eigenmodes of the
Laplacian). In a spherical universe the topology scale depends on the
curvature radius, but, in contrast to the hyperbolic case, as the
topology of a spherical manifold gets more complicated, the typical
distance between two images of a single cosmic source decreases. No
matter how close $\Omega$ is to $1$, only a finite number of spherical
topologies are excluded from detection. The particular case of the
detectability of lens spaces was studied in Ref.~\cite{gomero2}, which
also considers the detectability of hyperbolic topologies.\\

At present, the status of the constraints on the topology of the
universe is still very preliminary.  Regarding locally Euclidean
spaces, it was shown on the basis of the COBE data that in the case of
a vanishing cosmological constant the size of the fundamental domain
of a 3-torus has to be larger than $L \geq 4800 \,h ^{-1}
\UUNIT{Mpc}{}$~\cite{sokolov93,staro93,stevens93,costa95}, where the
length $L$ is related to the smallest wavenumber $2 \pi / L$ of the
fundamental domain, which induces a suppression of fluctuations on
scales beyond the size $L$ of the fundamental domain.  This constraint
does not exclude a toroidal universe since there can be up to eight
copies of the fundamental cell within our horizon.  This constraint
relies mainly on the fact that the smallest wavenumber is $2 \pi / L$,
which induces a suppression of fluctuations on scales beyond the size
of the fundamental domain.  This result was generalized to all
Euclidean manifolds in Ref.~\cite{levin99}.  A non-vanishing
cosmological constant induces more power on large scales, via the
integrated Sachs-Wolfe effect.  For instance if $\Omega_\Lambda = 0.9$
and $\Omega_\MAT = 0.1$, the constraint is relaxed to allow for 49
copies of the fundamental cell within our horizon.  The constraint is
also milder in the case of compact hyperbolic manifolds and it was
shown~\cite{aurich00,inoue00b,cornish00} that the angular power
spectrum was consistent with the COBE data on multipoles ranging from
2 to 20 for the Weeks and Thurston manifolds.  Another approach, based
on the method of images, was developed
in~\cite{bond1,bond2,bond3}. Only one spherical space using this
method of images was considered in the literature, namely projective
space~\cite{souradeep}. The tools developed in this article, as well
as in our preceding works~\cite{lwugl,glluw,luw02}, will let us fill
the gap regarding the simulation of CMB maps in spherical universes.

Technically, in standard relativistic cosmology, the universe is
described by a Friedmann-Lema\^{\i}tre spacetime with locally
isotropic and homogeneous spatial sections.  These spatial sections
can be defined as constant density or time hypersurfaces.  With such a
splitting, the equations of evolution of the cosmological
perturbations, which give birth to the large scale structures of the
universe, reduce to a set of coupled differential equations involving
the Laplacian.  This system is conveniently solved in Fourier space.
In the case of a multiply connected universe, we visualize space as
the quotient $X / \Gamma$ of a simply connected space $X$ (which is
just a $3$-sphere $\STR$, a Euclidean space $\RTR$, or a hyperbolic
space $\HTR$, depending on the curvature) by a group $\Gamma$ of
symmetries of $X$ that is discrete and fixed point free.  The group
$\Gamma$ is called the holonomy group.  To solve the evolution
equations we must first determine the eigenmodes
$\UPSTKS{\Gamma}{k}{}$ and eigenvalues $- \kappa_k^2$ of the Laplacian
on $X / \Gamma$ through the generalized Helmholtz equation
\begin{equation}
\label{Helmotz1}
\Delta \UPSTKS{\Gamma}{k}{} = -\kappa_k^2 \UPSTKS{\Gamma}{k}{} ,
\end{equation}
with
\begin{equation}
\kappa_k^2 = k^2 - K ,
\end{equation}
where $k$ indexes the set of eigenmodes, the constant $K$ is positive,
zero or negative according to whether the space is spherical, flat or
hyperbolic, respectively,\footnote{We work here in comoving units, but
when spatial section are not flat, the curvature of the comoving space
is {\em not} normalized: it is not assumed to be $+ 1$ in the
spherical case nor is it assumed to be $- 1$ in the hyperbolic case.
Thus our eigenvalues may differ numerically from those found in the
mathematical literature, although of course they agree up to a fixed
constant multiple $|K|^{-\frac{1}{2}}$.} and the boundary conditions
are compatible with the given topology. The Laplacian in
Eq.~(\ref{Helmotz1}) is defined as $\Delta \equiv D^i D_i$, $D_i$
being the covariant derivative associated with the metric $\gamma_{i
j}$ of the spatial sections ($i$, $j = 1$, $2$, $3$).  The eigenmodes
of $X / \Gamma$, on which any function on $X / \Gamma$ can be
developed, respect the boundary conditions imposed by the
topology. That is, the eigenmodes of $X / \Gamma$ correspond precisely
to those eigenmodes of $X$ that are invariant under the action of the
holonomy group $\Gamma$.  Thus any linear combination of such
eigenmodes will satisfy, by construction, the required boundary
conditions.  In this way we visualize the space of eigenmodes of $X /
\Gamma$ as a subspace of the space of eigenmodes of $X$, namely the
subspace that is invariant under the action of $\Gamma$.  The
computational challenge is to find this $\Gamma$-invariant subspace
and construct an orthonormal basis for it.  In the case of flat
manifolds the eigenmodes of $X / \Gamma = \RTR / \Gamma$ can be found
analytically.  In the case of hyperbolic manifolds, many numerical
investigations of the eigenmodes of $X / \Gamma = \HTR / \Gamma$ have
been
performed~\cite{inoue00b,aurich89a,aurich89b,inoue99,aurich96,cornish99}.
In the case of spherical manifolds, the eigenmodes of $X / \Gamma =
\STR / \Gamma$ have been found analytically for lens and prism
spaces~\cite{luw02} and otherwise can be found
numerically~\cite{lwugl}.

In the following, we will develop the eigenmodes of $X / \Gamma$ on
the basis $\YXKLM{X}{k}{\ell}{m}$ of the eigenmodes of the universal
covering space as
\begin{equation}
\label{eq:1}
\UPSTKS{\Gamma}{k}{s}
 = \sum_{\ell = 0}^\infty \sum_{m = -\ell}^{\ell}
   \XITKSLM{\Gamma}{k}{s}{\ell}{m} \YXKLM{X}{k}{\ell}{m} ,
\end{equation}
so that all the topological information is encoded in the
coefficients $\XITKSLM{\Gamma}{k}{s}{\ell}{m}$, where $s$ labels
the various eigenmodes sharing the same eigenvalue $-\kappa_k^2$,
both of which are discrete numbers\footnote{The spectrum is
discrete when the space is compact, e.g. the torus or any
spherical space. In a non-compact multi-connected space such as a
cylinder it will have a continuous component.}.  The sum over
$\ell$ runs from $0$ to infinity if the universal covering space
is non-compact (i.e., hyperbolic or Euclidean). The
$\XITKSLM{\Gamma}{k}{s}{\ell}{m}$ coefficients can be determined
analytically for Euclidean manifolds. In the case of a spherical
manifold, the modes are discrete,
\begin{equation}
\label{knu}
  k = (\nu + 1) \sqrt{K}
\end{equation}
where $\nu$ is a nonnegative integer (the cases $\nu = 0$ and $\nu =
1$ correspond to pure gauge modes~\cite{abott86}), so that $\kappa_k^2
= \nu (\nu + 2) K$, and the sum over $\ell$ is finite and runs from 0
to $\nu$ since the multiplicity of a mode $k$ is at most its
multiplicity in the universal cover, which is $(\nu + 1)^2$. Among
spherical spaces, the case of prism and lens spaces are the simplest
since one can determine these coefficients
analytically~\cite{luw02}. The worst situation is that of compact
hyperbolic manifolds, which is analogous to the Euclidean case since
the universal covering $\RTR$ is not compact but for which no
analytical forms are known for the eigenmodes. One then needs to rely
on numerical computations (see, e.g., Refs.~\cite{cornish99,glluw}).

Our preceding works provided a full classification of spherical
manifolds~\cite{glluw} and developed methods to compute the eigenmodes
of the Laplacian in them~\cite{lwugl}. Among all spherical manifolds,
we were able to obtain analytically the spectrum of the Laplacian for
lens and prism spaces~\cite{luw02} which form two countable families
of spherical manifolds. The goal of the present article is to simulate
CMB maps for these two families of spherical topologies, as well as
for the Euclidean topologies.

We first review briefly the physics of CMB anisotropies
(Sec.~\ref{sec_cmb}) mainly to explain how to take into account
the topology (Sec.~\ref{subsec_implement}) once the coefficients
$\XITKSLM{\Gamma}{k}{s}{\ell}{m}$ are determined. We then detail
the computation of these coefficients, focusing on the cases where
it can be performed analytically, that is for flat spaces and lens
and prism spaces.  We then present results of numerical
simulations (Sec.~\ref{sec_simulation}) as well as simulated maps.
We discuss these maps qualitatively and confirm that the expected
topological correlations (namely matching
circles~\cite{cornish98}) are indeed present. The effects of the
integrated Sachs-Wolfe and Doppler terms, as well as the thickness
of the last scattering surface, are discussed in order to give
some insight into the possible detectability of these
correlations. Figure~\ref{method} summarizes the different
independent steps of the computation as well as their interplay.

\section*{Notations}

The local geometry of the universe is described by a
Friedmann-Lema\^{\i}tre metric
\begin{equation}
\label{fl_metric}
\ddd s^2
 = - c^2 \ddd t^2
   + a^2 (t) \left[\ddd \CHU^2 + \SK{K}^2 (\CHU) \ddd \Omega^2 \right] ,
\end{equation}
where $a(t)$ is the scale factor, $t$ the cosmic time, $\ddd
\Omega^2 \equiv \ddd \theta^2 + \sin^2 \theta \, \ddd \varphi^2$
the infinitesimal solid angle, $\CHU$ is the comoving radial distance,
and where
\begin{equation}
\label{def_sk}
  \SK{K}(\CHU) =
   \left\lbrace
   \begin{array}{ll}
    \sinh (\sqrt{|K|}\;\CHU) / \sqrt{|K|} & \quad {\rm (hyperbolic \, case)} \\
    \CHU                                  & \quad {\rm (flat \, case)      } \\
    \sin (\sqrt{K}\;\CHU) / \sqrt{K}      & \quad {\rm (spherical \, case) }
   \end{array}
   \right.
\end{equation}
In the case of spherical and hyperbolic spatial sections, we also
introduce the dimensionless coordinate
\begin{equation}\label{def_chn}
\CHN=\sqrt{|K|}\CHU,
\end{equation}
which expresses radial distances in units of the curvature radius.

\begin{center}
\begin{figure}
\unitlength=1cm
\begin{picture}(18,18)
\thicklines
\put(0,0)
 {\framebox(9.5,8){}}
\put(0.25,7.5)
 {\SZFIG III: Eigenmodes of $X / \Gamma$}
\put(0.5,6.75)
 {\circle{0.8}
\makebox(0,0)
 {\SZFIG$\Gamma$}}
\put(0.9,6.7)
 {\vector(1,0){3}}
\thinlines
\put(4,6.35)
 {\framebox(4.25,.8){}}
\thicklines
\put(4.25,6.6)
 {\SZFIG $\UPSTKS{\Gamma}{k}{s}
           = \sum_{\ell, m} \XITKSLM{\Gamma}{k}{s}{\ell}{m}
                            \YXKLM{X}{k}{\ell}{m}$}
\put(0.5,3.25)
 {\circle{0.8}
\makebox(0,0)
 {\SZFIG$K$}}
\put(0.7,3.6)
 {\vector(2,3){0.9}}
\put(0.9,3.2)
 {\vector(1,0){0.7}}
\put(0.7,2.8)
 {\vector(2,-3){0.9}}
\put(1.6,5)
 {\dashbox{0.05}(0.8,0.8)
   {\SZFIG$< 0$}}
\put(2.75,5.25)
 {\SZFIG numerical~\cite{cornish99,glluw}}
\put(1.6,2.85)
 {\dashbox{0.05}(0.8,0.8)
   {\SZFIG$= 0$}}
\put(2.75,3.1)
 {\SZFIG analytical~\cite{lachieze95,levin02}\ldots}
\put(1.6,0.6)
 {\dashbox{0.05}(0.8,0.8)
   {\SZFIG$> 0$}}
\put(2.75,1.2)
 {\SZFIG general case: numerical~\cite{glluw}}
\put(2.75,0.5)
 {\SZFIG lens and prism: analytical~\cite{luw02,glluw}}
\put(7.1,0.15)
 {(see Sec.~\ref{subsubsec_lens})}
\put(0,10)
 {\framebox(9.5,8){}}
\put(0.25,17.5)
 {\SZFIG I: Universal cover ($X$) eigenmodes}
\thinlines
\put(0.1,16.5)
 {\framebox(9.25,.75){}}
\thicklines
\put(0.25,16.75)
 {$\Delta \YXKLM{X}{k}{\ell}{m} = - (k^2 - K) \YXKLM{X}{k}{\ell}{m}
    \;;\quad
    \YXKLM{X}{k}{\ell}{m}
     = \RXKL{X}{k}{\ell} (\CHU) \YLM{\ell}{m} (\theta, \varphi)$}
\put(0.5,13.25)
 {\circle{0.8}
    \makebox(0,0){$K$}}
\put(0.7,13.6)
 {\vector(2,3){0.9}}
\put(0.9,13.2)
 {\vector(1,0){0.7}}
\put(0.7,12.8)
 {\vector(2,-3){0.9}}
\put(1.6,15)
 {\dashbox{0.05}(0.8,0.8)
   {$< 0$}}
\put(2.75,15.25)
 {$X = \HTR:\,
   \RXKL{X}{k}{\ell}
    = \sqrt{\frac{N_{k \ell}}{k \SK{K} (\CHU)}}
      \PLM{- 1 / 2 + i \omega}{-1 / 2 - \ell}
        \left(\cosh\CHN \right)$}
\put(4.25,14.6)
  {$\omega \in[0,\infty[\quad\hbox{or}\quad
    i \omega \in[0,1]$}
\put(1.6,12.85)
 {\dashbox{0.05}(0.8,0.8)
   {$0$}}
\put(2.75,13.1)
 {$X = \RTR:\,
   \RXKL{X}{k}{\ell}
    = \sqrt{\frac{2}{\pi}} j_\ell(k \CHU)$}
\put(4.25,12.45)
  {$k\in[0,\infty[$}
\put(1.6,10.6)
 {\dashbox{0.05}(0.8,0.8)
   {$> 0$}}
\put(2.75,10.85)
 {$X = \STR:\,
   \RXKL{X}{k}{\ell}
     = \sqrt{\frac{M_{k \ell}}{k \SK{K} (\CHU)}}
       \PLM{1 / 2 + \nu}{- 1 / 2 - \ell}
         \left(\cos\CHN \right)$}
\put(4.25,10.2)
  {$\nu=2,3,\ldots$}
\put(7.5,10.15)
 {(see  App.~\ref{app_A})}
\put(11,8.5)
 {\framebox(7,1){}}
\put(11.25,8.9)
 {\SZFIG V: Output: $\ALM{\ell}{m}$, maps, $C_\ell$\ldots}
\put(16.1,8.65)
 {(see Sec.~\ref{sec_simulation})}
\put(11,10)
 {\framebox(7,8){}}
\put(11.25,17.5)
 {\SZFIG II: CMB in $X$}
\thinlines
\put(11.25,16.5)
 {Perturbation equations}
\put(11.8,16.2)
 {(local physics)}
\put(15.25,16.5)
 {Initial conditions}
\put(12,15)
 {\SZFIG $O_k^{[X]} \left(\RXKL{X}{k}{\ell}\right)$}
\put(12.25,14)
 {\SZFIG $G_\ell(k)$}
\put(15.9,14)
 {\SZFIG $\PS (k)$}
\put(12.5,12)
 {\SZFIG $\left<\ALM{\ell}{m} \ALM{\ell'}{m'}^* \right>
           = C_\ell \KRON{\ell}{\ell'} \KRON{m}{m'}, $}
\put(12.5,11)
 {\SZFIG $C_\ell = \frac{2}{\pi} \int k^2 \ddd k G_\ell^2(k) \PS (k)$}
\put(15.9,10.15)
 {(see Sec.~\ref{subsec_local})}
\thicklines
\put(12.25,10.75)
 {\dashbox{0.05}(4.75,1.75){}}
\put(12.8,16){\vector(0,-1){.6}}
\put(12.8,14.8){\vector(0,-1){.5}}
\put(16.2,16.3){\vector(0,-1){2}}
\put(13.2,14.15){\framebox(2.6,0){}}
\put(14.5,14.15){\vector(0,-1){1.6}}

\put(11,0)
 {\framebox(7,8){}}
\put(11.25,7.5)
 {\SZFIG IV: CMB in $X / \Gamma$}
\put(15.9,0.15)
 {(see Sec.~\ref{subsec_implement})}
\put(12.75,6.5)
 {\SZFIG $\left<\ALM{\ell}{m} \ALM{\ell'}{m'}^* \right>
           = \CLMLPMP{\ell}{m}{\ell'}{m'} ,$}
\put(11.75,5.5)
 {\SZFIG $\CLMLPMP{\ell}{m}{\ell'}{m'}
           = \frac{2}{\pi}
             \sum_{k, s} \XITKSLM{\Gamma}{k}{s}{\ell}{m}
                         \XITKSLM{\Gamma}{k}{s \; *}{\ell'}{m'}$}
\put(14.14,4.9)
 {\SZFIG $\quad G_\ell(k) G_{\ell'}(k) \PS (k)$}
\put(11.6,4.65)
 {\dashbox{0.05}(5.9,2.4){}}
\thinlines
\put(11.4,1)
 {Perturbation equations}
\put(11.15,0.7)
 {(local physics, unchanged)}
\put(15.25,1)
 {Initial conditions}
\put(15.5,0.7)
 {(unchanged)}
\put(12,2)
 {\SZFIG $O_k^{[X]} \left(\UPSTKS{\Gamma}{k}{s}\right)$}
\put(12.25,3.5)
 {\SZFIG $G_\ell(k)$}
\put(15.9,3.5)
 {\SZFIG $\PS (k)$}
\thicklines
\put(12.7,1.3){\vector(0,1){.6}}
\put(12.7,2.4){\vector(0,1){1}}
\put(16.2,1.3){\vector(0,1){2.05}}
\put(13.2,3.6){\framebox(2.6,0){}}
\put(14.5,3.6){\vector(0,1){1.0}}
\put(4.75,10)
 {\vector(0,-1){2}}
\put(9.5,14)
 {\vector(1,0){1.5}}
\put(9.5,4)
 {\vector(1,0){1.5}}
\put(14.5,10)
 {\vector(0,-1){0.5}}
\put(14.5,8)
 {\vector(0,1){0.5}}
\end{picture}
\caption{The computation of CMB anisotropies in multi-connected
spaces follows a series of independent steps. From the knowledge of
the spatial curvature $K$, one determines the universal covering
space, $X$, as well as the eigenmodes of the Laplacian in this
simply-connected space (upper left box); these functions are
well-known and frequently used in standard cosmology computations, and
are recalled in Appendix~\ref{app_A} [we have introduced $\omega = k /
\sqrt{|K|}$ for the hyperbolic case and $\nu = k / \sqrt{K} - 1$ for
the spherical case [recall Eq.~(\ref{knu})], and used $\CHN$ defined
by Eq.~(\ref{def_chn})]. Once some cosmological parameters and a
scenario of structure formation has been chosen (upper right box), the
CMB anisotropies in the universal covering space can be computed. This
step is also a standard step that can be achieved numerically by a
number of codes (see Sec.~\ref{subsec_local}). An independent
computation (lower left box) is the determination of the eigenmodes of
the Laplacian compatible with the topology of space, specified by the
choice of the holonomy group $\Gamma$. Our approach is to encode all
the topological information in a set of parameters
$\XITKSLM{\Gamma}{k}{s}{\ell}{m}$. Their computation is described in
Sec.~\ref{subsubsec_lens} and can be performed either numerically or
analytically according to the case at hand. The implementation of the
topology in a standard CMB code (lower right box) is described in
Sec.~\ref{subsec_implement} and yields the complete correlation matrix
of the $\ALM{\ell}{m}$ from which one can (center right box) compute
$C_\ell$, simulate maps, etc.}
\label{method}
\end{figure}
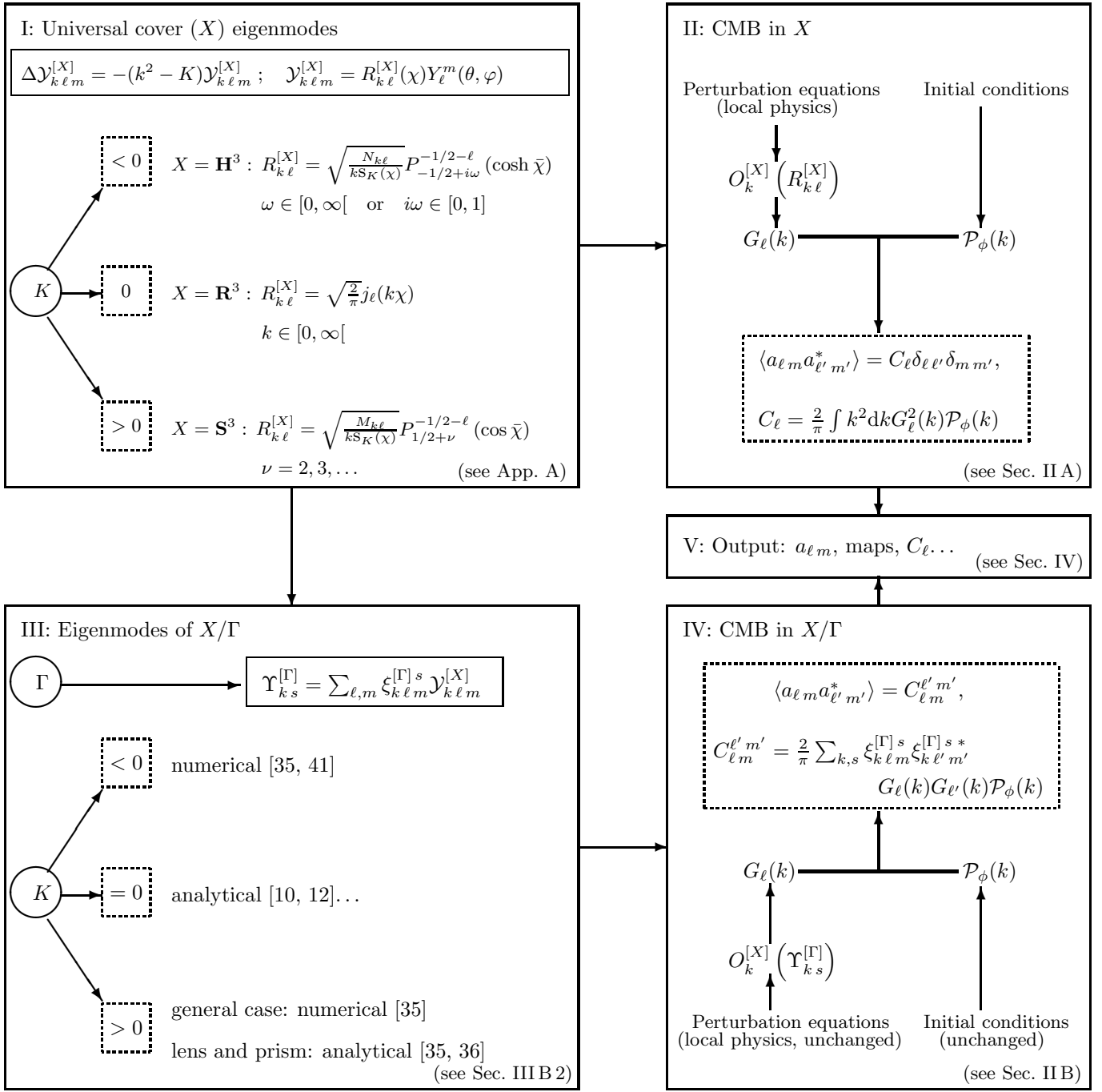
\end{center}

\section{CMB anisotropies}
\label{sec_cmb}

The equations dictating the evolution of cosmological
perturbations are local differential equations, and remain
unchanged by the topology of the spatial sections. Indeed the goal
of this section is not to derive the whole set of equations that
has to be solved (see, e.g., Refs.~\cite{cmb1,cmb2,cmb3,cmb4} for
reviews) but to explain the steps that must be changed to take
into account the topology.

\subsection{Local physics of cosmological perturbations}
\label{subsec_local}

The CMB is observed to a high accuracy as black-body radiation with
temperature $T_0 = 2.726 \pm 0.002 \UUNIT{K}{}$~\cite{fixsen}, almost
independently of the direction.  After accounting for the peculiar
motion of the Sun and Earth, the CMB has tiny temperature fluctuations
of order $\delta T / T_0 \sim 10^{- 5}$ that are usually decomposed in
terms of spherical harmonics
\begin{equation}
\label{dttalm}
\frac{\delta T}{T_0} (\theta, \varphi)
  = \sum_{\ell = 0}^\infty \sum_{m = - \ell}^\ell
    \ALM{\ell}{m}^\OBS \YLM{\ell}{m} (\theta, \varphi) .
\end{equation}
This relation can be inverted by using the orthonormality of the
spherical harmonics to get
\begin{equation}
\label{p4}
\ALM{\ell}{m}
 = \int \frac{\delta T}{T} \YLM{\ell}{m}{}^*
        \sin \theta \ddd \theta \, \ddd \varphi .
\end{equation}
The coefficients $\ALM{\ell}{m}$ obviously satisfy the conjugation
relation
\begin{equation}
\label{almetoile}
\ALM{\ell}{- m} = (- 1)^m \ALM{\ell}{m}^* .
\end{equation}
The angular correlation function of these temperature anisotropies
is observed on a $2$-sphere around us and can be decomposed on a
basis of the Legendre polynomials $P_\ell$ as
\begin{equation}
\label{dTT}
\left< \frac{\delta T}{T}(\hat \gamma_1)
       \frac{\delta T}{T}(\hat \gamma_2)
\right>_{\hat \gamma_1 . \hat \gamma_2 = \cos \theta_{12}}
 = C^\OBS (\theta_{12})
 = \frac{1}{4 \pi}
   \sum_\ell (2 \ell + 1) C_\ell^\OBS P_\ell(\cos \theta_{12}),
\end{equation}
where the brackets stand for an average on the sky, i.e., on all
pairs of directions $(\hat \gamma_1, \hat \gamma_2)$ subtending an
angle $\theta_{12}$. The coefficients $C_\ell^\OBS$ of the
development of $C^\OBS (\theta_{12})$ on the Legendre polynomials
are thus given by
\begin{equation}
C_\ell^\OBS
 = \frac{1}{2 \ell + 1}
   \sum_{m = - \ell}^\ell
   \left \langle \ALM{\ell}{m}^\OBS \ALM{\ell}{m}^\OBS {}^* \right\rangle .
\end{equation}
These $C_\ell^\OBS$ can be seen as estimators of the variance of the
$\ALM{\ell}{m}$\footnote{One can show that this is indeed the best
estimator of their variance when the fluctuations are isotropic and
Gaussian~\cite{grrr}.} and represent the rotationally invariant
angular power spectrum. They have therefore to be compared to the
values $C_\ell$ predicted by a given cosmological model, which is
specified by (i) a model of structure formation which fixes the
initial conditions for the perturbation (e.g., inflation, topological
defects, etc), (ii) the geometry and matter content of the universe
(via the cosmological parameters) and (iii) the topology of the
universe.

For the case of a simply connected topology, the $C_\ell$ are usually
computed as follows.  First, one starts from a set of initial
conditions given by an early universe scenario. Typically these are
the initial conditions predicted in the framework of an inflationary
model, although this is of no importance in the discussion that
follows. Second, the modes of cosmological interest are evolved from
an epoch before they enter into the Hubble radius till now.  Third,
one computes the variance for the $\ALM{\ell}{m}$. The details of this
procedure are well known~\cite{cmb1,cmb2,cmb3,cmb4}.

For simplicity let us sketch the case of a flat universe. The
temperature fluctuation in a given direction of the sky can be related
to (i) the eigenmodes $\exp (i \bk \cdot \bx)$ of the Laplacian (here
$\bx$ represents a vector in the usual Cartesian coordinate system) by
a linear convolution operator $O_k^{[\RTR]} \left(\exp (i \bk \cdot
\bx) \right)$ depending on the modulus $k$ only [as well as on the
universal cover (here, $\RTR$) and the cosmological parameters], and
(ii) a three-dimensional variable $\hat e_\bk$ related to the initial
conditions, by the formula
\begin{equation}
\frac{\delta T}{T}(\theta, \varphi)
 = \int \frac{\ddd^3 \bk}{(2 \pi)^{3 / 2}}
        O_k^{[\RTR]} \left(\ee^{i\bk\cdot\bx} \right)
        \sqrt{\PS (k)} \hat e_\bk ,
\end{equation}
where $\PS (k)$ is the gravitational initial power spectrum,
normalized so that $\PS (k) \propto k^{- 3}$ for a
Harrison-Zel'dovich spectrum, and where in most inflationary
models the random variable $\hat e_\bk$ describes a Gaussian
random field and satisfies
\begin{equation}
\left<\hat e_\bk \hat e_{\bk'}^* \right> = \DIR{\bk - \bk'}
\end{equation}
with $\hat e_{-\bk}=\hat e_{\bk}^*$ when the space is simply
connected. This relation stems from the fact that the temperature
fluctuation is real, but it may be expressed differently for
multi-connected spaces.  Decomposing the exponential by means of
Eq.~(\ref{C7}) and using Eq.~(\ref{C5}) allows to rewrite the
temperature fluctuation as
\begin{eqnarray}
\frac{\delta T}{T}(\theta, \varphi)
 & = & \sum_{\ell, m} i^\ell
      \int k^2 \ddd k \sqrt{\PS (k)}\,
      O_k^{[\RTR]} \left(\sqrt{\frac{2}{\pi}} j_\ell(k \CHU) \right)
      \left[\int \ddd \Omega_\bk \YLM{\ell}{m}{}^* (\theta_\bk, \varphi_\bk)
            \hat e_\bk \right] \YLM{\ell}{m} (\theta, \varphi)
\nonumber \\ \label{p44b} & = &
       \sum_{\ell, m} i^\ell \int k^2 \ddd k \sqrt{\PS (k)}\,
       O_k^{[\RTR]} \left(\YXKLM{\RTR}{k}{\ell}{m}\right) \hat e_{\ell m}(k) ,
\end{eqnarray}
where we have defined
\begin{equation}
\hat e_{\ell m}(k)
 \equiv \int \ddd \Omega_\bk
             \YLM{\ell}{m}{}^* (\theta_\bk, \varphi_\bk) \hat e_\bk ,
\end{equation}
which is the ``average'' of the random fields $\hat e_\bk$ over all
the $\bk$ of the same modulus. This quantity can therefore be
identified as a two-dimensional Gaussian random variable satisfying
$\left< \hat e_{\ell m}(k) \hat e_{\ell'm'}^*(k') \right> = \DIR{k -
k'} \KRON{\ell}{\ell'} \KRON{m}{m'} / k^2$.  It follows that the
coefficients $\ALM{\ell}{m}$ take the general form
\begin{equation}
\label{p4_2}
\ALM{\ell}{m}
 = i^\ell \int k^2 \ddd k \sqrt{\PS (k)}
               G_\ell (k) \hat{e}_{\ell m} (k) ,
\end{equation}
with
\begin{equation}
G_\ell (k) = O_k^{[\RTR]} \left(\RXKL{\RTR}{k}{\ell}\right) ,
\end{equation}
and here the function $G_\ell (k)$ can be approximated by (see, e.g.,
\cite{abott86,cmb4})
\begin{eqnarray}
\label{swdopisw}
G_\ell (k)
 & = &   j_\ell (k (\eta_0 - \eta_\LSS))
         \left(  \frac{\delta T}{T} (k, \eta_\LSS)
               + \Phi (k, \eta_\LSS) + \Psi (k, \eta_\LSS)
         \right)
       + j'_\ell (k (\eta_0 - \eta_\LSS)) \frac{v_\BAR (k, \eta_\LSS)}{k}
\nonumber \\ & &
       + \int_{\eta_\LSS}^{\eta_0} j_\ell (k (\eta_0 - \eta))
         \left(\dot \Phi (k, \eta) + \dot \Psi (k, \eta) \right) \; \ddd \eta ,
\end{eqnarray}
which is indeed a linear convolution operator acting on
$\RXKL{\RTR}{k}{\ell}$ as announced above.  Here, $\eta_\LSS$ and
$\eta_0$ are the conformal time at the last scattering epoch and
today, respectively, $j_\ell$ is the spherical Bessel function of
index $\ell$, $\Phi$ and $\Psi$ are the two Bardeen potentials, and
$v_\BAR$ is the velocity divergence of the baryons. The only
modification of note when one considers a non flat universe is that
the $j_\ell$ are to be replaced by their analog for non flat
geometries, the so-called ultraspherical Bessel
functions~\cite{cmb5,abott86}, and for a closed universe the integral
over $k$ is replaced by a discrete sum.

In conclusion, and without loss of generality, the temperature
fluctuation can be decomposed as in Eq.~(\ref{p44b}) whatever the
curvature of space. For a simply connected topology and Gaussian
initial conditions the addition property (\ref{C5}) of the spherical
harmonics imposes that
\begin{equation}
\label{corr_topo_triv}
\left< \ALM{\ell}{m} \ALM{\ell'}{m'}^* \right>
 = C_\ell \KRON{\ell}{\ell'} \KRON{m}{m'} ,
\end{equation}
and therefore the $C_\ell$ coefficients encode all the information
regarding the CMB anisotropies.

\subsection{Implementing the topology}
\label{subsec_implement}

The topology does not affect local physics, so the equations
describing the evolution of the cosmological perturbations are
left unchanged. As a consequence, quantities such as the Bardeen
potentials $\Phi$, $\Psi$, etc., are computed in the same way as
described above, and the operator $O_k^{[X]}$ is therefore the
same. However, a change of topology translates into a change of
the modes that can exist in the universe. In particular, the
functions $\YXKLM{X}{k}{\ell}{m}$ are typically not well defined
on the quotient space $X/\Gamma$. Therefore, the only change that
has to be performed is the substitution
\begin{equation}
\YXKLM{X}{k}{\ell}{m} \to \UPSTKS{\Gamma}{k}{s} ,
\end{equation}
where the $\UPSTKS{\Gamma}{k}{s}$ form an orthonormal basis for the
space of eigenmodes of the Laplacian on the given topology $X/\Gamma$.
One must then remember that the mode functions $\UPSTKS{\Gamma}{k}{s}$
can be decomposed uniquely by Eq.~(\ref{eq:1}) and that the
convolution operator $O_k^{[X]}$ is linear.  When the multi-connected
topology is compact, it follows that Eq.~(\ref{p44b}) will take the
form
\begin{equation}
\label{p44t}
\frac{\delta T}{T} (\theta, \varphi)
 = \frac{(2\pi)^3}{V}
   \sum_{k, s} O_k^{[X]} \left(\UPSTKS{\Gamma}{k}{s}\right)
              \sqrt{\PS (k)} \hat{e}_\bk ,
\end{equation}
where now $\hat{e}_\bk$ is a three-dimensional random variable which
is related to the discrete mode $\bk$. Equivalently one can write
$\hat{e}_\bk \equiv \hat{e}_{k\,s}$ where $k$ is the modulus of $\bk$
and the index $s$ describes all the eigenmodes of the Laplacian for
fixed modulus $k$ in the topology $X / \Gamma$ of volume $V$. These
random variables satisfy the normalization
\begin{equation}
\left< \hat{e}_\bk \hat{e}_{\bk'}^* \right>
 = \frac{V}{(2 \pi)^3} \KRON{k}{k'} \KRON{s}{s'} .
\end{equation}
For a given value of the wavenumber $k$, there are fewer
eigenmodes in the multi-connected case, so that $s$ has to
be seen as a ``subset''  of the set $\{\ell, m\}$.

By inserting the expansion of $\UPSTKS{\Gamma}{k}{s}$ in terms of the
covering space eigenmodes, as given by Eq.~(\ref{eq:1}), we obtain
\begin{equation}
\label{p44tbis}
\frac{\delta T}{T}(\theta, \varphi)
 = \frac{(2 \pi)^3}{V}
   \sum_{k, s} \sum_{\ell, m}
                    \XITKSLM{\Gamma}{k}{s}{\ell}{m}
                     O_k^{[X]} \left(\YXKLM{X}{k}{\ell}{m}\right)
               \sqrt{\PS (k)} \hat{e}_\bk .
\end{equation}
It follows that the $\ALM{\ell}{m}$, seen as random variables, are given
by
\begin{equation}
\label{p44t2}
\ALM{\ell}{m}
 = \frac{(2 \pi)^3}{V}
   \sum_k \sqrt{\PS (k)} O_k^{[X]} \left (\RXKL{X}{k}{\ell} \right)
          \sum_s \XITKSLM{\Gamma}{k}{s}{\ell}{m} \hat{e}_\bk.
\end{equation}
Note that the sum over $s$ is analogous to the sum over angles
defining the two-dimensional random variable $\hat e_{\ell m}$ in
Eq.~(\ref{p44b}). Since the $\ALM{\ell}{m}$ are linear functions of
the initial three-dimensional random variables, they are still
Gaussian distributed but they are not independent anymore (as
explained before, this is the consequence of the breakdown of global
isotropy and/or homogeneity). The correlation between the coefficients
$\ALM{\ell}{m}$ is given by
\begin{eqnarray}
\label{corr_mat}
\left< \ALM{\ell}{m} \ALM{\ell'}{m'}^* \right>
 = \frac{(2 \pi)^3}{V}
   \sum_k \PS (k) O_k^{[X]} \left(\RXKL{X}{k}{\ell} \right)
                  O_k^{[X]} \left(\RXKL{X}{k}{\ell'} \right)
          \sum_s \XITKSLM{\Gamma}{k}{s}{\ell}{m}
                 \XITKSLM{\Gamma}{k}{s \; *}{\ell'}{m'} .
\end{eqnarray}
Clearly these correlations can have non-zero off-diagonal terms,
reflecting the global anisotropy induced by the multi-connected
topology, so that Eq.~(\ref{corr_topo_triv}) no longer holds and the
observational consequences of a given topology on the CMB anisotropies
are given by the correlation matrix (\ref{eq:I:1}).  This means in
particular that for fixed $\ell$, the $\ALM{\ell}{m}$ might not have
the same variance, although they all follow Gaussian statistics as
long as the initial conditions do. This translates into an apparent
non Gaussianity in the sense that the $C_\ell$ will not follow the
usual $\chi^2$ distribution.  Strictly speaking, this is not a
signature of non Gaussianity but of anisotropy.

Note also that the correlation matrix~(\ref{corr_mat}) is not rotation
invariant. It will explicitly depend on the orientation of the
manifold with respect of the coordinate system. However, knowing how
the spherical harmonics transform under a rotation allows us to
compute the correlation matrix under any other orientation of the
coordinate system. To finish let us note that one can define the usual
$C_\ell$ coefficients in any topology by the formula
\begin{equation}
C_\ell \equiv \frac{1}{2\ell + 1} \sum_m \CLMLPMP{\ell}{m}{\ell}{m} ,
\end{equation}
which is easily shown to be rotationally invariant.

\section{Eigenmodes of multi-connected spaces}
\label{sec_modes}

\subsection{Flat spaces}
\label{subsec_flat}

The purpose of this subsection is to compute in detail the
coefficients $\XITKSLM{\Gamma}{k}{s}{\ell}{m}$ in the case of a
cubic 3-torus of comoving size $L$, referred to as $T_1$. The
method generalized easily to any compact flat manifold.

In the case at hand, the topology implies the ``quantization'' of
the allowed wave vectors
\begin{equation}
\label{eq:bk}
\bk = \frac{2 \pi}{L}
      \left(n_1\hat{\bf x} + n_2 \hat {\bf y} + n_3 \hat {\bf z} \right)
    = \frac{2 \pi}{L} \bn
    = k \bhn ,
\end{equation}
with
\begin{eqnarray}
\bn & \equiv & (n_1, n_2, n_3) , \\
n & \equiv & \sqrt{\bn\cdot\bn} , \\
\bhn & \equiv & (n_1, n_2, n_3) / n , \\
k & \equiv & \sqrt{\bk\cdot\bk} = \frac{2 \pi}{L} n ,
\end{eqnarray}
so that the label $s$ can be chosen to be the ``unit'' integer vector
$\bhn$. The label multiplicity ${\rm mult}(k) \equiv {\rm card} \{s\}$
of a mode $k$ is given in this case by the number of representations
of $n^2$ by 3 squares, i.e., by ${\rm card} \{\bhn\}$ (see
Fig.~\ref{figmult} below). The corresponding normalized eigenmodes in
Cartesian coordinates are thus simply given by
\begin{equation}
\label{xit1}
\UPSTKS{T_1}{k}{\bhn} ({\bf r})
 = \frac{\ee^{i\bk\cdot{\bf r}}}{(2 \pi)^{3 / 2}} ,
\end{equation}
with $\bk$ given by Eq.~(\ref{eq:bk}). Using the
decomposition~(\ref{C7}) of the exponential and plugging in the
closure relation~(\ref{C5}), one gets that
\begin{equation}
\XITKSLM{T_1}{k}{s}{\ell}{m}
 = \XITKSLM{T_1}{2 \pi n / L}{\bhn}{\ell}{m}
 = i^\ell \YLM{\ell}{m}{}^*(\bhn) ,
\end{equation}
where $\bhn$ can also be defined by the two spherical angles
$(\theta_{\bn}$, $\varphi_{\bn})$ which are explicitly given by
\begin{eqnarray}
\tan \theta_\bn
 & = & \frac{\sqrt{n_1^2 + n_2^2}}{n_3} , \\
\tan \varphi_\bn
 & = & \frac{n_2}{n_1} .
\end{eqnarray}
This expression could also have been obtained by simply
considering the Fourier transform of $\YXKLM{\RTR}{k}{\ell}{m}$ as
given by Eq.~(\ref{C6}).  One can check that the normalization of
the basis $\UPSTKS{T_1}{k}{\bhn}$ [i.e., $\int
\UPSTKS{T_1}{k}{\bhn} \UPSTKS{T_1}{k'}{\bhn'}{}^* \ddd^3 \bx =
\DIR{\bk - \bk'}$] implies that the coefficients
$\XITKSLM{T_1}{k}{\bhn}{\ell}{m}$ satisfy the closure relation
\begin{equation}
\label{closure_flat}
\sum_{\ell = 0}^{\infty} \sum_{m = - \ell}^{\ell}
      \XITKSLM{T_1}{k}{\bhn}{\ell}{m}\, \XITKSLM{T_1}{k'}{\bhn'\;*}{\ell}{m}
 = \DIR{\cos \theta_\bn - \cos \theta_{\bn'}}
   \DIR{\varphi_\bn - \varphi_{\bn'}}
   \KRON{k}{k'} .
\end{equation}
The Dirac distribution in the above expression can be shown, by
using either Eqs.~(\ref{C4},\ref{C5}) or Eq.~(\ref{C8}) alone, to
be
\begin{equation}
\DIR{\cos\theta_\bn - \cos\theta_{\bn'}}
\DIR{\varphi_\bn - \varphi_{\bn'}}
 = \sum_{\ell = 0}^\infty \frac{2 \ell + 1}{4 \pi} P_\ell (\bhn . \bhn') .
\end{equation}
>From these results, we deduce that the correlation matrix of the
$\ALM{\ell}{m}$ is given by
\begin{equation}
\label{sum30}
\CLMLPMP{\ell}{m}{\ell'}{m'}
 = \left( \frac{2 \pi}{L} \right)^3
   \sum_{n} i^{\ell - \ell'} \PS \left(\frac{2\pi n}{L} \right)
            G_\ell \left(\frac{2 \pi n}{L} \right)
            G_{\ell'} \left(\frac{2 \pi n}{L} \right)
            \sum_\bhn \YLM{\ell}{m}{}^* (\bhn) \YLM{\ell'}{m'} (\bhn) .
\end{equation}
Using Eq.~(\ref{C5}) and the fact that $\sum_{n, \bhn} =
\sum_{\bn}$, the $C_\ell$ coefficients are simply
\begin{equation}
C_{\ell}
 = \frac{1}{2 \pi} \left(\frac{2 \pi}{L} \right)^3
   \sum_{\bn} \PS \left(\frac{2 \pi n}{L} \right)
              G_\ell^2\left(\frac{2 \pi n}{L}\right) ,
\end{equation}
which was used in many earlier
works~\cite{sokolov93,staro93,stevens93,costa95} on the influence of
topology on the CMB.

Note that spherical harmonics satisfy the following symmetry relation
\begin{equation}
\YLM{\ell}{m} (\theta, - \varphi)
 = \ee^{- 2 i m \varphi} \YLM{\ell}{m}(\theta, \varphi) .
\end{equation}
Due to the symmetry of the torus with respect to the $y = 0$
plane, in the sum over $\bhn$ in Eq.~(\ref{sum30}) a term
$(\theta_\bn, \varphi_\bn)$ will always be associated to a term
$(\theta_\bn, - \varphi_\bn)$ leading to a term of the form $4
\cos \left( (m' - m) \varphi_\bn \right) \YLM{\ell}{m}{}^*
(\theta_\bn, 0) \YLM{\ell'}{m'} (\theta_\bn, 0)$, the only
exception being the term arising when $n_2 = 0$, which is real.
>From this result, one easily shows that the correlation matrix
satisfies
\begin{eqnarray}
\label{rel}
\CLMLPMP{\ell}{m}{\ell'}{m'} \in {\bb R} .
\end{eqnarray}
Along similar lines, the relations
\begin{eqnarray}
\YLM{\ell}{m} (\theta, \pi)
 & = & \hbox{e}^{im\varphi} \YLM{\ell}{m}(\theta, 0) , \\
\YLM{\ell}{m} (\pi - \theta, \varphi)
 & =& (-1)^{\ell + m} \YLM{\ell}{m} (\theta, \varphi) ,
\end{eqnarray}
imply
\begin{equation}
\CLMLPMP{\ell}{m}{\ell'}{m'}
 = \frac{1}{4} \left[1 + (-1)^{m - m'} \right]
               \left[1 + (-1)^{\ell - \ell'} \right]
   \CLMLPMP{\ell}{m}{\ell'}{m'} ,
\end{equation}
so that
\begin{equation}
\label{toregen}
\CLMLPMP{\ell}{m}{\ell'}{m'} \neq 0
\quad \Rightarrow \quad
m - m' \equiv 0 \MOD{2}
\quad{\rm and}\quad
\ell - \ell' \equiv 0 \MOD{2} .
\end{equation}
Furthermore Eqs.(\ref{rel},\ref{C2}) imply
\begin{equation}
\label{toregen2}
\CLMLPMP{\ell}{m}{\ell'}{m'} = \CLMLPMP{\ell}{- m}{\ell'}{- m'} .
\end{equation}
Let us emphasize that these properties of the correlation matrix
still hold even if the torus is not cubic. However, a cubic torus
is invariant under a $\pi / 2$-rotation about the $z$ axis, so if
$(n_1, n_2, n_3)$ corresponds to a wavenumber then so does $(n_2,
- n_1, n_3)$, and one has
\begin{equation}
\label{torecub}
\CLMLPMP{\ell}{m}{\ell'}{m'} \neq 0
\quad \Rightarrow \quad
m - m' \equiv 0 \MOD{4} .
\end{equation}

\subsection{Spherical spaces}
\label{subsec_spherical}

The goal of this section is to recall the basic analytical results
concerning the lens and prism spaces (Sec.~\ref{subsubsec_lens}).  In
these spaces, the eigenmodes and eigenvalues of the Laplacian operator
can be determined analytically using toroidal
coordinates~\cite{luw02}. CMB computations use spherical coordinates,
so we must perform a change of coordinates and a change of basis
(detailed in Appendix~\ref{app_B}). Fortunately, this can also be
achieved analytically to compute the coefficients
$\XITKSLM{\Gamma}{k}{s}{\ell}{m}$.

\subsubsection{Generalities}
\label{subsubsec_general}

In our preceding article~\cite{glluw}, we presented in a pedestrian
way the complete classification of three-dimensional spherical
topologies and we described how to compute their holonomy
transformations.

The isometry group of the $3$-sphere is $\GRP{SO}{4}$. Every isometry
in $\GRP{SO}{4}$ can be decomposed as the product of a right-handed
and a left-handed Clifford translation, and the factorization is
unique up to simultaneous multiplication of both factors by -1.
Furthermore, the space $\STR$ itself enjoys a group structure as the
set ${\cal S}^3$ of unit length quaternions. Each right-handed (resp.\
left-handed) Clifford translation corresponds to left (resp.\ right)
quaternion multiplication of ${\cal S}^3$, so the group of
right-handed (resp. left-handed) Clifford translations is isomorphic
to ${\cal S}^3$. It follows that $\GRP{SO}{4}$ is isomorphic to ${\cal
S}^3 \times {\cal S}^3 / \lbrace \pm({\bf1}, {\bf1}) \rbrace$ and thus
the classification of the subgroups of $\GRP{SO}{4}$ can be deduced
from the classification of subgroups of ${\cal S}^3$.  There is a
two-to-one homomorphism from ${\cal S}^3$ to $\GRP{SO}{3}$; the finite
subgroups of $\GRP{SO}{3}$ are the cyclic, dihedral, tetrahedral,
octahedral and icosahedral groups, so the finite subgroups of ${\cal
S}^3$ are their lifts, namely
\begin{itemize}

\item the cyclic groups $Z_n$ of order $n$,

\item the binary dihedral groups $D_m^*$ of order $4m$, $m \ge 2$,

\item the binary tetrahedral group $T^*$ of order 24,

\item the binary octahedral group $O^*$ of order 48,

\item the binary icosahedral group $I^*$ of order 120,

\end{itemize}
where a binary group is the two-fold cover of the corresponding
plain group.

>From this classification, it can be shown that there are three
categories of spherical 3-manifolds.
\begin{itemize}

\item The {\it single action manifolds} are those for which a subgroup
$R$ (resp. $L$) of ${\cal S}^3$ acts as pure right-handed (resp.
pure left-handed) Clifford translations. They are thus the
simplest spherical manifolds and can all be written as $\STR /
\Gamma$ with $\Gamma = Z_n, D_m^*, T^*, O^*, I^*$.

\item The {\it double-action manifolds} are those for which subgroups
$R$ and $L$ of ${\cal S}^3$ act simultaneously as right- and
left-handed Clifford translations, and every element of $R$ occurs
with every element of $L$. There are obtained for the groups $\Gamma =
\Gamma_1 \times \Gamma_2$ with $(\Gamma_1, \Gamma_2) = (Z_m, Z_n)$,
$(D^*_m, Z_n)$, $(T^*, Z_n)$, $(O^*, Z_n)$, $(I^*, Z_n)$, with $\GCD
(m, n) = 1$, $\GCD (4 m, n) = 1$, $\GCD (24, n) = 1$, $\GCD(48, n)=1$,
and $\GCD(120, n) = 1$, respectively.

\item The {\it linked-action manifolds} are similar to the double
action manifolds, except that each element of $R$ occurs with only
some of the elements of $L$.

\end{itemize}
The classification of these manifolds is summarized in Fig.~8 of
Ref.~\cite{glluw}.

\subsubsection{Lens and prism spaces}
\label{subsubsec_lens}

In this article, we focus on prism spaces $\STR / D_m^*$ and lens
spaces $L(p, q)$ . The latter are defined by identifying the lower
surface of a lens-shaped solid to the upper surface with a
rotation of angle $ 2 \pi q / p$ for $p$ and $q$ relatively prime
integers with $0 < q < p$. Furthermore, we may restrict our
attention to $0 < q \leq p / 2$ because for values of $q$ in the
range $p / 2 < q < p$ the twist $2 \pi q / p$ is the same as $- 2
\pi (p - q) / p$, thus $L(p, q)$ is the mirror image of $L(p, p -
q)$. Lens spaces can be single action, double action, or linked
action;  Fig.~9 of Ref.~\cite{glluw} summarizes their classification.\\

The eigenmodes and eigenvalues of prism and lens spaces can be
obtained analytically by working in toroidal
coordinates~\cite{luw02}. Starting from Cartesian coordinates, in
which the equation for the $3$-sphere is $x^2 + y^2 + z^2 + w^2 = 1$,
the toroidal coordinates $(\chi_\ST, \theta_\ST, \varphi_\ST)$ are
defined via the equations
\begin{eqnarray}
\label{CoordinateDefinition}
x & = & \cos \chi_\ST \, \cos \theta_\ST , \\
y & = & \cos \chi_\ST \, \sin \theta_\ST , \\
z & = & \sin \chi_\ST \, \cos \varphi_\ST , \\
w & = & \sin \chi_\ST \, \sin \varphi_\ST ,
\end{eqnarray}
with
\begin{eqnarray}
0 & \leq & \chi_\ST  \leq \pi / 2 , \\
0 & \leq & \theta_\ST  \leq 2 \pi , \\
0 & \leq & \varphi_\ST  \leq 2 \pi .
\end{eqnarray}
Ref.~\cite{luw02} gives the eigenmodes of $\STR$ explicitly as
\begin{equation}
\label{PsiSolution} \QKLM{\nu}{\ell_\ST}{m_\ST}
 = \BKLM{\nu}{\ell_\ST}{m_\ST}
   \cos^{|\ell_\ST|}\chi_\ST \;
   \sin^{|m_\ST |} \chi_\ST \;
   \JAC{|m_\ST|}{|\ell_\ST|}{d} (\cos 2 \chi_\ST)
   f (|\ell_\ST| \theta_\ST) f (|m_\ST| \varphi_\ST) ,
\end{equation}
where $\nu$ is the integer parameterizing $k = (\nu + 1) \sqrt{K}$
as in Eqn.~(\ref{knu}), $\JAC{|m_\ST|}{|\ell_\ST|}{d}$ is the
Jacobi polynomial, and $f$ stands for the cosine (resp. sine)
function when $\ell_\ST$ or $m_\ST \geq 0$ (resp.\ $\ell_\ST$ or
$m_\ST < 0$). For each value of $\nu$, the indices $\ell_\ST$ and
$m_\ST$ range over all integers satisfying
\begin{eqnarray}
|\ell_\ST | + |m_\ST | & \leq & \nu \qquad {\rm and}\label{Eigenbasis} \\
|\ell_\ST | + |m_\ST | & \equiv & \nu \MOD{2} ,\label{Eigenbasis2}
\end{eqnarray}
and for convenience we define
\begin{equation}
\label{titi}
d = \frac{1}{2} (\nu - |\ell_\ST| - |m_\ST|) .
\end{equation}
The normalization coefficients $\BKLM{\nu}{\ell_\ST}{m_\ST}$ are
given by\footnote{Note the factor $1/(\nu+1)$ which differs from
Ref.~\cite{luw02} due to a different choice of normalization.}
\begin{equation}
\BKLM{\nu}{\ell_\ST}{m_\ST}
 = \frac{\sigma_{\ell_\ST} \sigma_{m_\ST}}{\pi(\nu+1)}
   \sqrt{\frac{2 (\nu + 1)\, d !\, (|\ell_\ST| + |m_\ST| + d) !}
        {(|\ell_\ST| + d) !\, (|m_\ST| + d) !}}
\end{equation}
with $\sigma_{i} = 1 / \sqrt{2}$ if $i = 0$ and $\sigma_{i} = 1$
otherwise.

Using these definitions, Ref.~\cite{luw02} shows that for lens
spaces the explicit set of coefficients
$\ETAKSLM{\Gamma}{s}{\nu}{\ell_\ST}{m_\ST}$ such that
\begin{equation}
\UPSTKS{\Gamma}{k}{s}
 = \sum_{\ell_\ST, m_\ST}
        \ETAKSLM{\Gamma}{\nu}{s}{\ell_\ST}{m_\ST}
        \QKLM{\nu}{\ell_\ST}{m_\ST} (\chi_\ST ,\theta_\ST, \varphi_\ST) ,
\end{equation}
can be obtained as follows:\\

\section*{Theorem 1: lens spaces}

{\it The eigenspace of the Laplacian on the lens space $L(p, q)$ has
an orthonormal basis that, when lifted to $Z_p$-invariant eigenmodes
of the $3$-sphere, comprises those eigenmodes in the left column for
which the corresponding condition in the right column is satisfied,
subject to the restriction that an eigenmode
$\QKLM{\nu}{\ell_\ST}{m_\ST}$ exists if and only if the integers
$\nu$, $\ell_\ST$, and $m_\ST $ satisfy $|\ell_\ST| + |m_\ST| \leq
\nu$ and $|\ell_\ST| + |m_\ST| \equiv \nu \MOD{2}$.}

\begin{center}
\begin{tabular}{|lcc|}
\hline
basis vectors & \qquad\qquad & condition \\
\hline
$\QKLM{\nu}{0}{0}$ & & always \\
$\QKLM{\nu}{\ell_\ST}{0}$ & & $\ell_\ST \equiv 0 \MOD{p}$ \\
$\QKLM{\nu}{0}{m_\ST}$ & & $q m_\ST  \equiv 0 \MOD{p}$ \\
$(\QKLM{\nu}{\ell_\ST}{m_\ST} + \QKLM{\nu}{- \ell_\ST}{- m_\ST}) /
\sqrt{2}, \quad
 (\QKLM{\nu}{- \ell_\ST}{m_\ST} - \QKLM{\nu}{\ell_\ST}{- m_\ST}) / \sqrt{2}$
 & & $\ell_\ST \equiv q m_\ST \MOD{p}$ \\
$(\QKLM{\nu}{\ell_\ST}{m_\ST} - \QKLM{\nu}{- \ell_\ST}{- m_\ST}) /
\sqrt{2}, \quad
 (\QKLM{\nu}{- \ell_\ST}{m_\ST} + \QKLM{\nu}{\ell_\ST}{- m_\ST}) / \sqrt{2}$
 & & $\ell_\ST \equiv - q m_\ST \MOD{p}$ \\
\hline
\end{tabular}
\end{center}
An analogous theorem was demonstrated for prism spaces and can be
found in Ref.~\cite{luw02}.\\

Unfortunately, for practical purposes the eigenmodes of the lens
and prism spaces are needed in spherical coordinates, while they
are most easily obtained in toroidal coordinates. As explained in
the Introduction, one needs the coefficients
$\XITKSLM{\Gamma}{k}{s}{\ell}{m}$ of the decomposition
(\ref{eq:1}). Since $\YXKLM{\STR}{k}{\ell}{m}$ and
$\QKLM{\nu}{\ell_\ST}{m_\ST}$ are two orthogonal bases of
dimension $(\nu + 1)^2$, all of whose elements have the same norm,
there is an orthogonal transformation taking one to the other
\begin{equation}
\label{3} \QKLM{\nu}{\ell_\ST}{m_\ST}
 = \sum_{\ell, m} \ALKLMLTMT{\nu}{\ell}{m}{\ell_\ST}{m_\ST}
                  \YXKLM{\STR}{k}{\ell}{m}.
\end{equation}
The ``transpose'' of this transformation $\alpha$ takes a given
eigenmode's ${\cal Q}$-based coefficients~$\eta$ to its ${\cal Y}$-based
coefficients~$\xi$:
\begin{equation}
\label{4}
\XITKSLM{\Gamma}{k}{s}{\ell}{m}
 = \sum_{\ell_\ST=-\nu}^\nu
     \sum_{m_\ST=-\nu}^\nu
          \ALKLMLTMT{\nu}{\ell}{m}{\ell_\ST}{m_\ST}
          \ETAKSLM{\Gamma}{s}{\nu}{\ell_\ST}{m_\ST} .
\end{equation}
The orthonormality of the basis $\UPSTKS{\Gamma}{k}{s}$ implies
that the coefficients $\XITKSLM{\Gamma}{k}{s}{\ell}{m}$ satisfy
\begin{equation}
\sum_{\ell = 0}^{\nu}
  \sum_{m = - \ell}^\ell
       \XITKSLM{\Gamma}{k}{s}{\ell}{m}
       \XITKSLM{\Gamma}{k}{s' \; *}{\ell}{m}
 = \KRON{s}{s'} .
\end{equation}
This relation is simpler than the closure
relation~(\ref{closure_flat}) obtained in the flat case because, for a
given $k$, the space of modes is finite-dimensional. The computation
of the coefficients $\ALKLMLTMT{\nu}{\ell}{m}{\ell_\ST}{m_\ST}$
appears in Appendix~\ref{app_B}.  Because both the ${\cal Q}$ basis
and the ${\cal Y}$ basis are orthonormal, the transformation $\alpha$
is orthogonal:
\begin{equation}
\sum_{\ell = 0}^{\nu}
  \sum_{m = - \ell}^\ell
       \ALKLMLTMT{\nu}{\ell}{m}{\ell_\ST}{m_\ST}
       \ALKLMLTMT{\nu}{\ell}{m}{\ell_\ST'}{m_\ST'}^*
 = \KRON{\ell_\ST}{\ell_\ST'} \KRON{m_\ST}{m_\ST'}
   \varepsilon_\nu (\ell_\ST, m_\ST)
   \varepsilon_\nu (\ell_\ST', m_\ST')
\end{equation}
where $\varepsilon_\nu (\ell_\ST, m_\ST) = 1$ if the conditions
(\ref{Eigenbasis},\ref{Eigenbasis2}) are satisfied and 0 otherwise.

With these coefficients, the CMB computation goes as in the flat
case, except for the fact that some integrals have to be replaced
by discrete sums. One easily gets that
\begin{equation}
\label{p44tbissph}
\frac{\delta T}{T}(\theta, \varphi)
 = \frac{(2 \pi)^3}{V}
   \sum_{\nu = 2}^\infty \sum_s
   \sum_{\ell, m}
   \XITKSLM{\Gamma}{k}{s}{\ell}{m}
   O_k^{[\STR]} \left(\YXKLM{\STR}{k}{\ell}{m}\right)
   \sqrt{\PS (k)}\,\hat{e}_\bk ,
\end{equation}
so that
\begin{equation}
\label{p44t2sph}
\ALM{\ell}{m}
 = \frac{(2 \pi)^3}{V}
   \sum_{\nu = 2}^\infty \sqrt{\PS (k)}\,
                         O_k^{[\STR]} \left(\RXKL{\STR}{k}{\ell} \right)
        \sum_s \XITKSLM{\Gamma}{k}{s}{\ell}{m} \hat{e}_\bk ,
\end{equation}
and
\begin{equation}
\CLMLPMP{\ell}{m}{\ell'}{m'}
 = \frac{(2 \pi)^3}{V}
   \sum_{\nu = 2}^\infty \PS (k) G_\ell (k) G_{\ell'} (k)
        \sum_s \XITKSLM{\Gamma}{k}{s}{\ell}{m}
               \XITKSLM{\Gamma}{k}{s \; *}{\ell'}{m'} ,
\end{equation}
as first obtained in Ref.~\cite{lwugl}. Note also that, following
Refs.~\cite{ls90,wb95}, a scale invariant spectrum will in that case
be defined as
\begin{equation}
\PS (k) \propto \frac{1}{k (k^2 - K)} .
\end{equation}

To finish, let us discuss the properties of the random variable $\hat
e_\bk$. Since the eigenmodes in toroidal coordinates,
$\QKLM{\nu}{\ell_\ST}{m_\ST}$, and the coefficients
$\ETAKSLM{\Gamma}{\nu}{s}{\ell_\ST}{m_\ST}$ are real valued, it
follows from Eq.~(C2) that
\begin{equation}
\ALKLMLTMT{\nu}{\ell}{m}{\ell_\ST}{m_\ST}^* =(-1)^m
 \ALKLMLTMT{\nu}{\ell}{-m}{\ell_\ST}{m_\ST} .
\end{equation}
It follows that
\begin{equation}
{\XITKSLM{\Gamma}{k}{s}{\ell}{m}}^* =(-1)^m
 \XITKSLM{\Gamma}{k}{s}{\ell}{-m} ,
\end{equation}
whatever $s$ and thus that the eigenmodes $\UPSTKS{\Gamma}{k}{s}$ are
real-valued. This implies that $\hat e_\bk$ is a real random variable,
contrary to the preceding example of the torus.

\begin{figure}
\centerline{\psfig{file=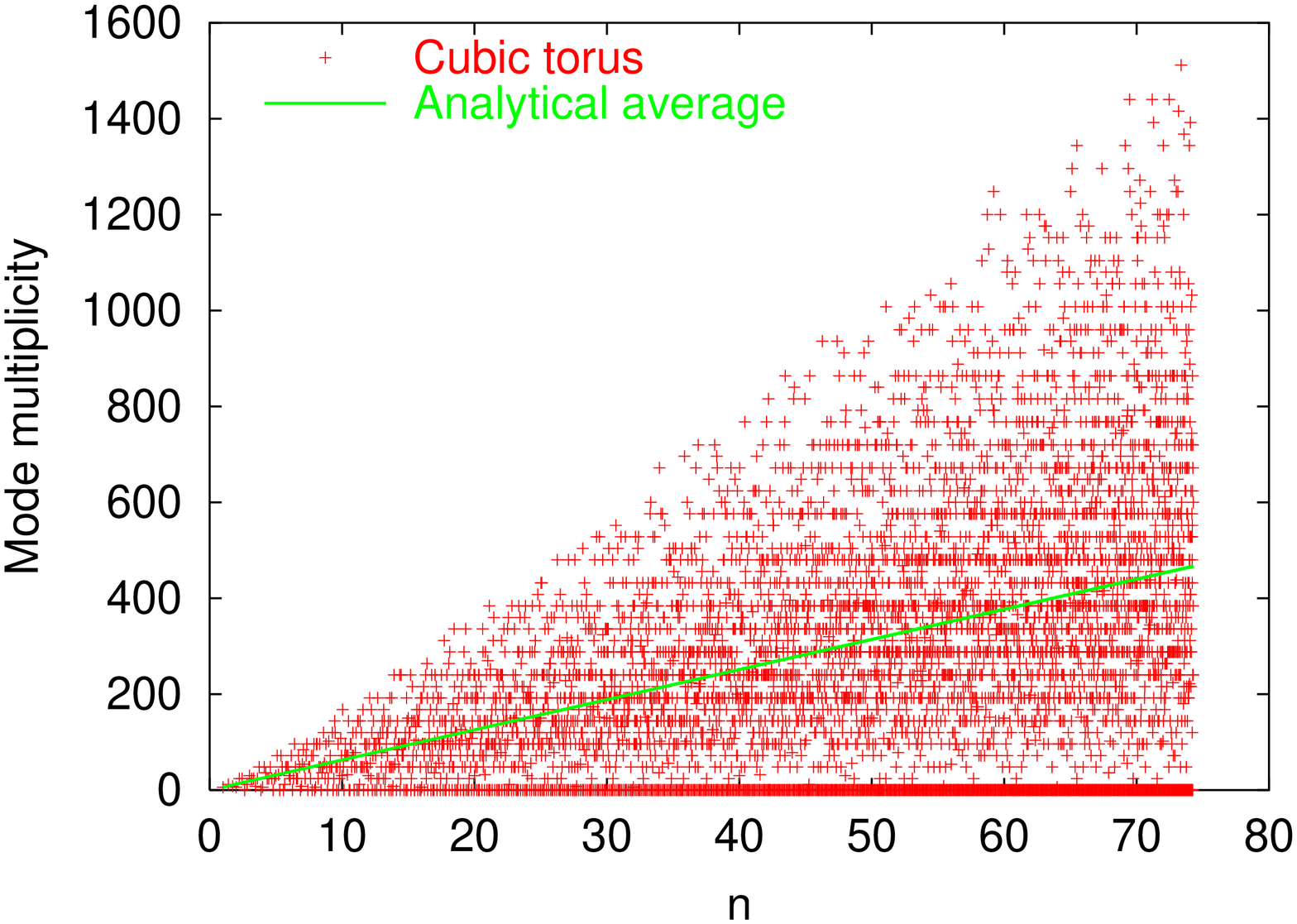,angle=270,width=3.5in}
            \psfig{file=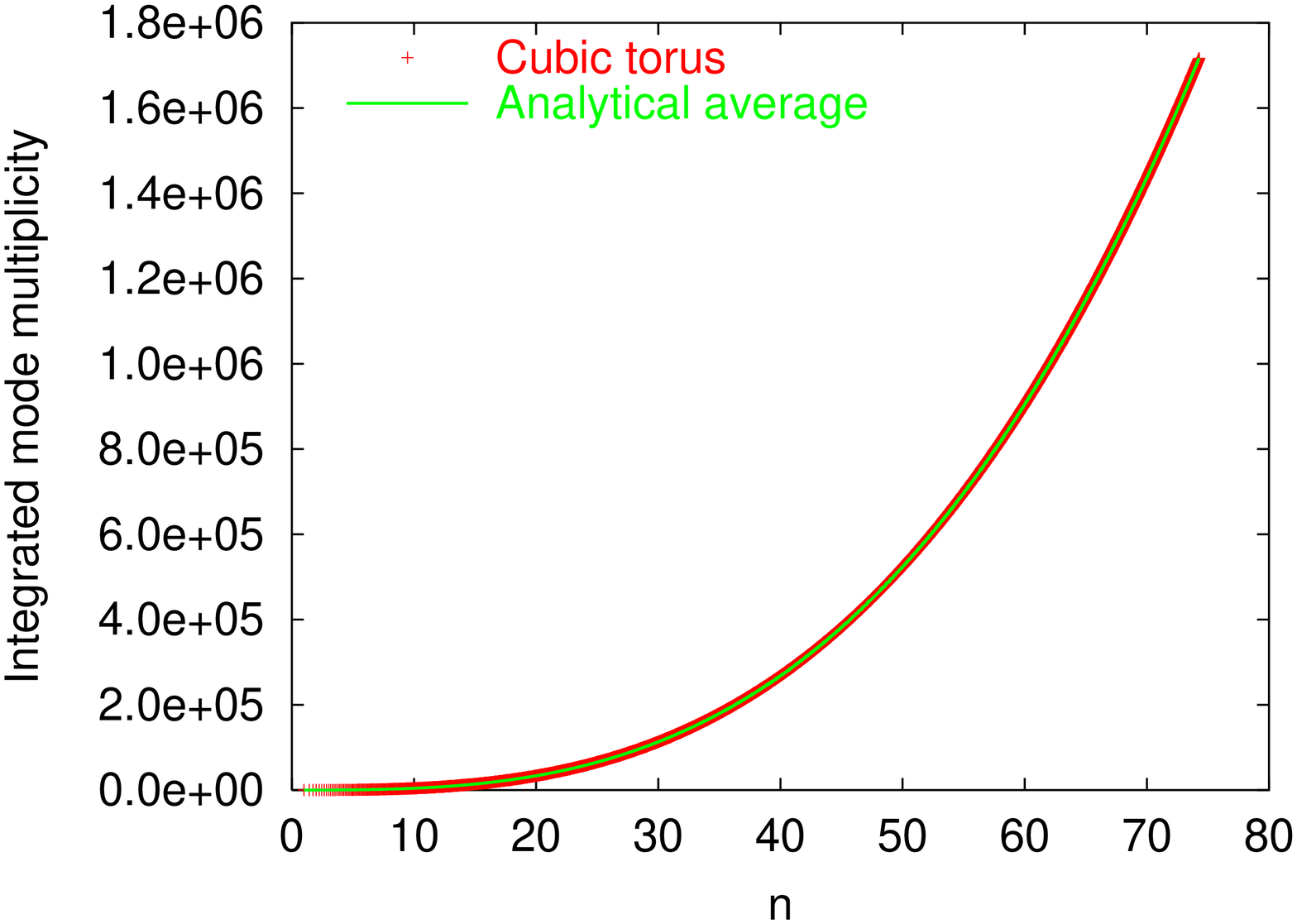,angle=270,width=3.5in}}
\centerline{\psfig{file=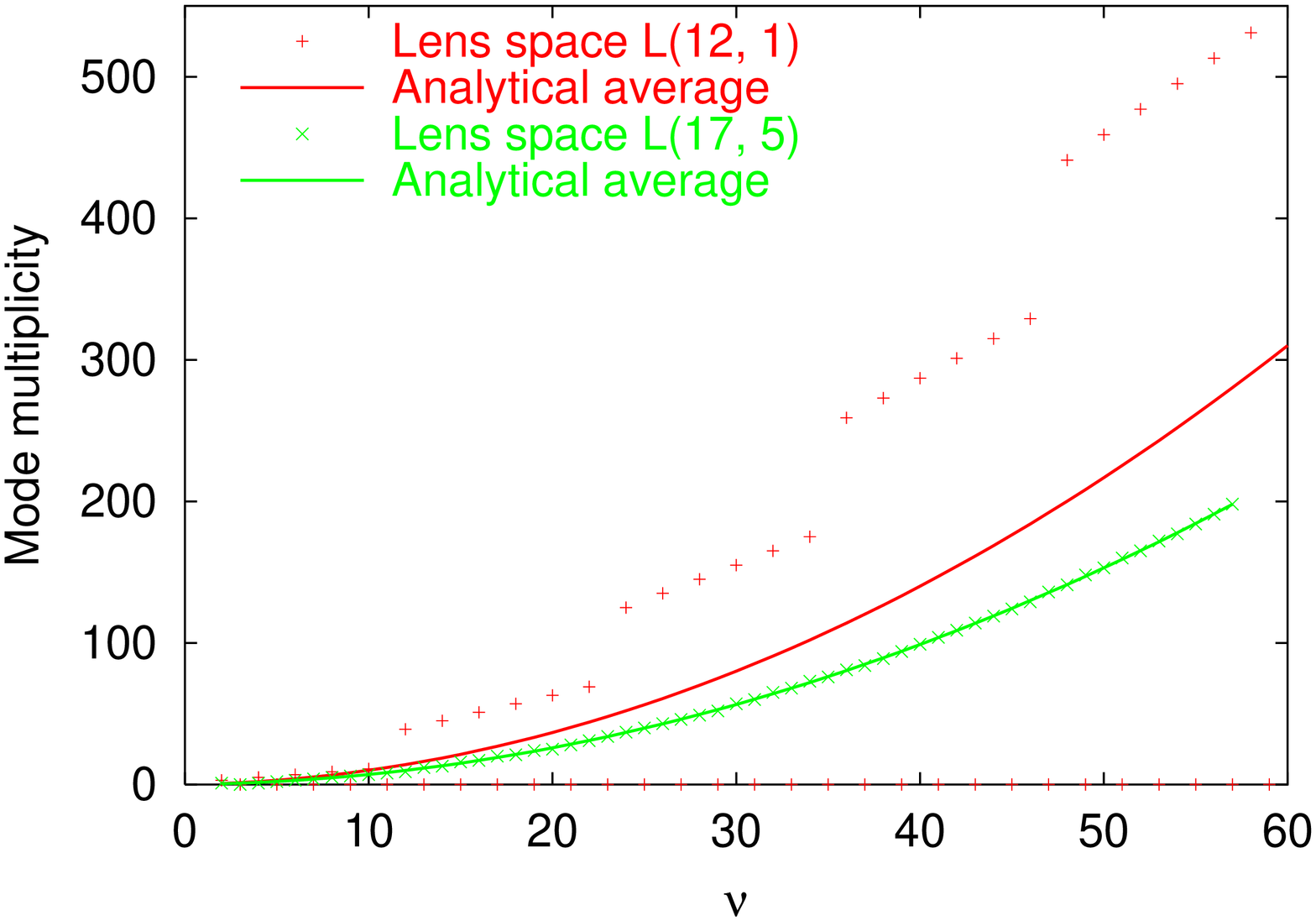,angle=270,width=3.5in}
            \psfig{file=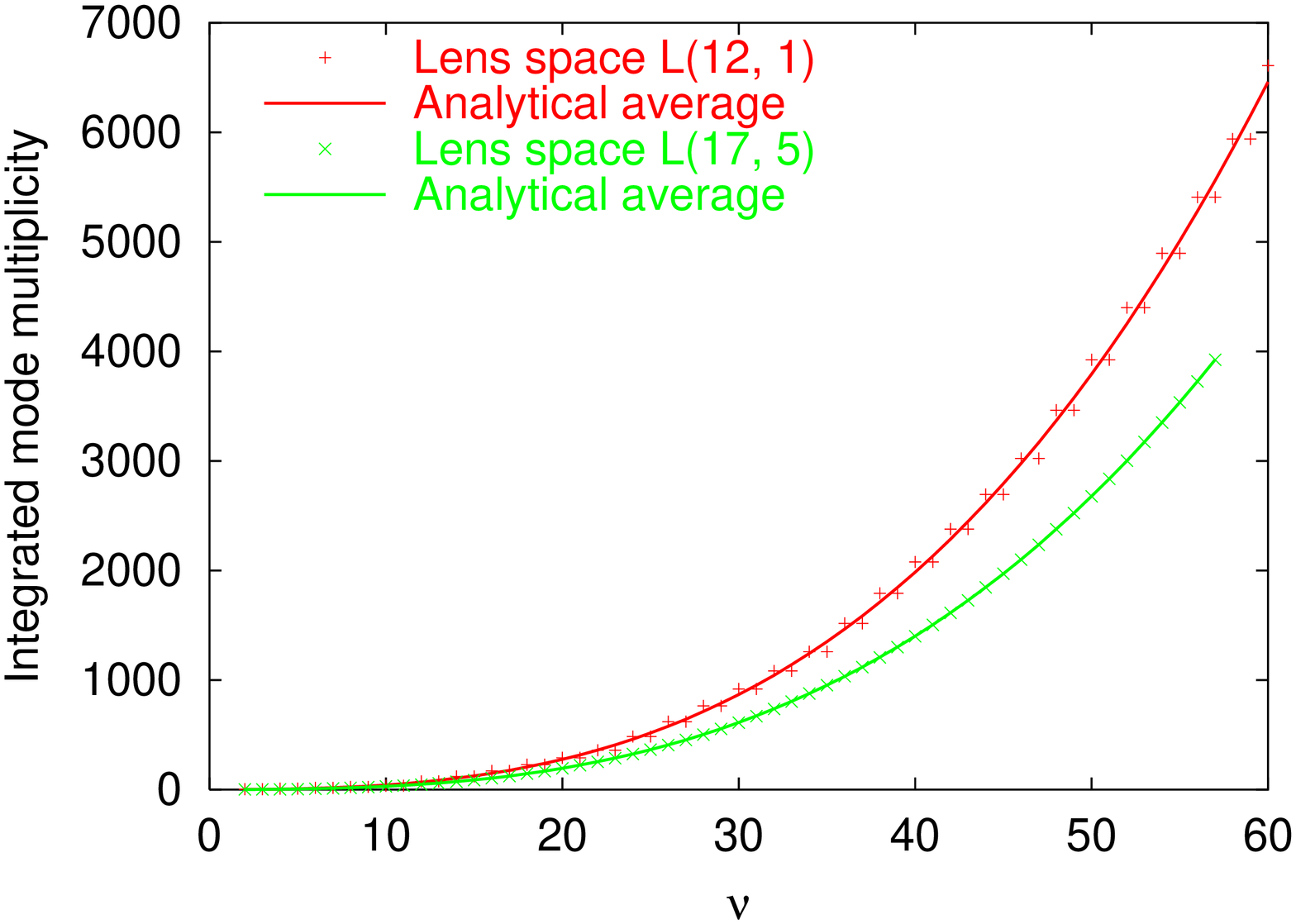,angle=270,width=3.5in}}
\caption{Multiplicity of the mode $k$ as a function of $n$ (cubic
torus, top panels) or $\nu$ (lens space $L (p, q)$, bottom
panels). For the cubic torus, $n$ is of the form $\sqrt{N}$, where $N$
is an integer, and for lens spaces (and more generally spherical
spaces), $\nu$ is an integer. Left panels show the multiplicity of
each mode for a given value of either $n$ of $\nu$. We have also given
an estimate of the ``average'' number of modes, given by $2 \pi n$ for
the torus and $(\nu + 1)^2 / p$ for the lens spaces, respectively. For
lens space $L (17, 5)$, the mode multiplicity closely follows the
analytical estimate, whereas for lens space $L (12, 1)$, modes exist
only for even values of $\nu$. For the cubic torus, the mode
multiplicity is even more irregular, and varies between $0$ and $\sim
20 n$. For example, it is always $0$ when $n = \sqrt{8 m + 7}$. Right
panels show the integrated mode multiplicity, that is the number of
modes smaller or equal to some value ($2 \pi n / L$ and $(\nu + 1)
\sqrt{K}$, respectively). Here, the analytical estimates ($4 \pi n^3 /
3$ and $(\nu + 1) (\nu + 2) (2 \nu + 3) / 6 p$, respectively) provide
a much better estimation (see Section~5 of Ref.~\cite{lwugl} for a
detailed discussion of these properties).}
\label{figmult}
\end{figure}

\section{Numerical computations}
\label{sec_simulation}

\subsection{Implementation}
\label{sub_qualitatif}

The correlation matrix for $\ell \leq \ell_\MAX$ has $\ell_\MAX^4$
coefficients. However, the parity and symmetry relations
\begin{eqnarray}
\CLMLPMP{\ell}{m}{\ell'}{m'}
 & = & (- 1)^{m - m'} \CLMLPMP{\ell}{-m}{\ell'}{- m'}{}^*  \\
\CLMLPMP{\ell}{m}{\ell'}{m'}
 & = & \CLMLPMP{\ell'}{m'}{\ell}{m}{}^*
\end{eqnarray}
reduce the problem to computing only a quarter of them. Then, for
a given topology, symmetries can further reduce the number of
coefficients to compute. For example, with a cubic torus,
Eqs.~(\ref{toregen},\ref{torecub}) insure that only one
coefficient out of eight is nonzero, and the
symmetries~(\ref{toregen2}) also give the coefficients when one
changes the sign of both $m$ and $m'$. This leaves only
$\ell_\MAX^4 / 64$ coefficients to compute. For example, a COBE
scale map ($\ell_\MAX \sim 30$) requires $12500$ coefficients,
while a Planck scale map ($\ell_\MAX \sim 1500$) requires $\sim 8
\times 10^{10}$ coefficients.

Each coefficients is computed using Eq.~(\ref{sum30}), which
involves a sum over all the wavemodes $\bk$. For a given
resolution $\ell_\MAX$ the modulus of the largest wavemode is
given by $k_\MAX \sim 3 \ell_\MAX / \eta_0$. Moreover, the density
of wavemodes is proportional to the size of the torus, so that we
have $O(\ell_\MAX^3 L^3)$ modes. Therefore, the computational time
and the memory requirement scale as $\ell_\MAX^7 L^3$ and
$\ell_\MAX^3 L^3$, respectively. This is obviously a serious
limitation of our algorithm. For example, computing the
correlation matrix for a COBE scale map on a relatively small
torus ($L = 2 R_H$, where $R_H$ is the Hubble radius) takes around
$10$ hours on a $900 \UUNIT{MHz}{}$ CPU and allocates $60$
Megabytes of memory.

When the topology is simply connected, it is a well-known fact that
the $C_\ell$ are in general a smooth function of the multipole
$\ell$. This reduces the computational time because for $\ell_\MAX =
1500$ one needs to compute only $\sim 50$ coefficients. For the
topologies we have studied, we did not find any evidence for a smooth
structure of the correlation matrix, at least at the relatively large
scales we considered. At which scale one can reliably approximate the
correlation matrix by its isotropic diagonal part (the $C_\ell$)
remains an open question.

Also, if one wants to simulate CMB maps from the correlation
matrix, one needs to diagonalize it. This procedure can also take
a lot of time because it is an $\ell_\MAX^6$ process. For the case
of the torus, however, this problem is not serious as the
symmetries of the torus insure that the matrix is block diagonal,
with eight blocks if the torus is cubic or four blocks otherwise.

Strictly speaking, one does not need the correlation matrix to compute
maps. One can do it directly by using Eq.~(\ref{p4_2}).  This amounts
to performing a realization of the three-dimensional random field
describing the cosmological perturbations, and projecting it onto the
sphere. In this case, one has only $\ell_\MAX$ coefficients to compute
(the $\ALM{\ell}{m}$) instead of the correlation matrix, so that the
memory requirements are roughly the same (one only needs to store the
value of the random field for each mode), but the computational time
scales as $\ell_\MAX^5 L^3$. In this case, computing maps for a cubic
torus of size $L = 2 R_H$ till $\ell_\MAX = 120$ takes $3$ hours on a
$1.7 \UUNIT{GHz}{}$ CPU and allocates 300 Megabytes of memory.

In the case of spherical spaces, the coefficients
$\XITKSLM{\Gamma}{k}{s}{\ell}{m}$ must also be computed
numerically. This involves determining both the coefficients
$\ETAKSLM{\Gamma}{s}{\nu}{\ell_\ST}{m_\ST}$ and
$\ALKLMLTMT{\nu}{\ell}{m}{\ell_\ST}{m_\ST}$. This computation can be
reduced by taking into account their symmetries, as described in
Appendix~\ref{app_B}, which imply that $\nu - |m| + |\ell_\ST|$ is
odd, $|m| = |m_\ST|$, $|\ell_\ST| + |m_\ST| \leq \nu$, and $|\ell_\ST|
+ |m_\ST| \equiv \nu \MOD{2}$. The computation can be performed
analytically with the use of symbolic computation software such as
Mathematica. In the case of the lens space $L (17, 5)$, the
computation up to $\nu = 43$ and $\nu = 55$ takes $3$ and $12$ hours
respectively on a $1 \UUNIT{Ghz}{}$ CPU with a negligible amount of
memory.

\subsection{Expected results}
\label{sub_theoriquement}

The three main effects that are expected on a CMB map computed in a
multi-connected topology are (i) the appearance of $\ell$--$\ell'$ and
$m$--$m'$ correlations reflecting the breakdown of global isotropy,
(ii) the existence of a cutoff in the CMB angular power spectrum on
large angular scales (low $\ell$), and (iii) the existence of pattern
correlations such as pairs of circles where the temperature
fluctuations are strongly correlated as they represent the
intersection of the last scattering surface with itself and therefore
show the temperature of the same emission region from different
directions. Note that the effects (ii) and (iii) will show up only if
the topological scale is smaller than the radius of the last
scattering surface while the effect (i) may be present even if the
topological scale is a bit larger than the diameter of the last
scattering surface.

So far, the main constraints that have been given on multi-connected
topologies come from the absence of a cutoff at large angular scales
in the COBE spectrum. This gives strong constraints on the minimal
size of the topology as the cutoff is given by the angular size of the
torus projected on the last scattering surface.  However, as
previously discussed, this cutoff in the ``true'' temperature
fluctuations can be compensated, at least partially, by an integrated
Sachs-Wolfe effect which arises, for example, when the cosmological
constant is large.

The third, and up-to-now never computed, effect of a multi-connected
topology is the appearance of pairs of circles which are correlated in
temperature. This correlation, however, is not perfect. It would be
perfect if the temperature fluctuation were a pure scalar function on
the last scattering surface $2$-sphere around us which would be the
case only if (i) the temperature anisotropies were given only by the
Sachs-Wolfe effect [first term of Eq.~(\ref{swdopisw})] and (ii) the
last scattering surface were infinitely thin.

It is well-known that the temperature fluctuations observed in a
given direction are in fact a combination of several effects:
first, one has the intrinsic temperature fluctuations of the
emitting region, which is eventually affected by a gravitational
redshift. These two contributions form the so-called Sachs-Wolfe
effect [first term in the righthand side of Eq.~(\ref{swdopisw})].
Second, if the emission region is not at rest with respect to the
observer, one will observe some apparent temperature fluctuations
which in fact result from a Doppler shift [second term in the
righthand side of Eq.~(\ref{swdopisw})]. Third, several events can
alter the photons energy and trajectory while travelling toward
us. In particular, they can be slightly disturbed from their
trajectory (lensing) and, more importantly, they can exchange
energy when they cross time-varying potential wells. This last
effect is usually referred to as the integrated Sachs-Wolfe effect
[third term in the righthand side of Eq.~(\ref{swdopisw}), see
Fig.~\ref{figdec} below].  Obviously, the Sachs-Wolfe effect is a
scalar quantity that depends only on the emission region.
Therefore, it should be the same whatever the direction of
observation. By contrast, the Doppler effect will explicitly
depend on the direction of observation.  If one observes two
directions which correspond to the same point of the last
scattering surface and which form a small angle, then one expects
that the Doppler contribution will be almost the same. If the
matching points are $90$ degrees from each other, then one expects
on average no correlation at all, whereas the Doppler effect
between two antipodal points will become anti-correlated. Finally,
since photons originating from the same emission region but
observed from different directions will travel through different
regions of space, they will undergo different integrated
Sachs-Wolfe effects, so that no significant correlations are
expected from this effect, which is therefore considered a noise
term for our purposes.

Actually, one aim of this work is precisely to compute the typical
amount of correlation one can expect on pairs of circles. Note
that this correlation is likely to depend on scale: on large
scales, one should be annoyed by the late integrated Sachs-Wolfe
effect; between the Sachs-Wolfe plateau and the first Doppler peak
(and, at a lesser extent, at every dip between two Doppler peaks),
the Doppler effect dominates; at the first peak, there is usually
(especially when the matter content is low) a significant
contribution of the early integrated Sachs-Wolfe effect (see
Fig.~\ref{figdec}); at very small scales, one feels the finite
width of the last scattering surface (see below), etc. Also, as
explained above, the relative position of the circles will play a
role because of the Doppler contribution. It is therefore
interesting to look at the best way to find matching circles on a
realistic CMB map. We leave this important point to future
work~\cite{future}.

The second (and probably less important) effect that reduces the
correlation between the circles is the finite width of the last
scattering surface. As far as we know, this effect has not yet
been carefully analyzed. It plays a role when one looks at
fluctuations on scales smaller than the projected width of the
last scattering surface. In this case when looking in a given
direction, one picks up fluctuations which are situated ``on one
side'' of the last scattering surface, but for pairs of circles,
one sees opposite sides of the last scattering surface. On larger
scales, the effect is negligible as one averages temperature
fluctuations on regions much larger than the thickness of the last
scattering surface.

\section{Results}
\label{sub_resultats}

We now outline some of the results we have already obtained from
our simulations. The main aim of this Section is to provide a
series of tests to check our simulations. A more detailed analysis
of the structure of the correlation matrix
$\CLMLPMP{\ell}{m}{\ell'}{m'}$ as well as a search for accurate
tests to detect the topology are left for future
work~\cite{future}.

\subsection{Flat case: cubic torus}

In all the simulations we performed, we have considered a flat
$\Lambda$CDM model with $\Omega_\Lambda = 0.7$, a Hubble parameter
of $H_0 \equiv 100 h \UUNIT{km}{} \UUNIT{s}{- 1} \UUNIT{Mpc}{- 1}$
with $h = 0.62$, a baryon density $\omega_\BAR \equiv \Omega_\BAR
h^2 = 0.019$ and a spectral index $n_{\rm S} = 1$. With this
choice of cosmological parameters, the Hubble radius is $R_H \sim
4.8 \UUNIT{Gpc}{}$, the ``horizon'' radius (under the hypothesis
of a radiation dominated universe at early times) is $R_h \sim
15.6 \UUNIT{Gpc}{}$, and the radius of the last scattering surface
is $R_\LSS \sim 15.3 \UUNIT{Gpc}{}$. The volume of the observable
universe is therefore $V_\OBS \sim 15 \times 10^3 \UUNIT{Gpc}{3}$.

\begin{figure}
\centerline{\psfig{file=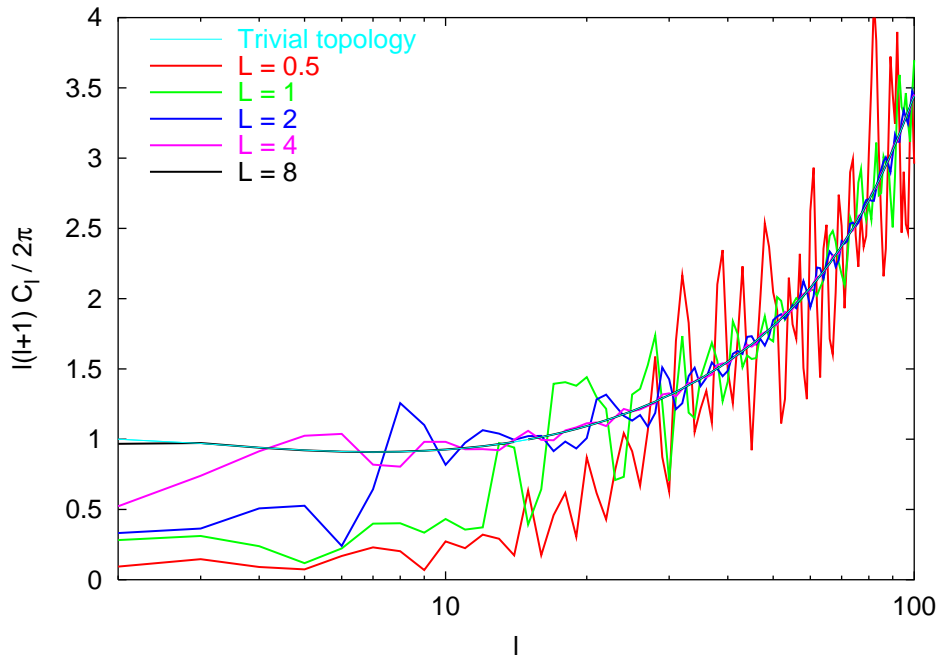,angle=270,width=5in}}
\caption{CMB anisotropies in the simply connected (i.e., usual)
topology and in toroidal universes based on cubic fundamental domains
of various sizes, expressed here in units of the Hubble radius. With
our choice of cosmological parameters, the situation $L = 8$
corresponds to a torus larger than the observable universe, which
shows a small depletion of power on large scales. For smaller tori,
the cutoff is much sharper.}
\label{fig1}
\end{figure}

Let us first compare the $C_\ell$ in the simply connected topology to
the $C_\ell$ in a torus (Fig.~\ref{fig1}). As expected, we see a
cutoff at some angular size which corresponds to the angular size of
the torus on the last scattering surface. This corresponds to the
multipole
\begin{equation}
\label{cutoffl} \ell_c \sim \frac{2 \pi R_\LSS}{L} ,
\end{equation}
where $L$ is the length of the cubic torus' fundamental
domain~\cite{jpp1,jpp2}.  Note that even when the torus is larger than
the size of the observable universe, the spectrum exhibits a loss of
power on large scales. This is because the Harrison-Zel'dovich
spectrum exhibits a significant amount of power at large scales (by
definition, it is scale invariant), and in practice, the modes that
contribute to the quadrupole of the CMB anisotropies can be as large
as ten times the size of the observable universe (the exact number
depends mostly on the spectral index and on the amplitude of the
integrated Sachs-Wolfe term). Therefore, this leaves hope to detect
the topology ``beyond the horizon'' where the circles method would
fail.

It is not easy to predict the amplitude of the power at scales larger
than the cutoff because it depends mostly on the amplitude of the
integrated Sachs-Wolfe effect, which is difficult to estimate even
when the topology is multiconnected. Another consequence of a
multiconnected topology are oscillations in the spectrum. These come
both from the fact that there is a sharp cutoff in the spectrum (which
causes oscillations in Fourier/Legendre space) and that the spectrum
is ``spiky'' on large scales. Should we consider a simply connected
universe with a cutoff at some scales, then the corresponding $C_\ell$
would be less irregular. Finally, note that on small angular scales,
the spectrum tends to behave as in the simply connected case, but
computing this scale remains an open problem at the moment.

\begin{figure}
\centerline{\psfig{file=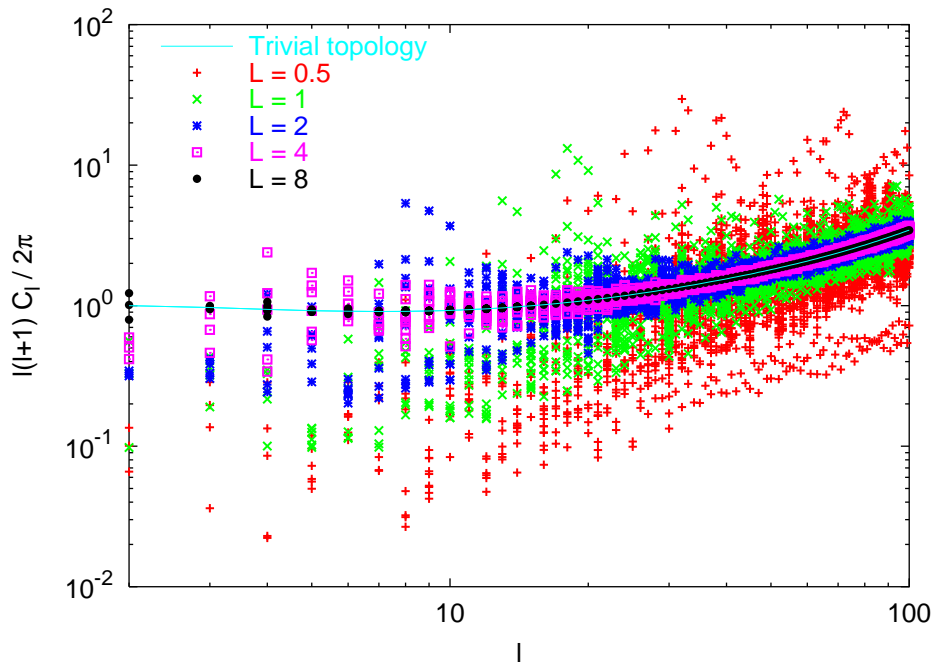,angle=270,width=5in}}
\caption{Variance of the $\ALM{\ell}{m}$ for fixed values of
$\ell$. The average of these values give the $C_\ell$ coefficients
shown in Fig.~\ref{fig1}. We insist that this does not correspond
to a realization of the random variables describing CMB
anisotropies, but to the variance of the $\ALM{\ell}{m}$. These
coefficients are far from sufficient to build maps of CMB
anisotropies as they do not include the correlations between
different $\ALM{\ell}{m}$.} \label{fig2}
\end{figure}

So far, we have considered only the $C_\ell$, which represent only
some average of the diagonal part of the correlation matrix. The true
diagonal part of the correlation matrix is given by the
$\CLMLPMP{\ell}{m}{\ell}{m}$ which represent the variance of the
$\ALM{\ell}{m}$. An example of their behavior is shown in
Fig.~\ref{fig2} for the same topologies as in Fig.~\ref{fig1}.
Several features appear on this figure. First and most importantly,
the dispersion in the variance of the $\ALM{\ell}{m}$ at fixed $\ell$
is very large. It appears that it is maximal at the cutoff scale
$\ell_c$, where the dispersion in the variance of the $\ALM{\ell}{m}$
can be as large as two orders of magnitude. This dispersion slowly
decays at larger multipoles, where one ``tends'' (in the sense of
observable quantities) towards the simply connected case, and
surprisingly also decays at scales larger than the cutoff. With the
hypothesis of a multi-connected universe, this dispersion would be
(incorrectly) interpreted as non Gaussianity. Since at present no non
Gaussianity or anisotropy was observed in the data, this allows new
constraints of the size of the fundamental domain.

\begin{figure}
\centerline{\psfig{file=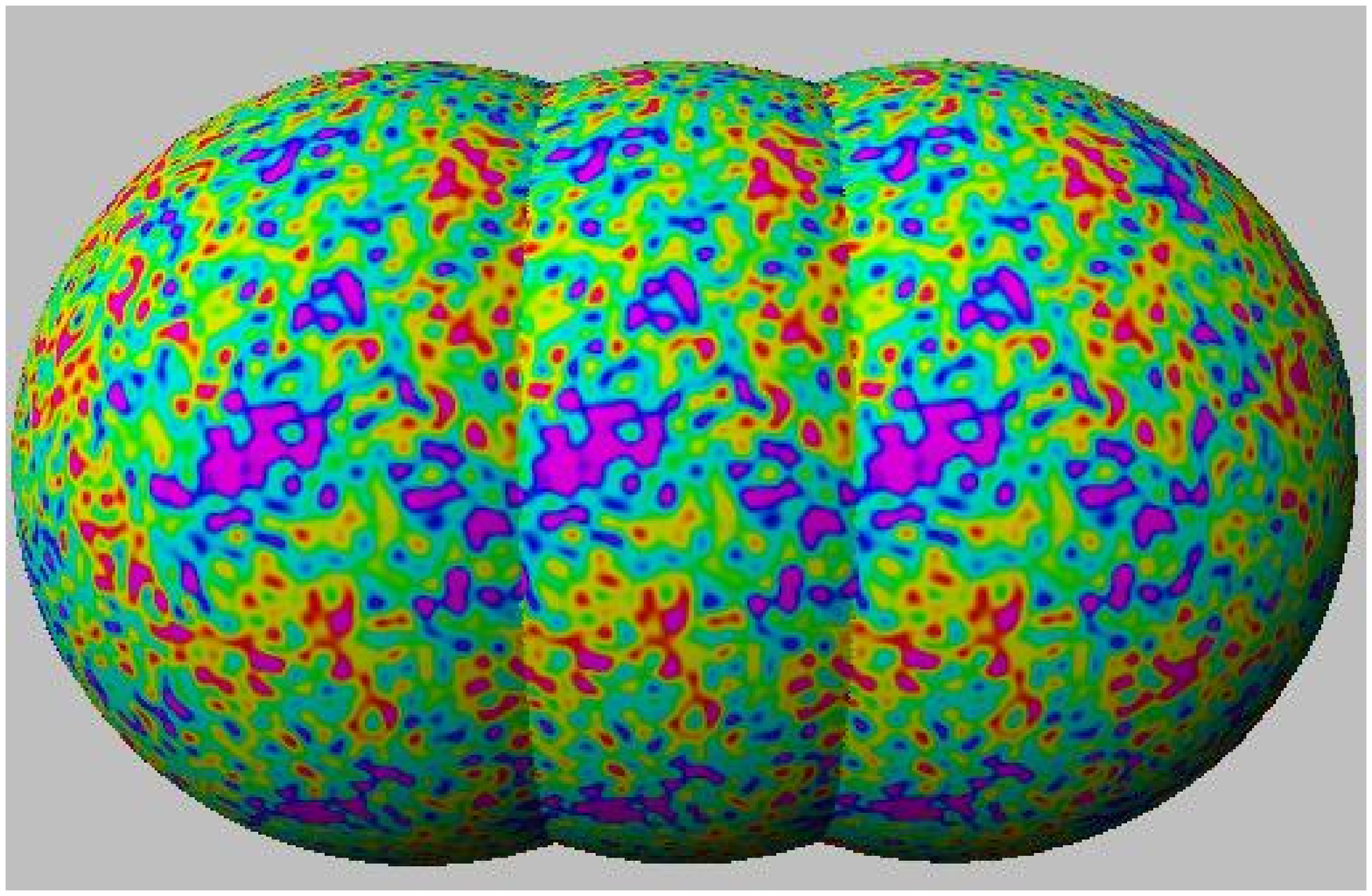,angle=0,width=3.5in}
            \psfig{file=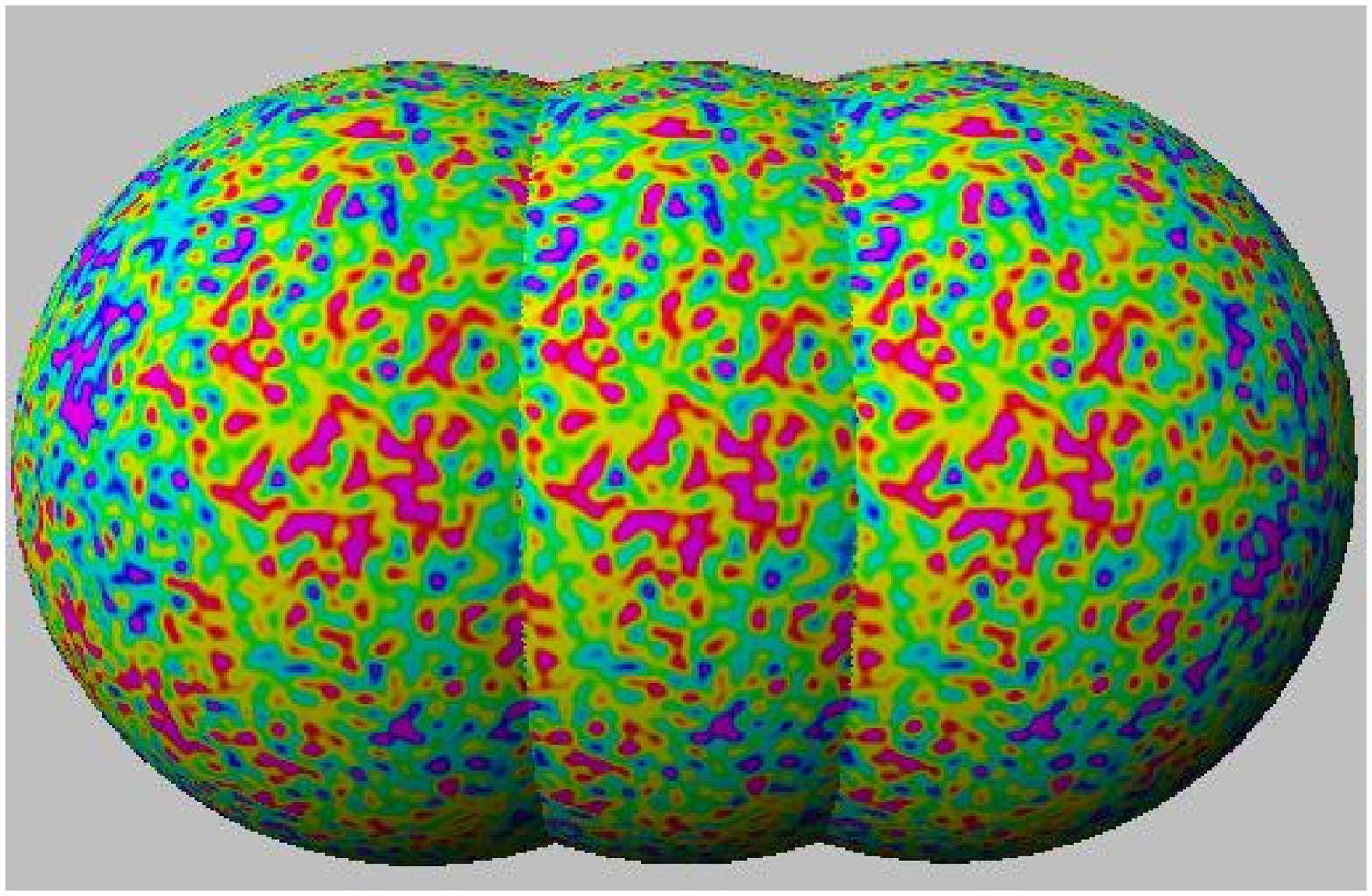,angle=0,width=3.5in}}
\centerline{\psfig{file=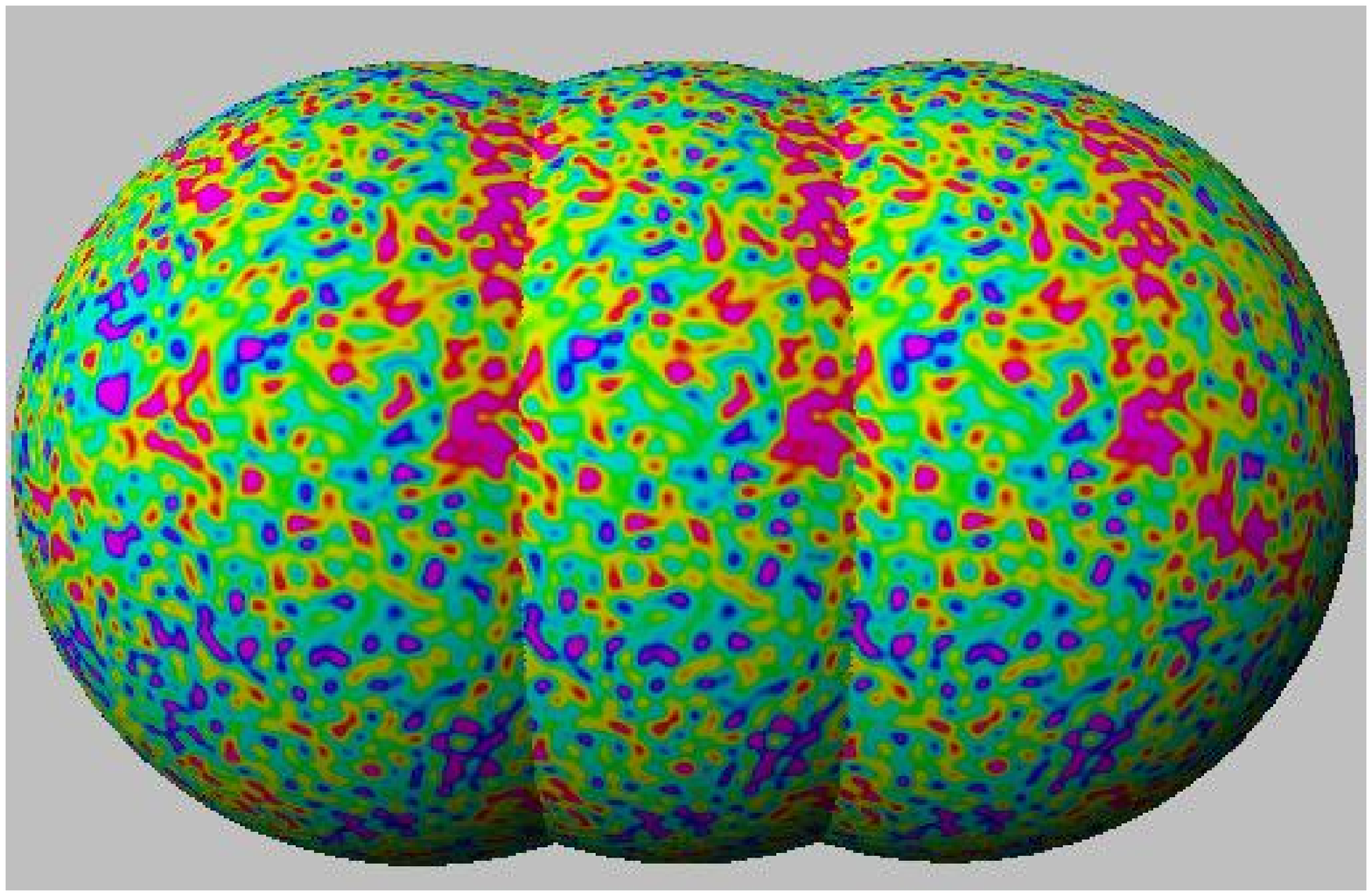,angle=0,width=3.5in}
            \psfig{file=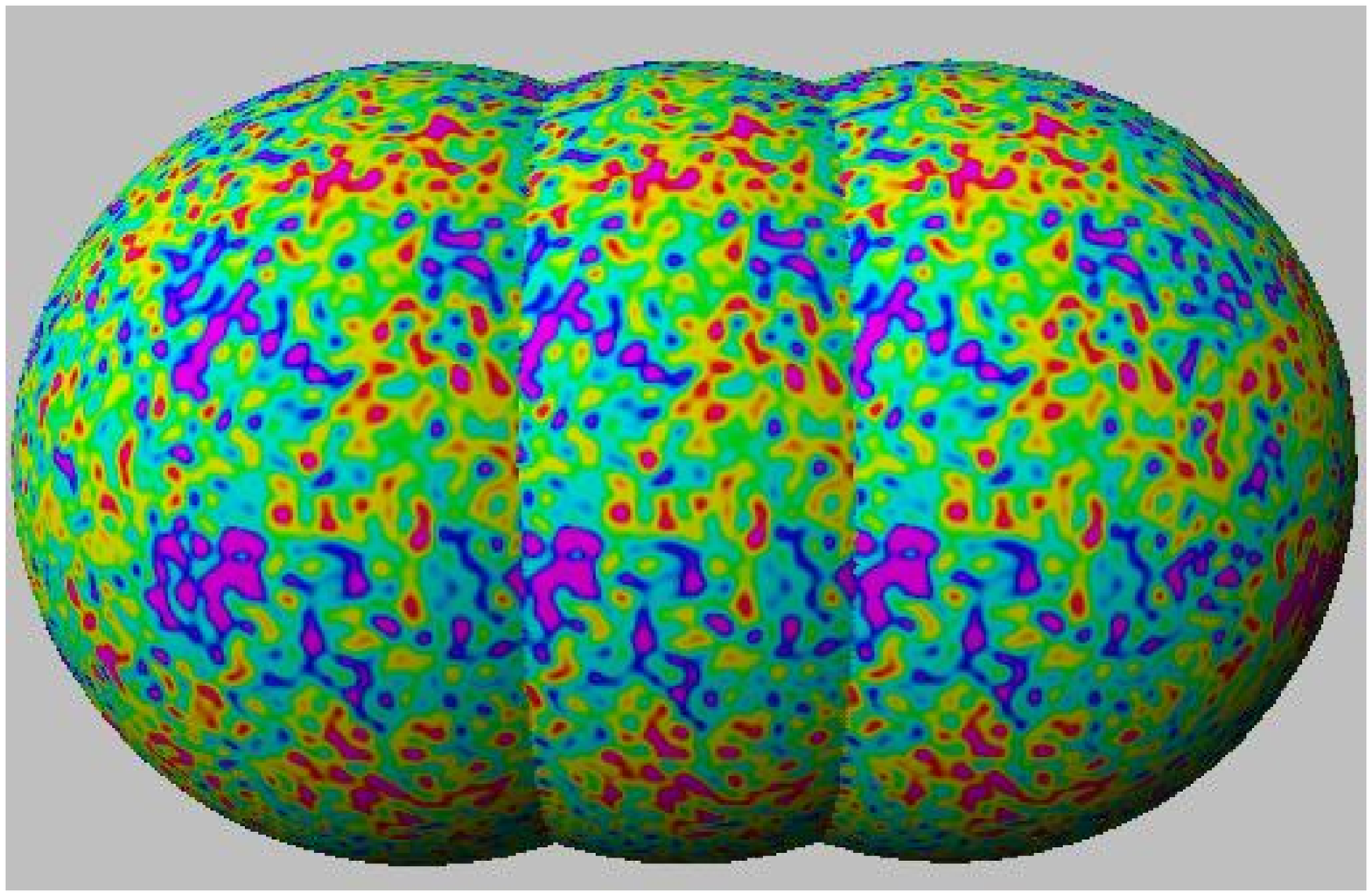,angle=0,width=3.5in}}
\caption{Four realizations of CMB maps of the whole temperature
anisotropies for a cubic torus. The resolution of each map is
$\ell_\MAX = 120$ . The last scattering surface is seen from the
outside as well as two of its closest topological images after
translation of $L$ and $- L$ along one axis of the torus. One can
check by eye that the temperature fluctuations are well (but not
perfectly) correlated along matching circles (i.e., along the
intersection between the last scattering surface and its
neighbors) located at latitudes $\theta = \pm 19^\circ$. Note that
as expected, there are very few fluctuations on scales larger than
the size of the torus (which is given here by the distance between
the circles).} \label{map2a}
\end{figure}

\begin{figure}
\centerline{\psfig{file=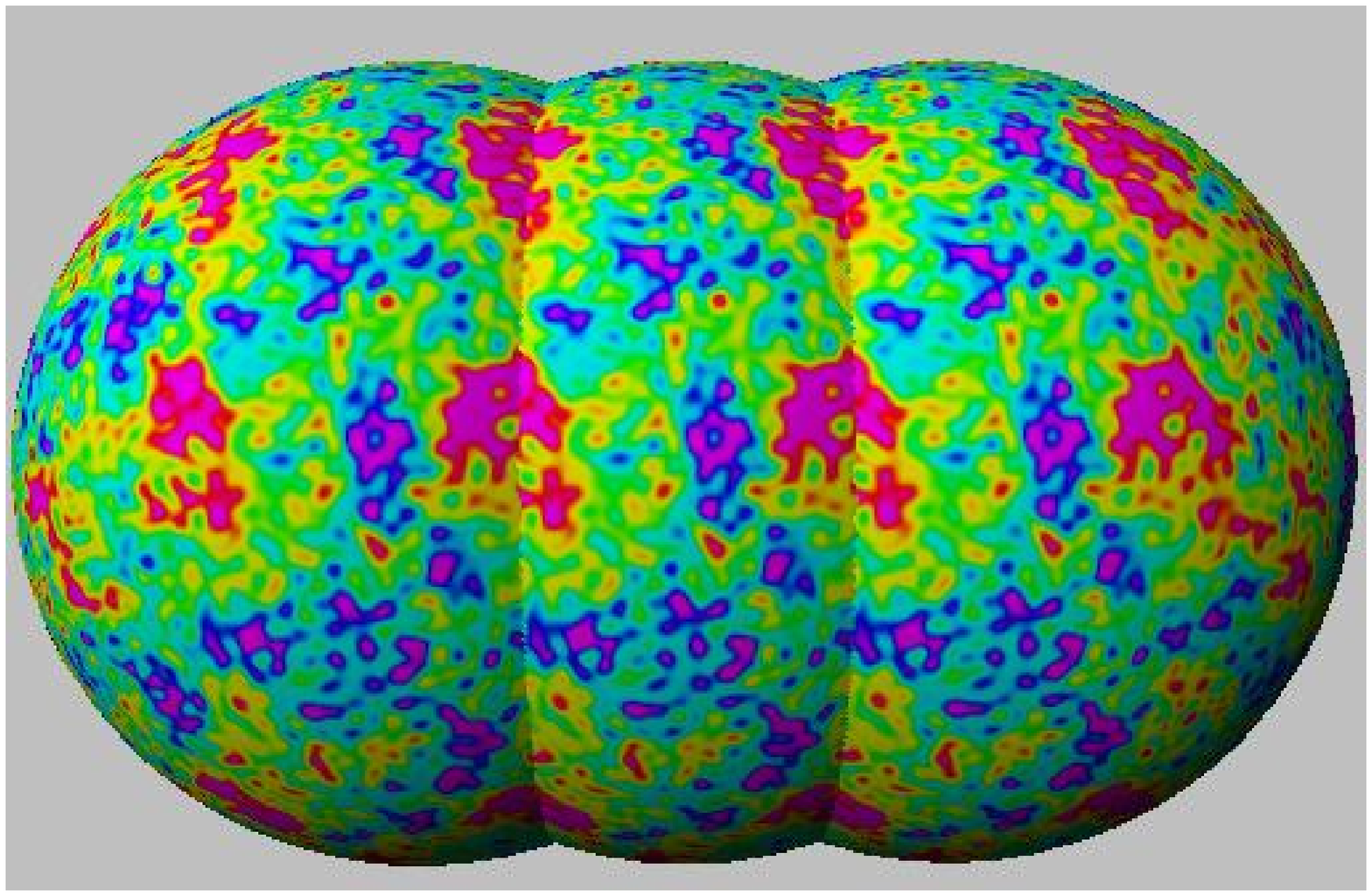,angle=0,width=3.5in}
            \psfig{file=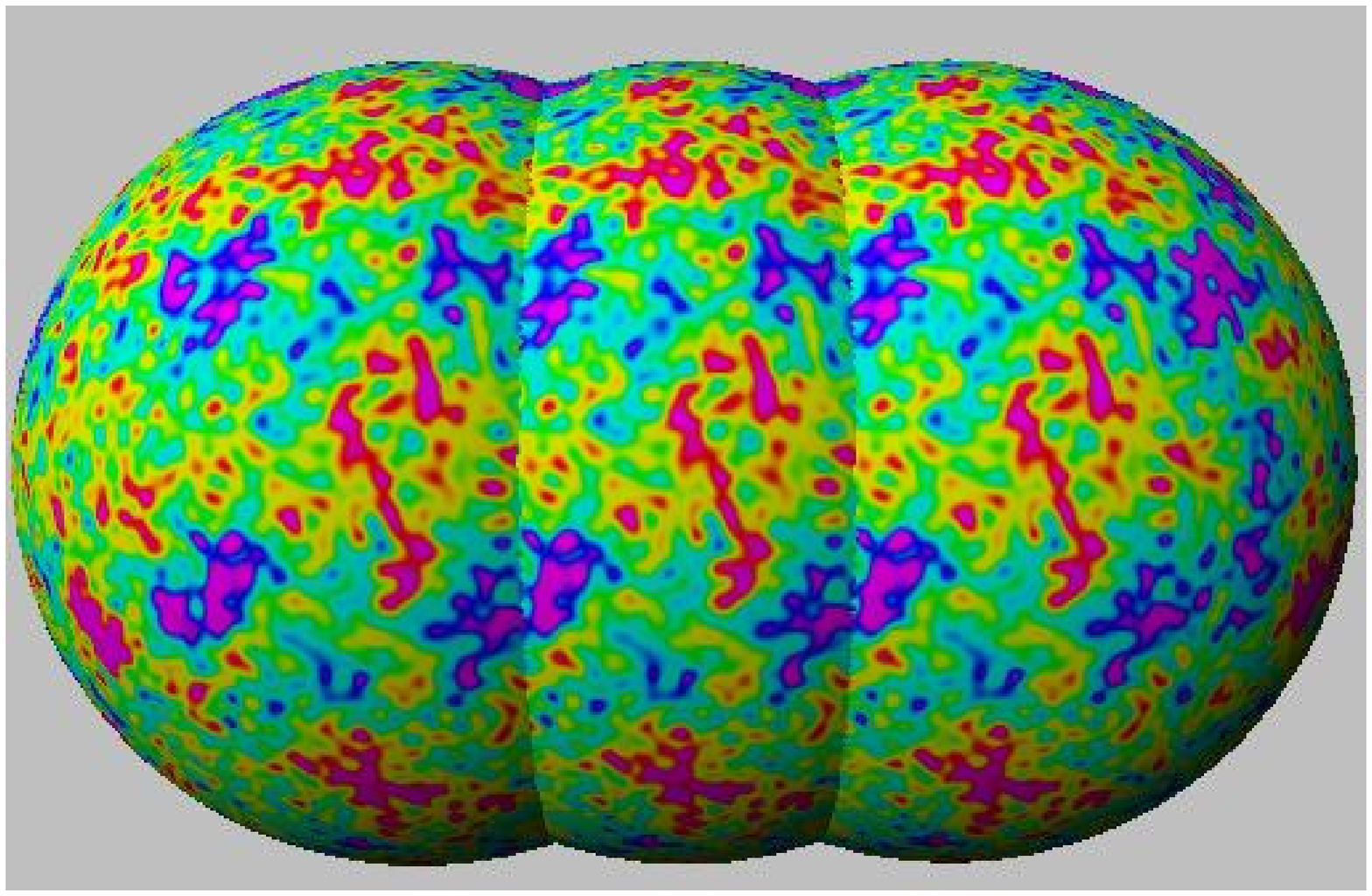,angle=0,width=3.5in}}
\centerline{\psfig{file=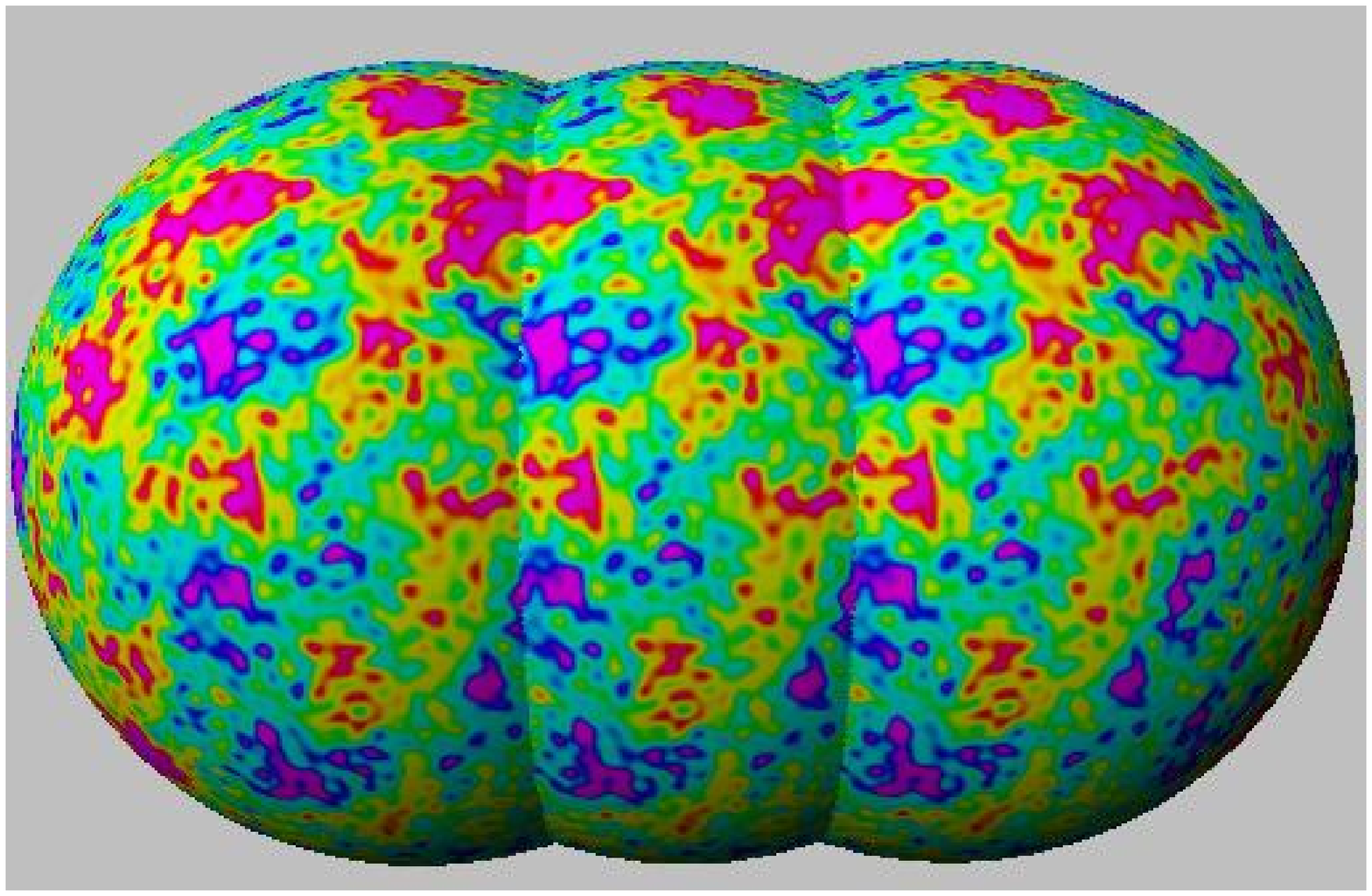,angle=0,width=3.5in}
            \psfig{file=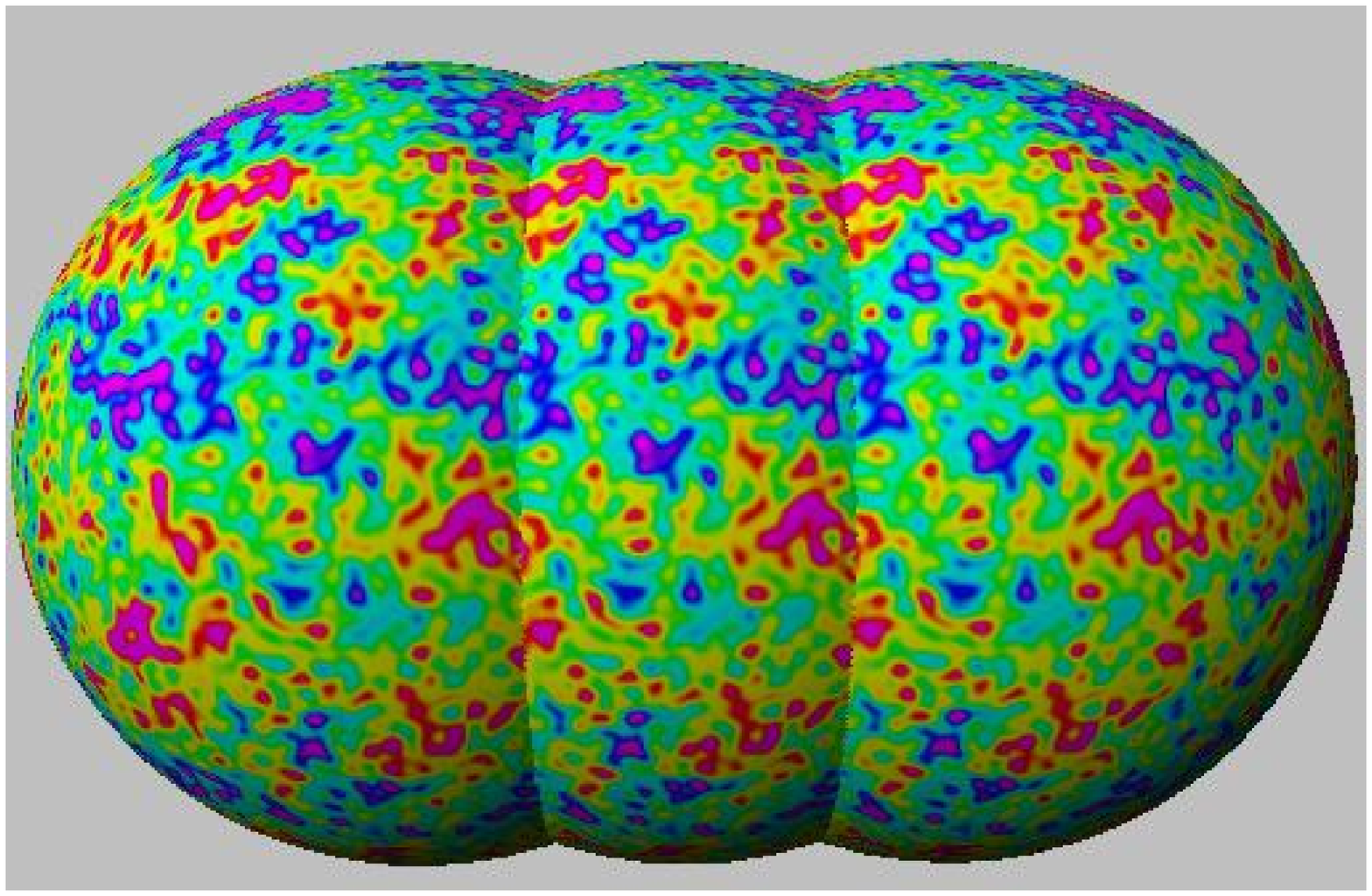,angle=0,width=3.5in}}
\caption{Same as in Fig.~\ref{map2a}, but with the Sachs-Wolfe
contribution only. These maps have comparatively less small scale
power than the previous one as $\ell = 120$ is close to the first dip
in the Sachs-Wolfe spectrum, so that one sees the large scales (up to
the torus angular size) better. Note that the matching between circles
is almost perfect here.}
\label{map2b}
\end{figure}

\begin{figure}
\centerline{\psfig{file=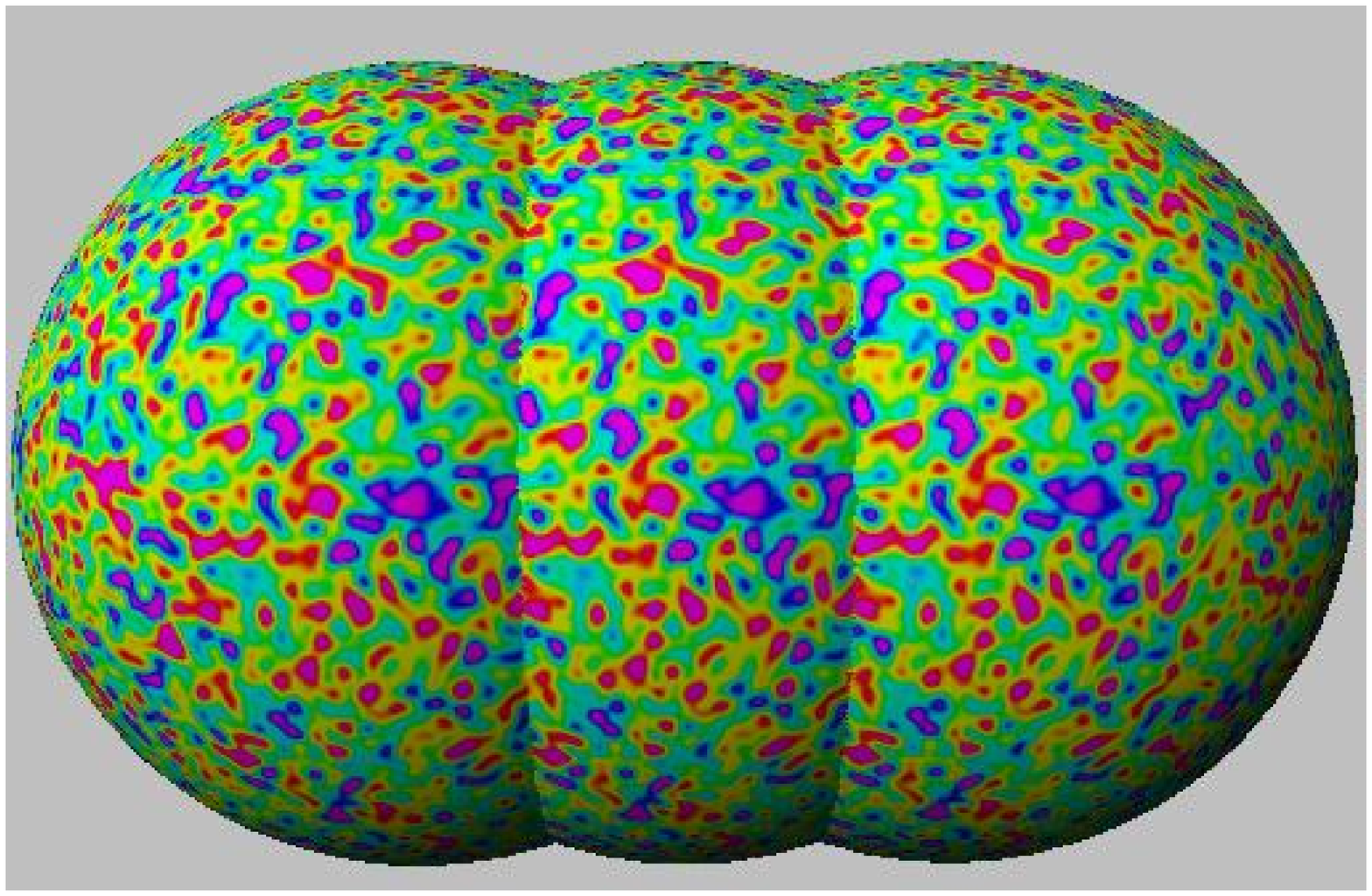,angle=0,width=3.5in}
            \psfig{file=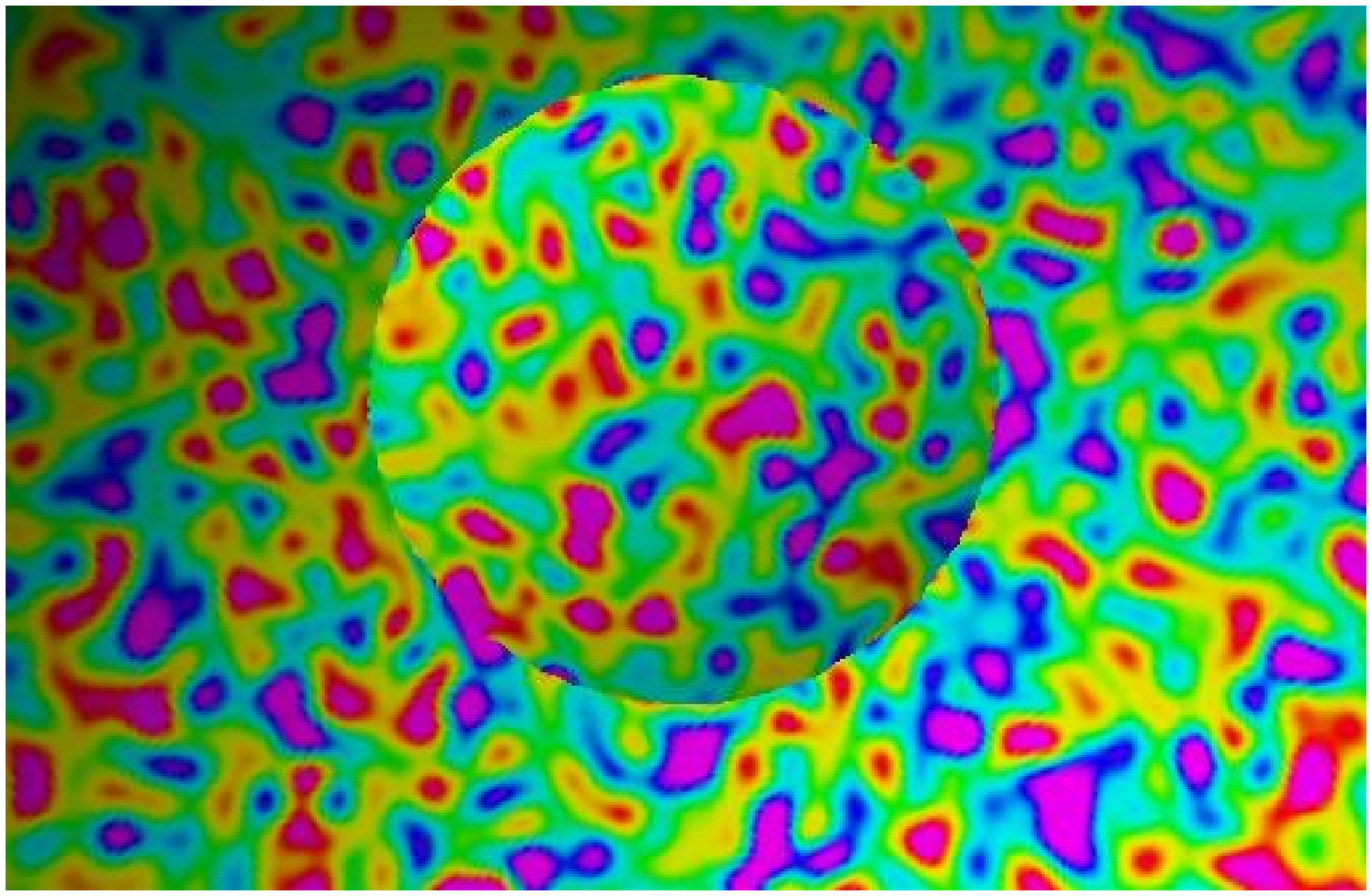,angle=0,width=3.5in}}
\centerline{\psfig{file=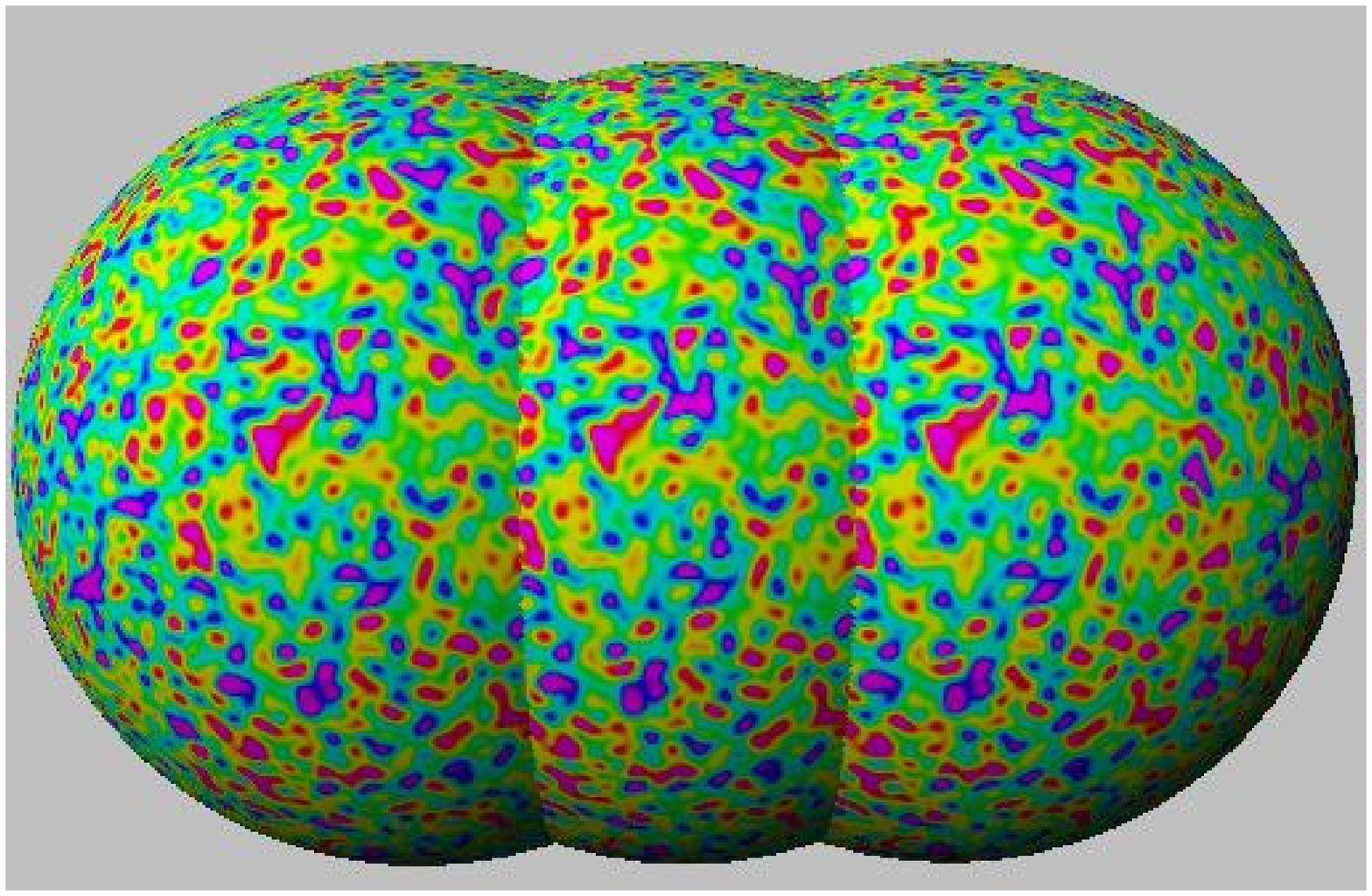,angle=0,width=3.5in}
            \psfig{file=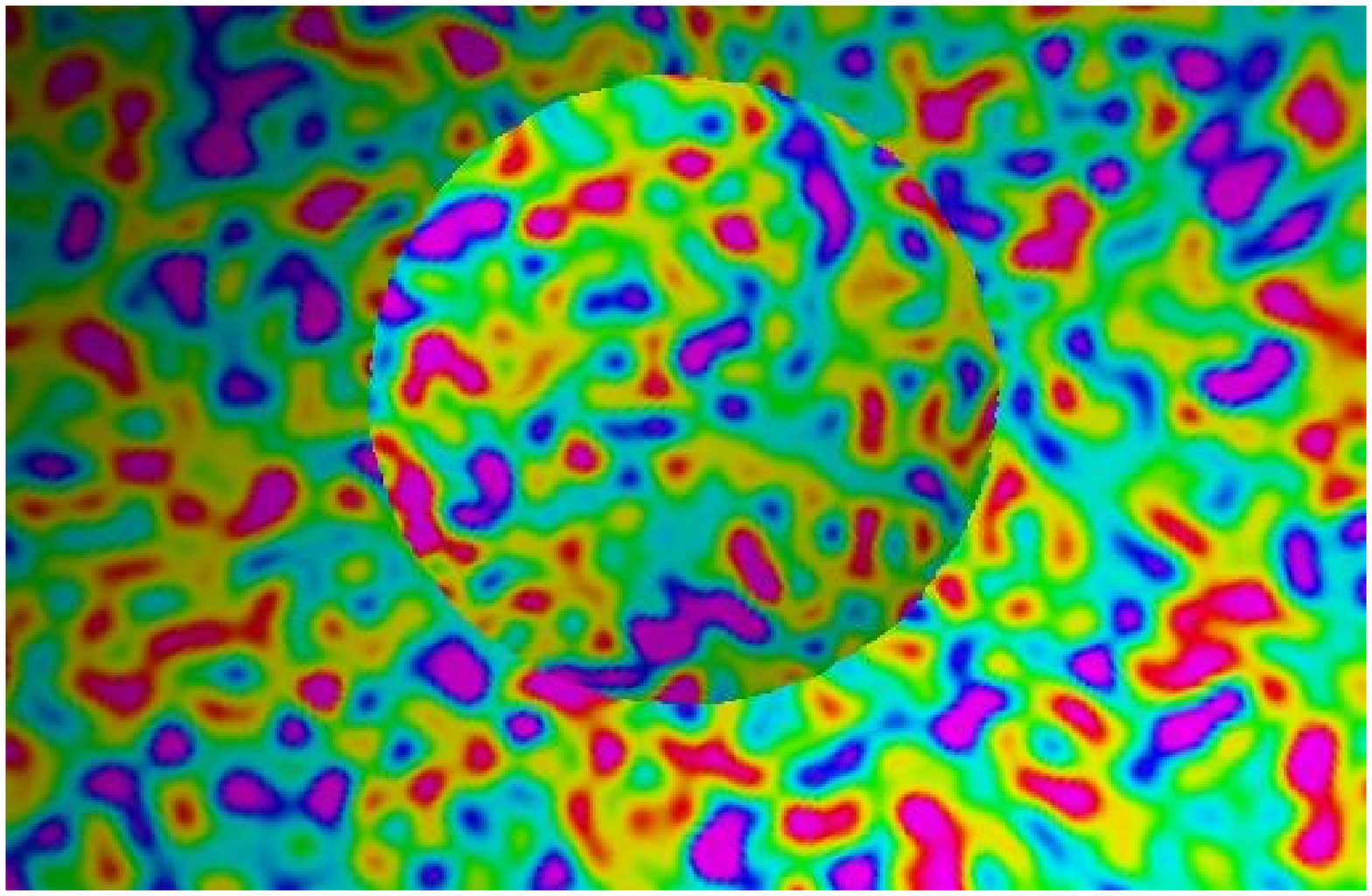,angle=0,width=3.5in}}
\caption{Same as in Fig.~\ref{map2a}, but for the Doppler
component only. Left panels show as previously the last scattering
surface and two of its images from the outside, with large
circles, which are therefore well correlated (but not as well as
for the Sachs-Wolfe or total contributions, however). On the right
panel, we show smaller matching circles ($\theta = \pm 71^\circ$),
which are more conveniently shown from the inside of the last
scattering surface. Here, the anticorrelation between circles is
obvious.} \label{map2c}
\end{figure}

\begin{figure}
\centerline{\psfig{file=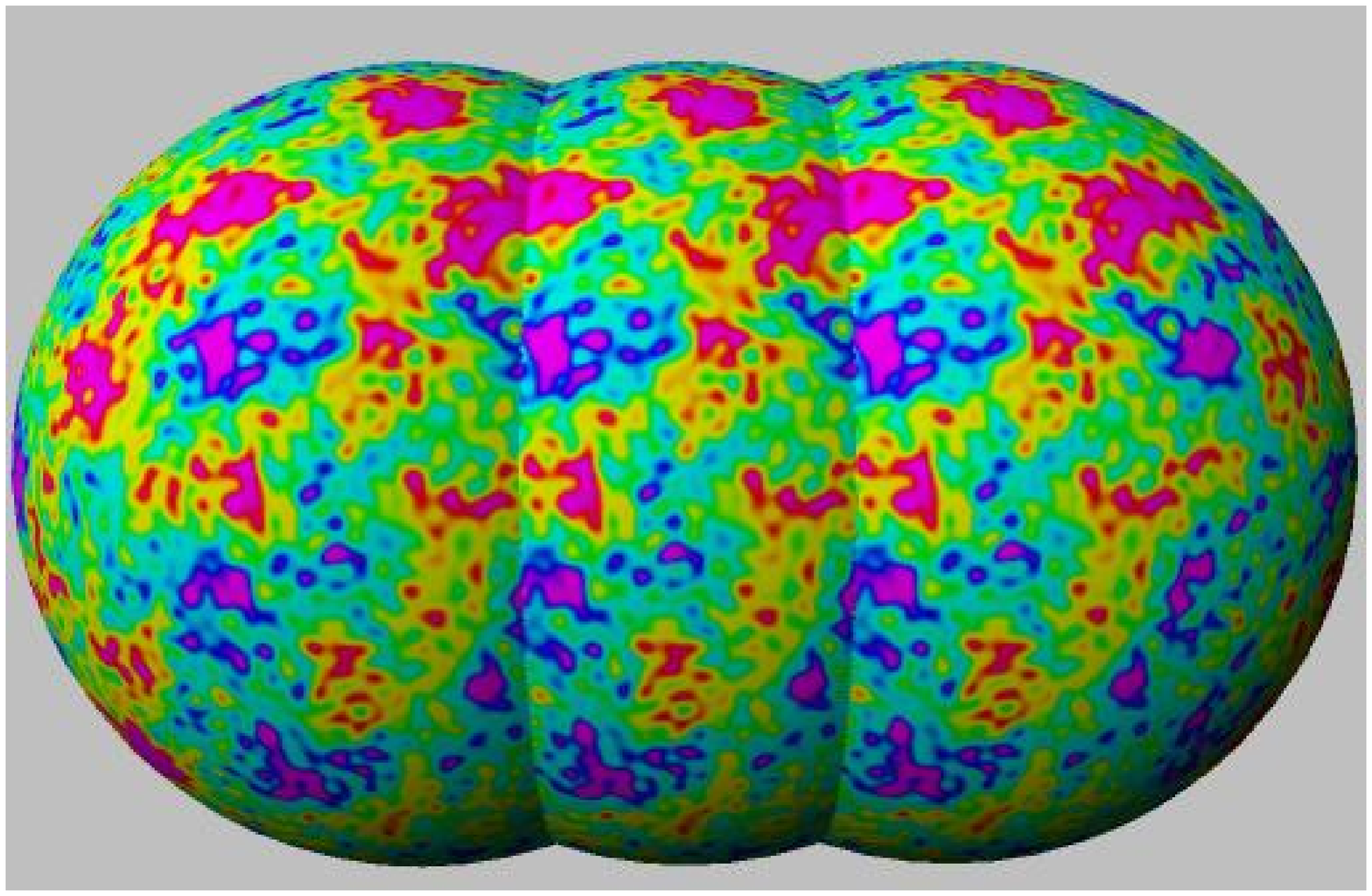,angle=0,width=3.5in}
            \psfig{file=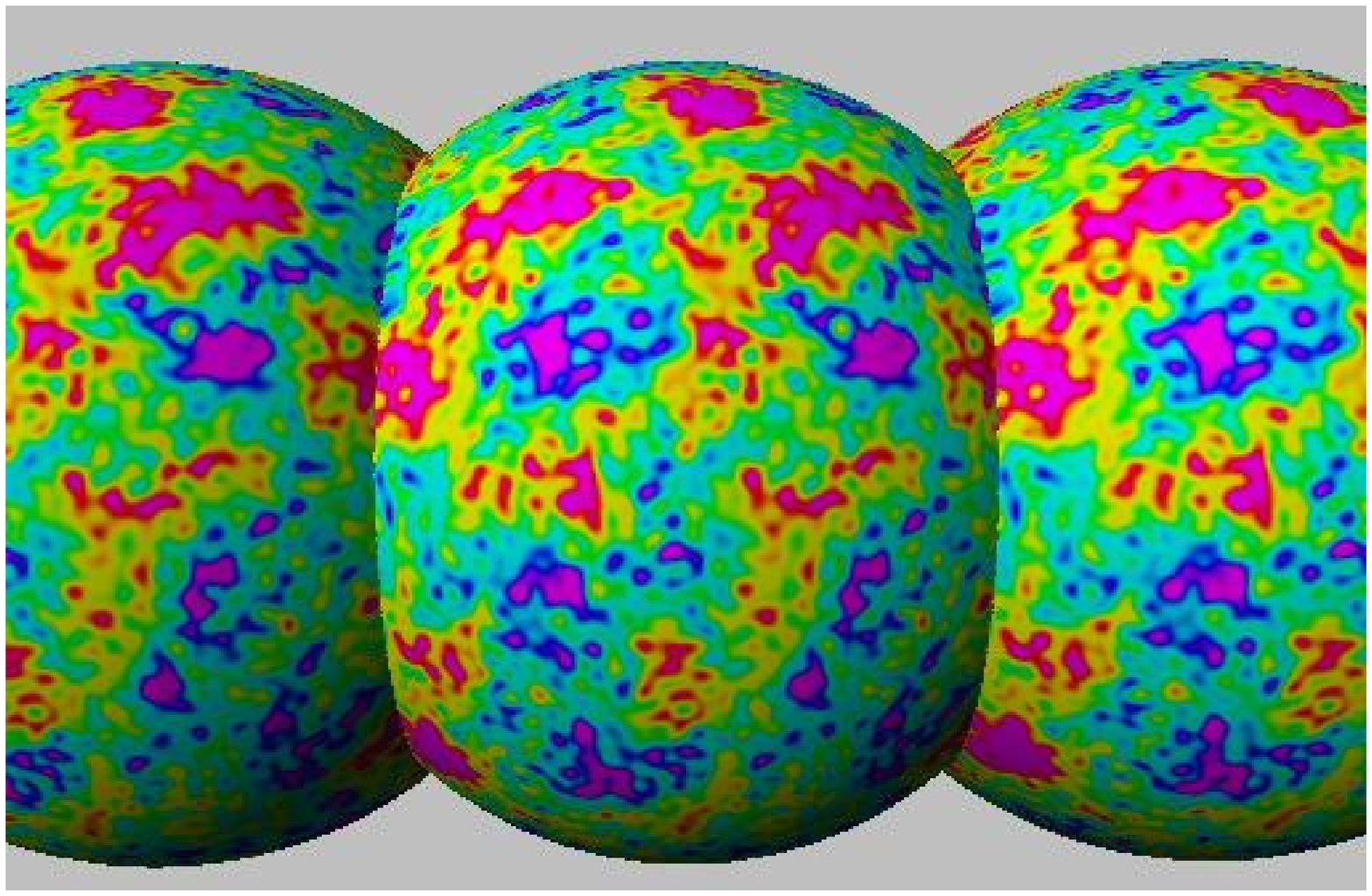,angle=0,width=3.5in}}
\centerline{\psfig{file=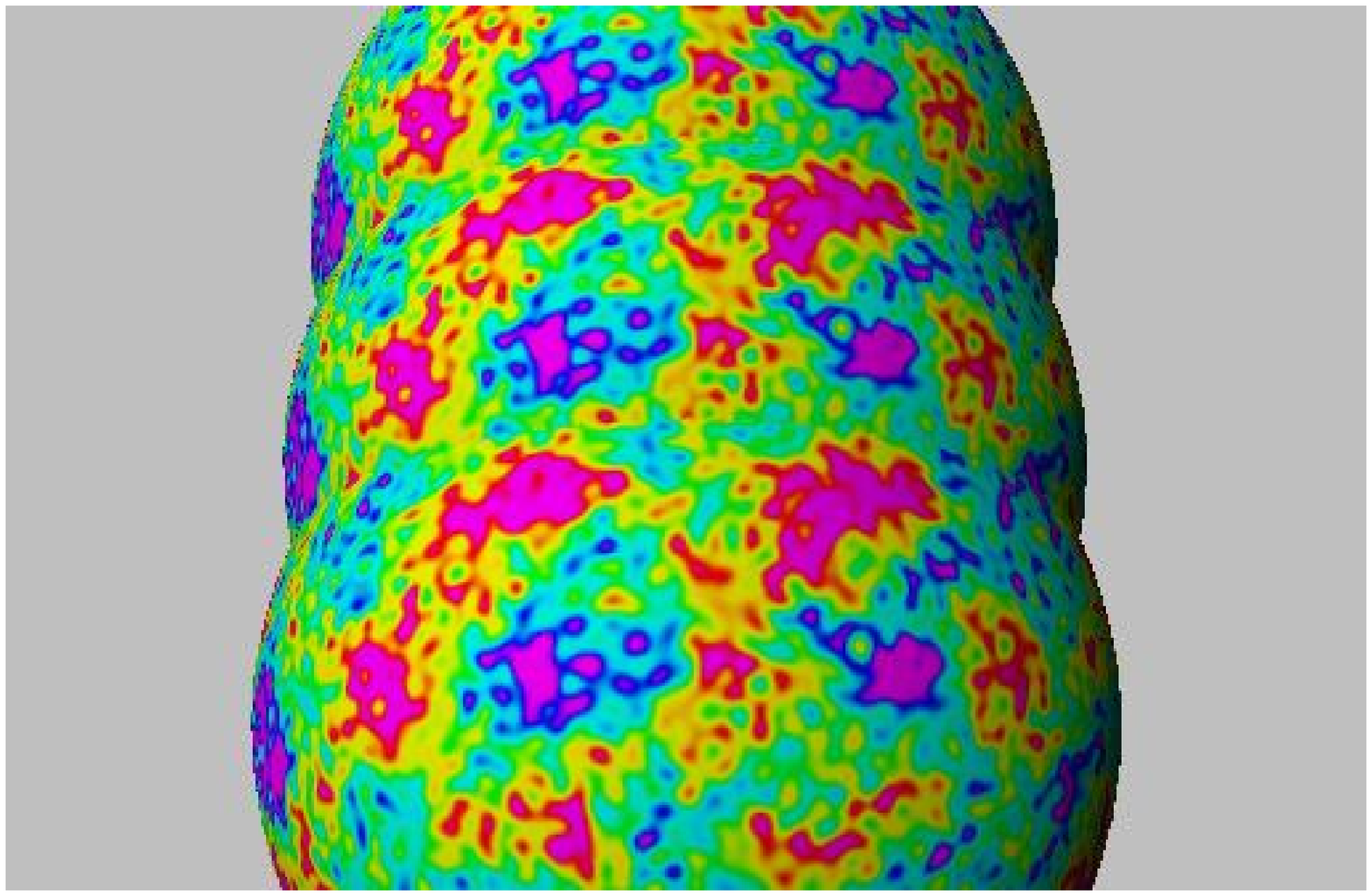,angle=0,width=3.5in}
            \psfig{file=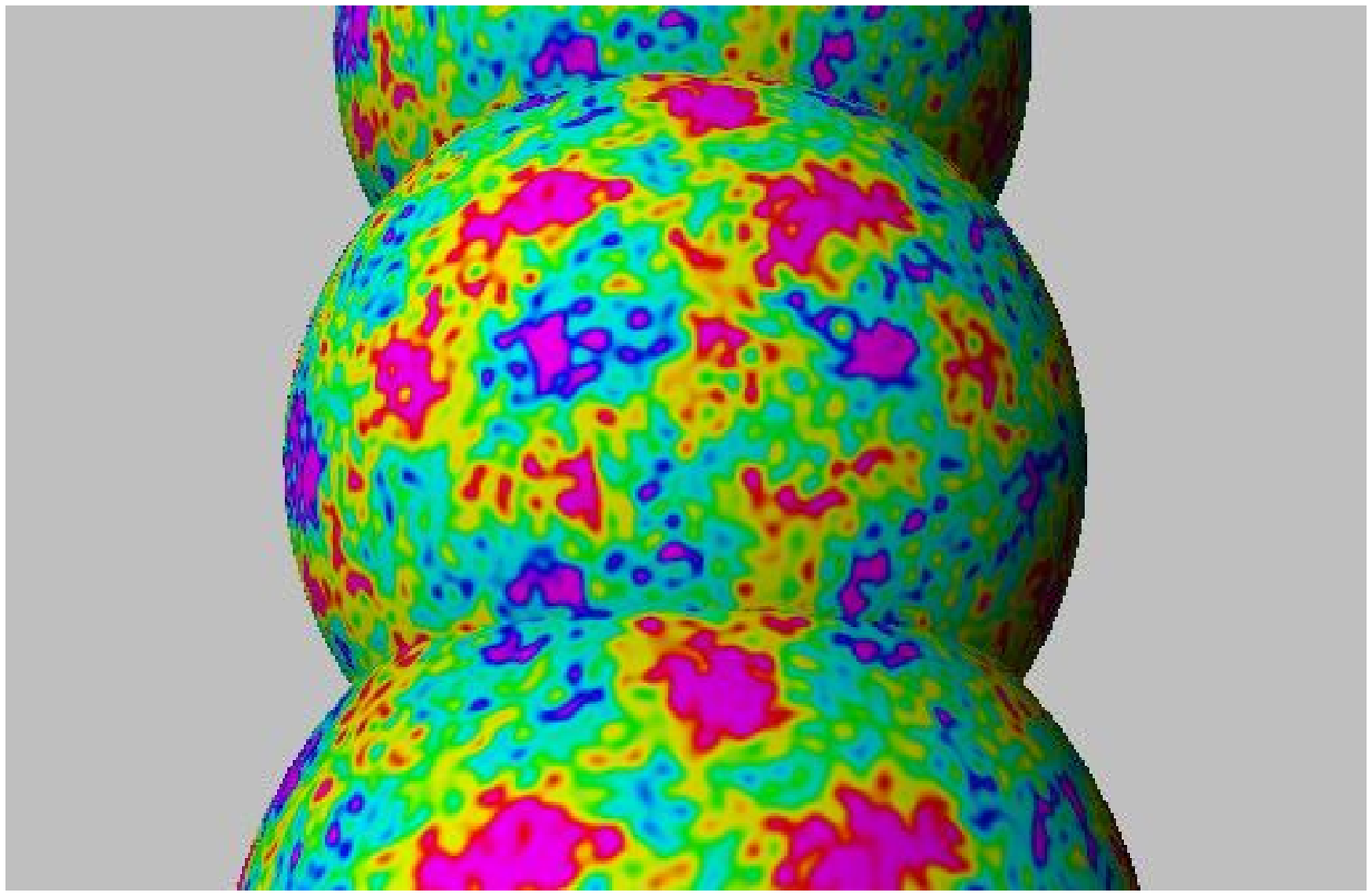,angle=0,width=3.5in}}
\centerline{\psfig{file=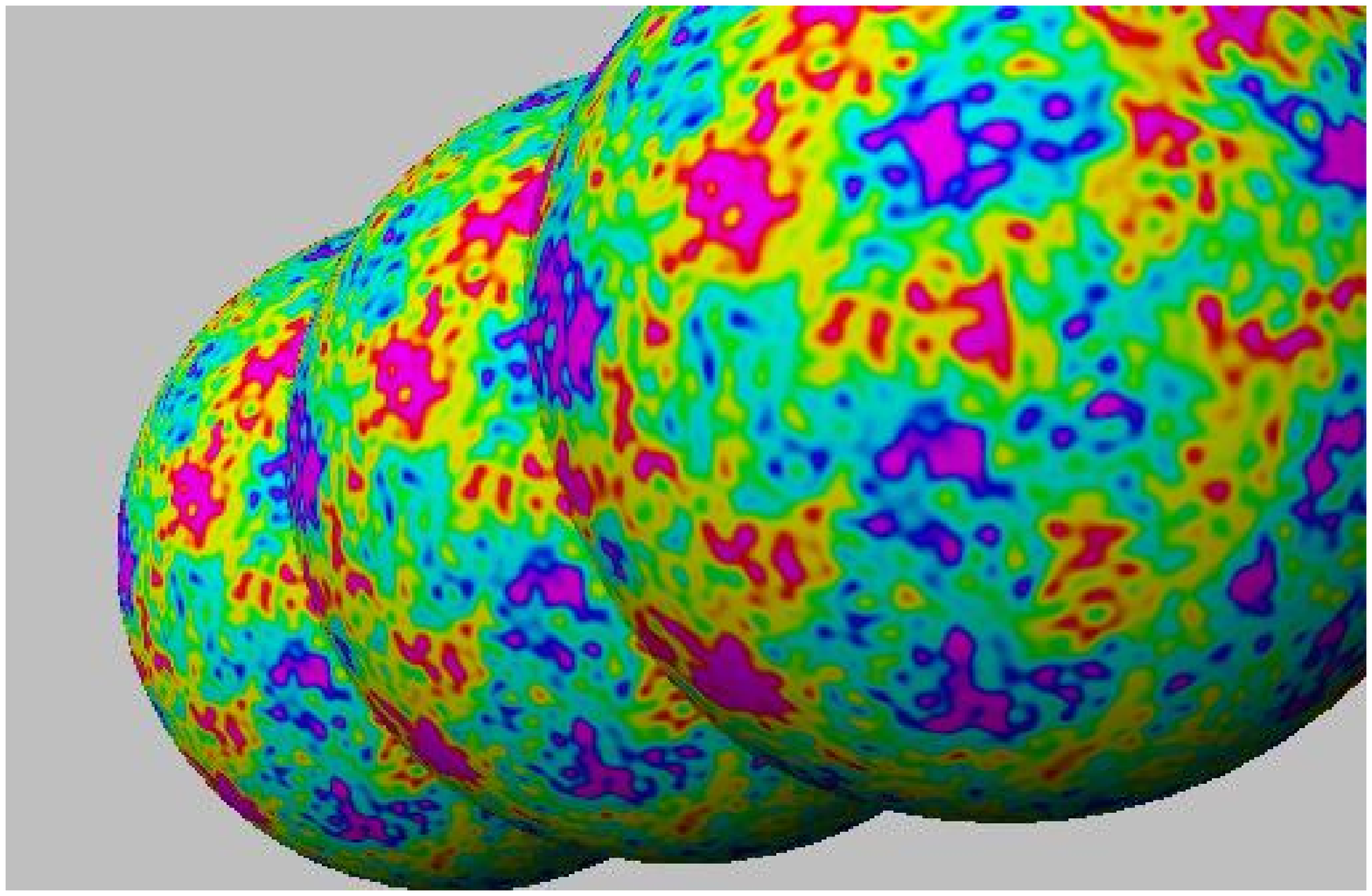,angle=0,width=3.5in}
            \psfig{file=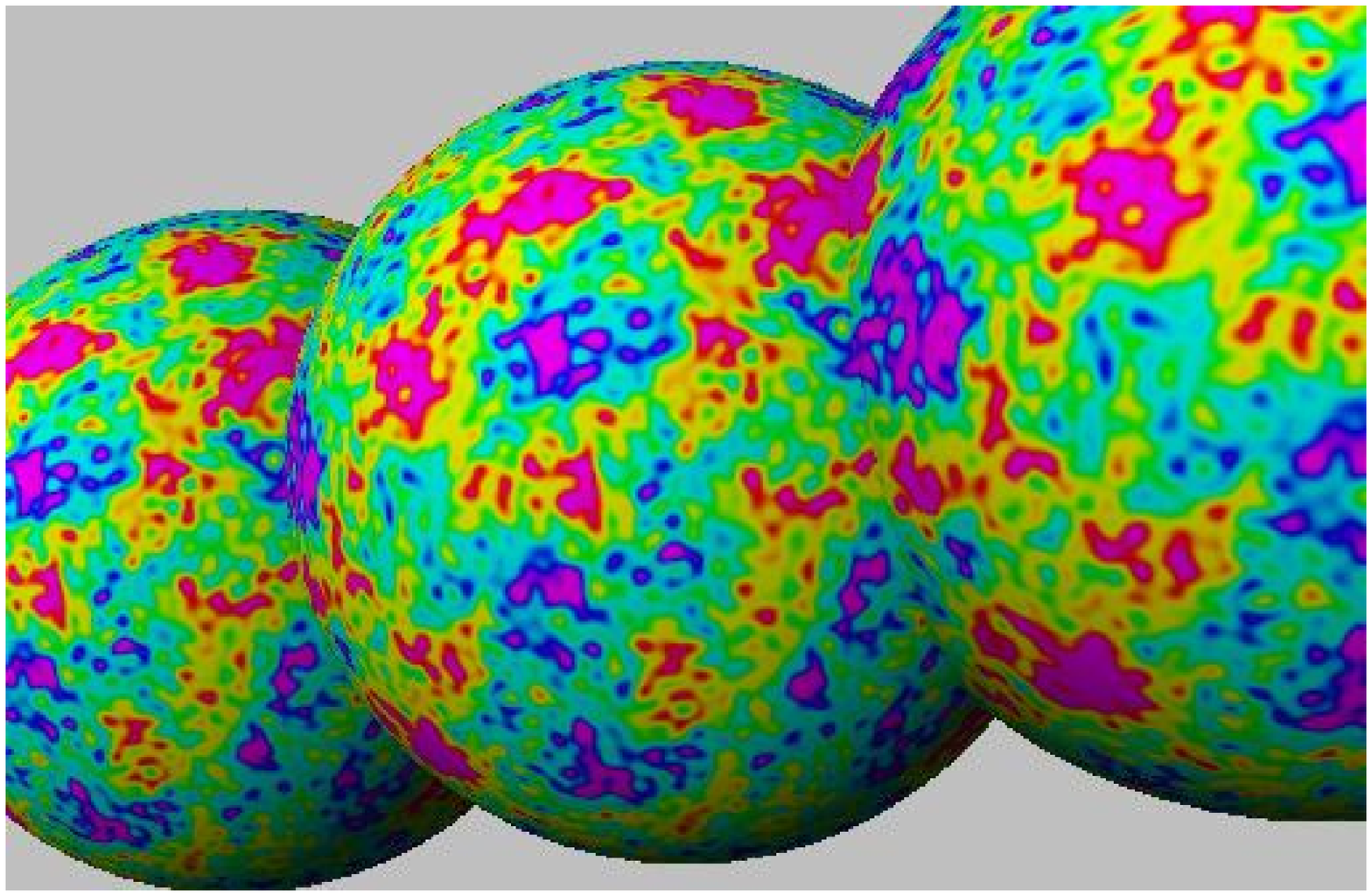,angle=0,width=3.5in}}
\centerline{\psfig{file=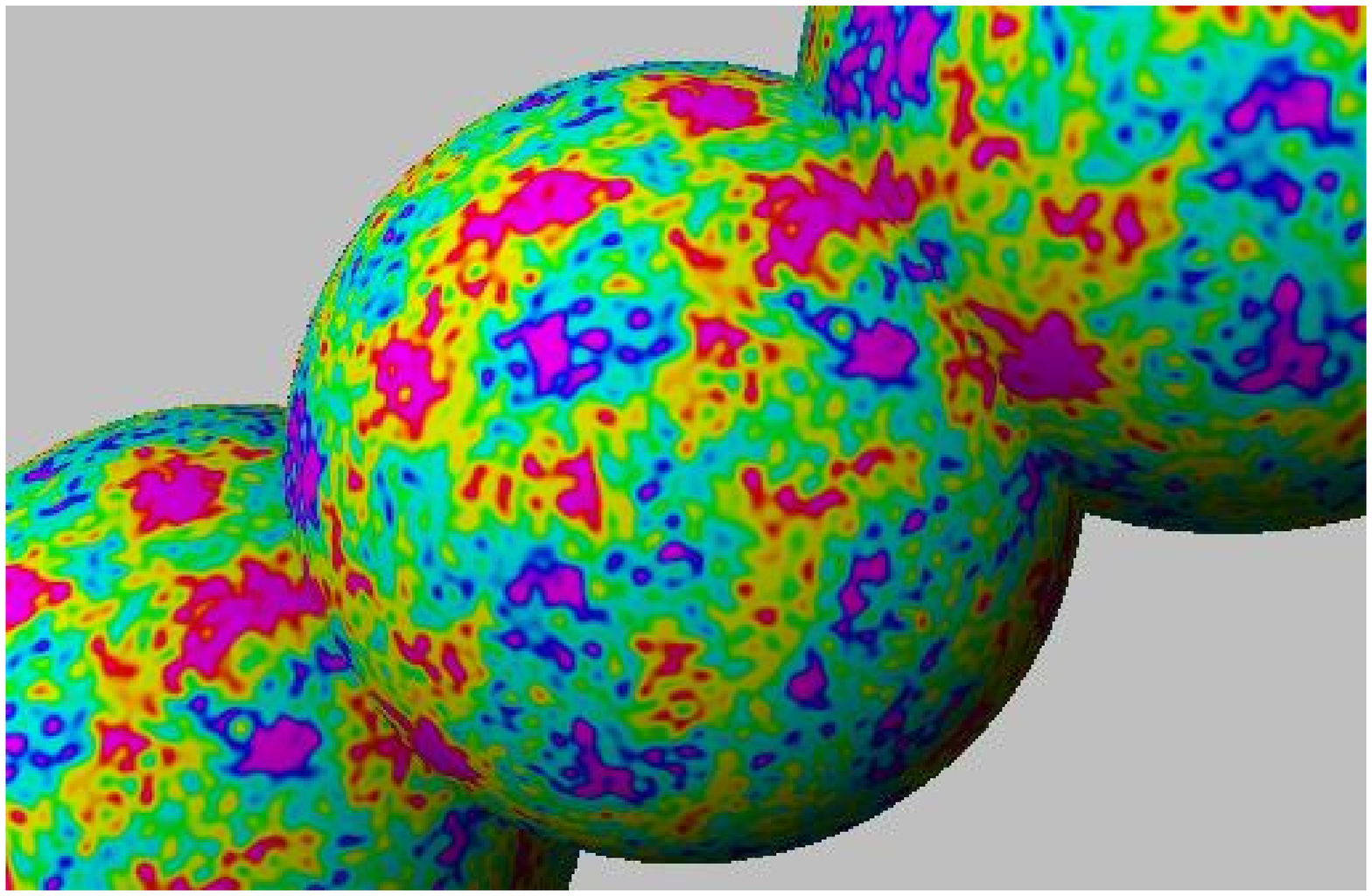,angle=0,width=3.5in}
            \psfig{file=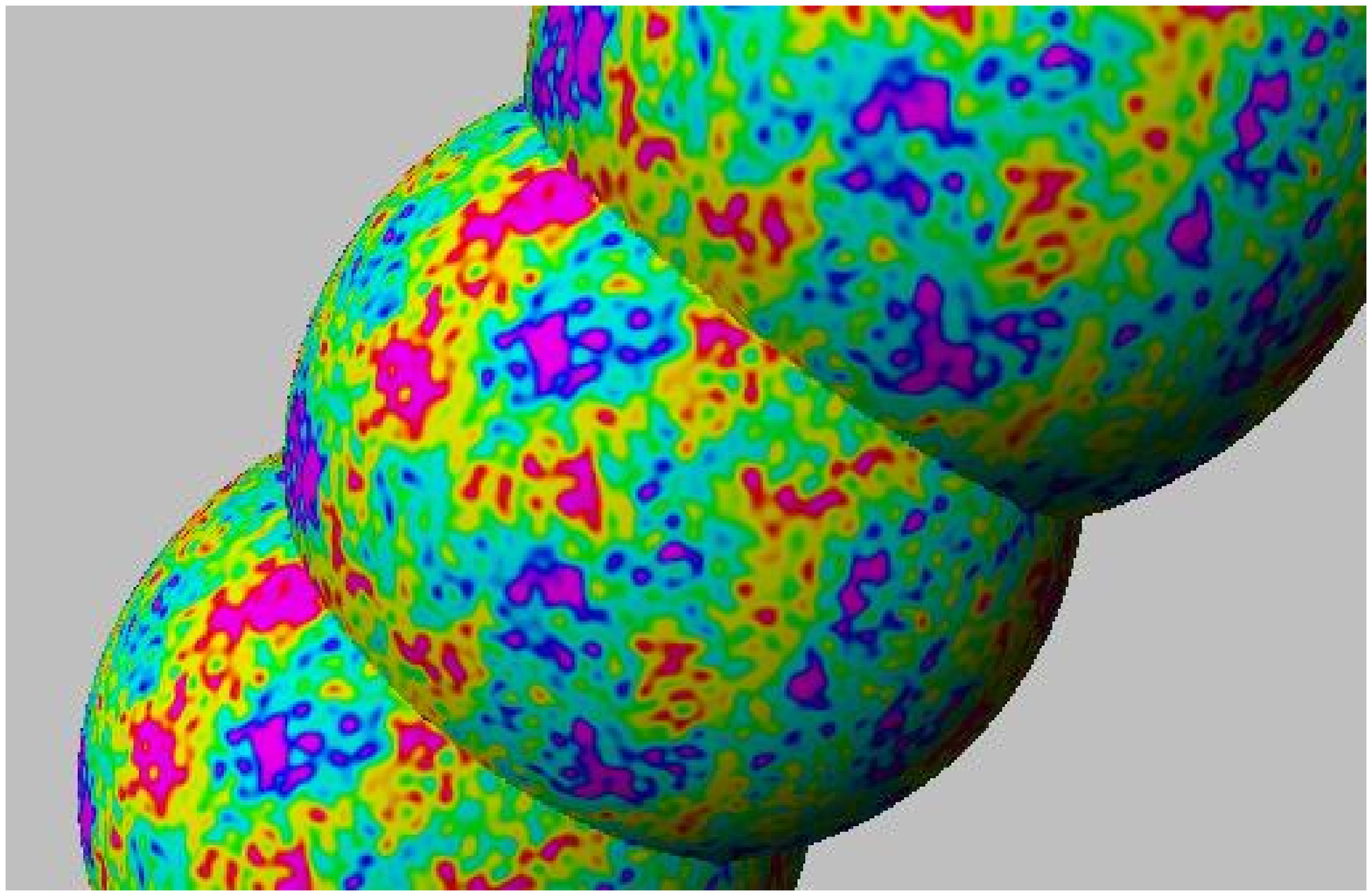,angle=0,width=3.5in}}
\caption{Eight pairs of matching circles among the $61$ existing
pairs for a single realization of the density fields. For clarity,
the orientation of the last scattering surface is the same in all
panels.} \label{map2d}
\end{figure}

We did not find any convenient way to represent the off-diagonal
terms of the correlation matrix. We therefore switch to showing
and analyzing some realizations corresponding to the numerically
computed correlation matrix. In what follows, we have fixed the
size of the torus to $L = 2 R_H$. We therefore have $N \sim 16$
copies of the torus in the observable universe, and from
Eq.~(\ref{cutoffl}) this corresponds to $\ell_c \sim 9$, as can be
checked in Fig.~\ref{fig1}. Although such a model is now excluded
by the data~\cite{sokolov93,staro93,stevens93,costa95}, we analyze
it in detail mostly for pedagogical purposes (and also because the
computing time scales as $L^3$).  In the case where the torus, or
more generally the fundamental domain, is smaller than the last
scattering surface, one expects to see pairs of circles where the
temperature is correlated ~\cite{cornish98}. Seeing these circles
at their expected position is therefore the most crucial test of
the procedure outlined in
Secs.~\ref{sec_cmb},\ref{sec_modes},\ref{sec_simulation}.  As
already announced, the aim here is not to derive a detailed
procedure to detect these circles, but to check our algorithm and
to explore some of the properties of these matching circles. Here,
we have $2 R_\LSS / L = 3.17 \sim \sqrt{10}$. One therefore
expects to have $61$ pairs of circles and $12$ pairs of points
having correlated temperature\footnote{The circles correspond to
the intersection of the last scattering surface with translates of
the form $L\cdot(n_1, n_2, n_3)$, where $(n_1, n_2, n_3) = (0, 0,
1)$, $(0, 0, 2)$, $(0, 0, 3)$, $(0, 1, 1)$, $(0, 1, 2)$, $(0, 2,
2)$, $(1, 1, 1)$, $(1, 1, 2)$, and $(1, 2, 2)$ plus all
permutations and sign changes among each triplet $(n_1, n_2,
n_3)$. The pairs of points correspond to the case where the
intersection between the last scattering surface and its translate
reduces to almost a single point, as is the case for $(n_1, n_2,
n_3) = (0, 1, 3)$ and its permutations and sign changes.}.  In
order to see the circles, it is convenient to show the last
scattering surface as a sphere seen from the outside and to look
at its intersection with itself after a translation of the form
$L\cdot(n_1, n_2, n_3)$, as shown in
Figs.~\ref{map2a}--\ref{map2d}.

\begin{figure}
\centerline{\psfig{file=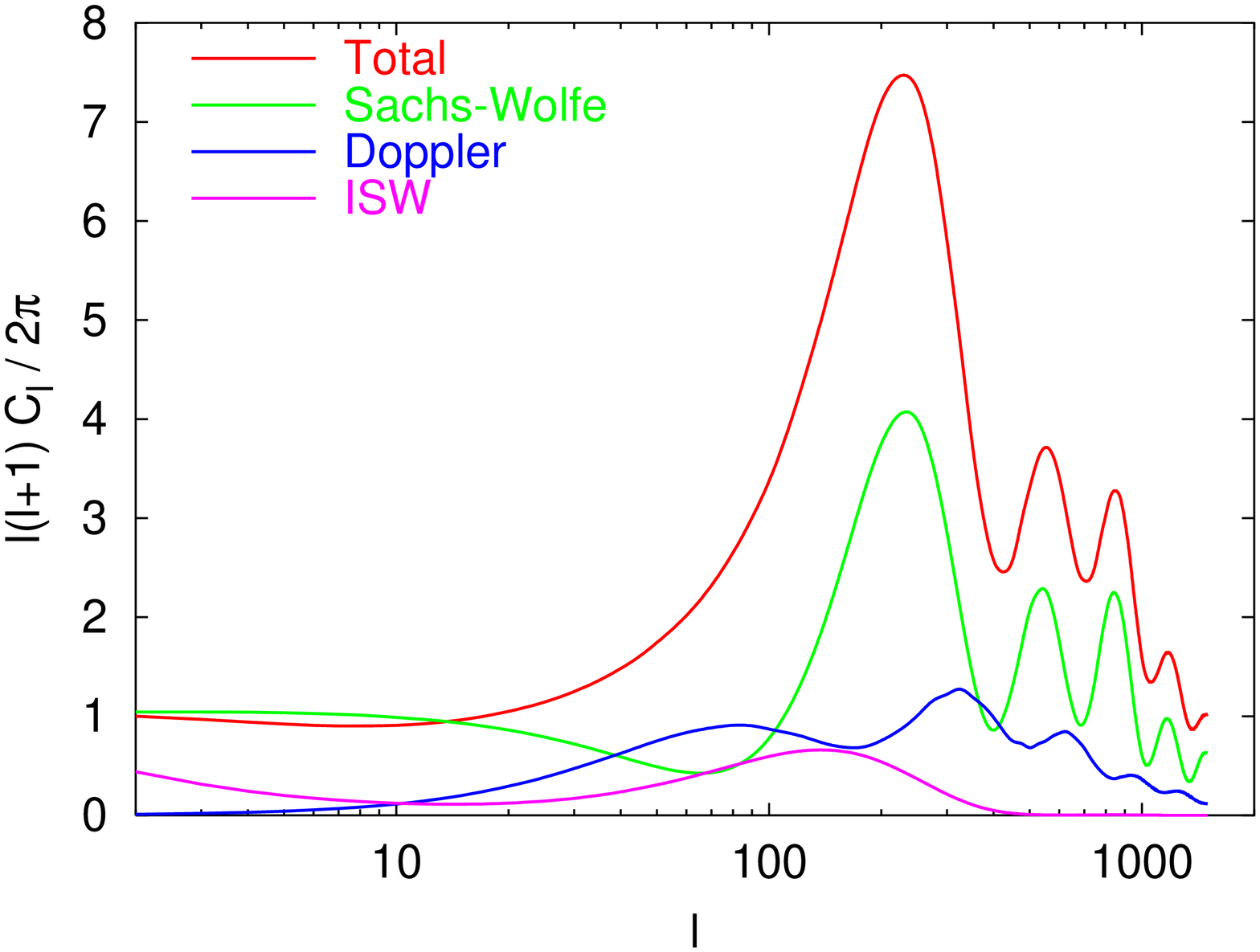,angle=270,width=3.5in}
            \psfig{file=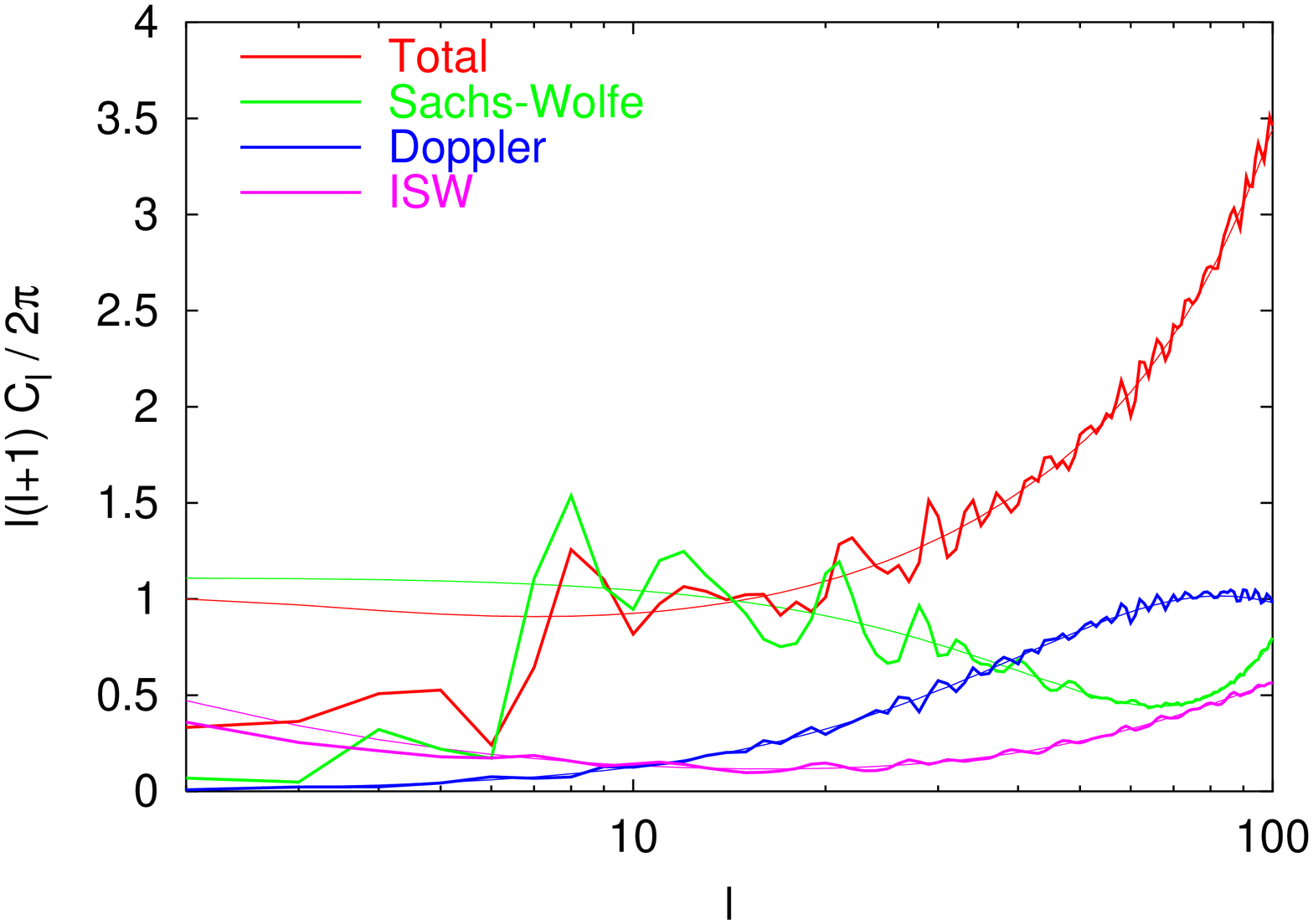,angle=270,width=3.5in}}
\caption{Decomposition of CMB anisotropies into the Sachs-Wolfe,
Doppler and ISW contributions. When the topology is simply
connected [left panel], the Sachs-Wolfe contribution is dominant
at the peaks and usually on the largest scales. The Doppler term
is dominant only before the first peak, and has a significant
contribution between peaks. The ISW term appears mostly at the
first peak (this is referred to as the early ISW effect) when the
radiation-to-matter transition occurs late (typically at low $h$),
and at large scales (late ISW effect) when the universe is not
matter dominated at $z = 0$. For standard values of the
cosmological parameters, it is not dominant. The right panel shows
the initial portion of the spectrum ($\ell \leq 100$) for a model
with the same cosmological parameters but with a toroidal
topology. The Sachs-Wolfe contribution shows a sharp cutoff at the
expected scale. Note that neither the Doppler nor ISW
contributions show a similar cutoff. This is due to two different
projection effects: for the Doppler term, this comes from the fact
that it is always negligible on large scales in $k$ space (it
scales as $k^2$), and when one goes to $\ell$ space, the
convolution~(\ref{swdopisw}) always transfers some power from the
small scales (where there is power in $k$ space) to large scales;
for the ISW, the presence of power comes from the fact that it is
generating long after the last scattering epoch, so that it
appears on large scales simply because it describes phenomena that
occur near us. However, since the Sachs-Wolfe contribution is
usually largest on large scales, the final spectrum still exhibits
a sharp cutoff.} \label{figdec}
\end{figure}

In the last Section, we pointed out that the correlation would depend
on the amount of Sachs-Wolfe, Doppler and integrated Sachs-Wolfe
effects. The decomposition of the temperature anisotropies both in the
simply connected topology and in the toroidal cases are shown in
Fig.~\ref{figdec}. Note that we show only the relative amplitude of
these effects and not their cross correlation.

We now turn to the correlation between pairs of circles, as
introduced in ~\cite{cornish98}.

If the topology is not known in advance, the relative position between
matching circles can be arbitrary, so that in general the search for
circles is a six parameter problem: two parameters for the center of
the first circle, two more for the center of the second circle, one
for their common radius, and one for their relative phase (i.e. the
twist with which they are identified). In the case of a torus, the
circles sit directly opposite each other on the sky (eliminating two
parameters) and there is no twist (eliminating another parameter), so
the problem reduces to a three parameters search\footnote{Another way
to see that in a torus it's a three parameter search is to visualize
the situation in the universal covering space.  Place one copy of the
last scattering surface with its center at the origin, and imagine a
translated copy with its center at some point $(x,y,z)$.  Each choice
of $(x,y,z)$ uniquely determines a circle of intersection (assuming $0
< x^2 + y^2 + z^2 < R_\LSS^2$), and conversely each pair of circles
arises from exactly two points $(x,y,z)$ and $(-x,-y,-z)$ and no
others.  Thus the point $(x,y,z)$ serves to parameterize the circle
search.}. We are not going to perform such a study, but rather focus
on some features of matching circles in a toroidal universe.

A simple estimator for the correlation between pairs of circles
that are horizontal with respect to the coordinate system is
obviously
\begin{equation}
\label{ct1t2} C (\theta_1, \theta_2) \equiv \frac{1}{2\pi} \int
 \frac{2 \Theta (\theta_1, \varphi) \Theta (\theta_2, \varphi)}
      {\Theta^2 (\theta_1, \varphi) + \Theta^2 (\theta_2, \varphi)}
 \ddd \varphi ,
\end{equation}
where we have set $\Theta \equiv \delta T / T$.  If the temperature
fluctuations on a pair of circles are completely uncorrelated, then on
average $C = 0$. If they are completely correlated, then $C = 1$, and
if they are anti correlated, then $C = - 1$. Each $\ALM{\ell}{m}$ can
be seen either as the theoretical expectation for a given model, or an
observed quantity that can be measured from the CMB sky. In the first
case, it represents a feature that one can expect from a given model,
and in the second case it represents an estimator of some features
predicted by the topology. Here, we shall concentrate on the {\em
observed} $C (\theta_1, \theta_2)$ that we compute from simulated
maps, first to check the validity of our procedure to compute the
correlation matrix $\CLMLPMP{\ell}{m}{\ell'}{m'}$, and second to
convince us that it is possible to see the presence of matching
circles using simple techniques (although we do not pretend that this
method is optimal).

In principle, two matching circles have the same angular diameter, so
that only the case $\theta_2 = - \theta_1$ is relevant, but we have
chosen to leave $\theta_2$ as a free parameter to see to what extent
uncorrelated circles might happen to seem correlated by chance.

\begin{figure}
\centerline{\psfig{file=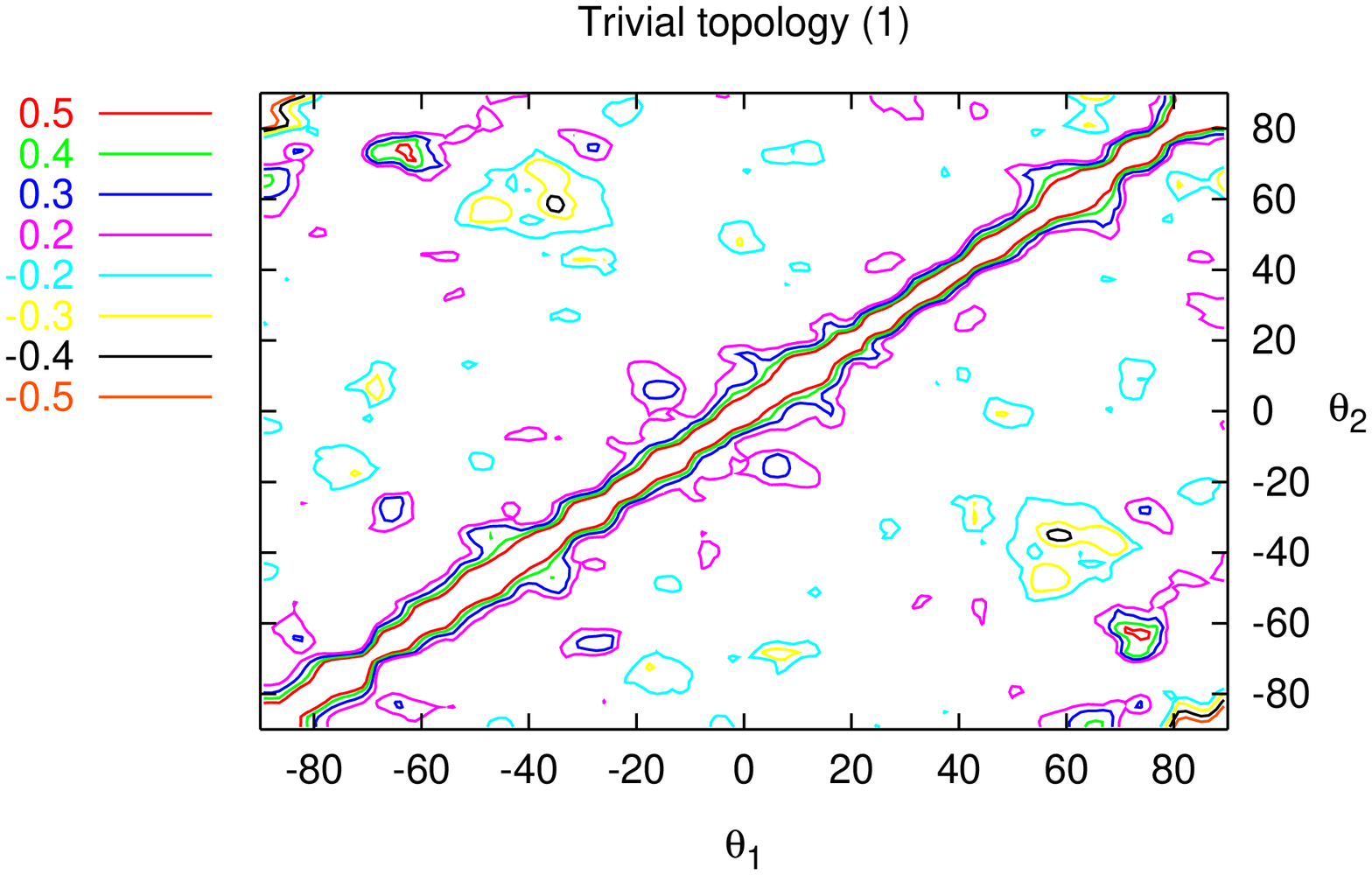,bbllx=135pt,bblly=120pt,
                   bburx=500pt,bbury=660pt,angle=270,width=3.5in}
            \psfig{file=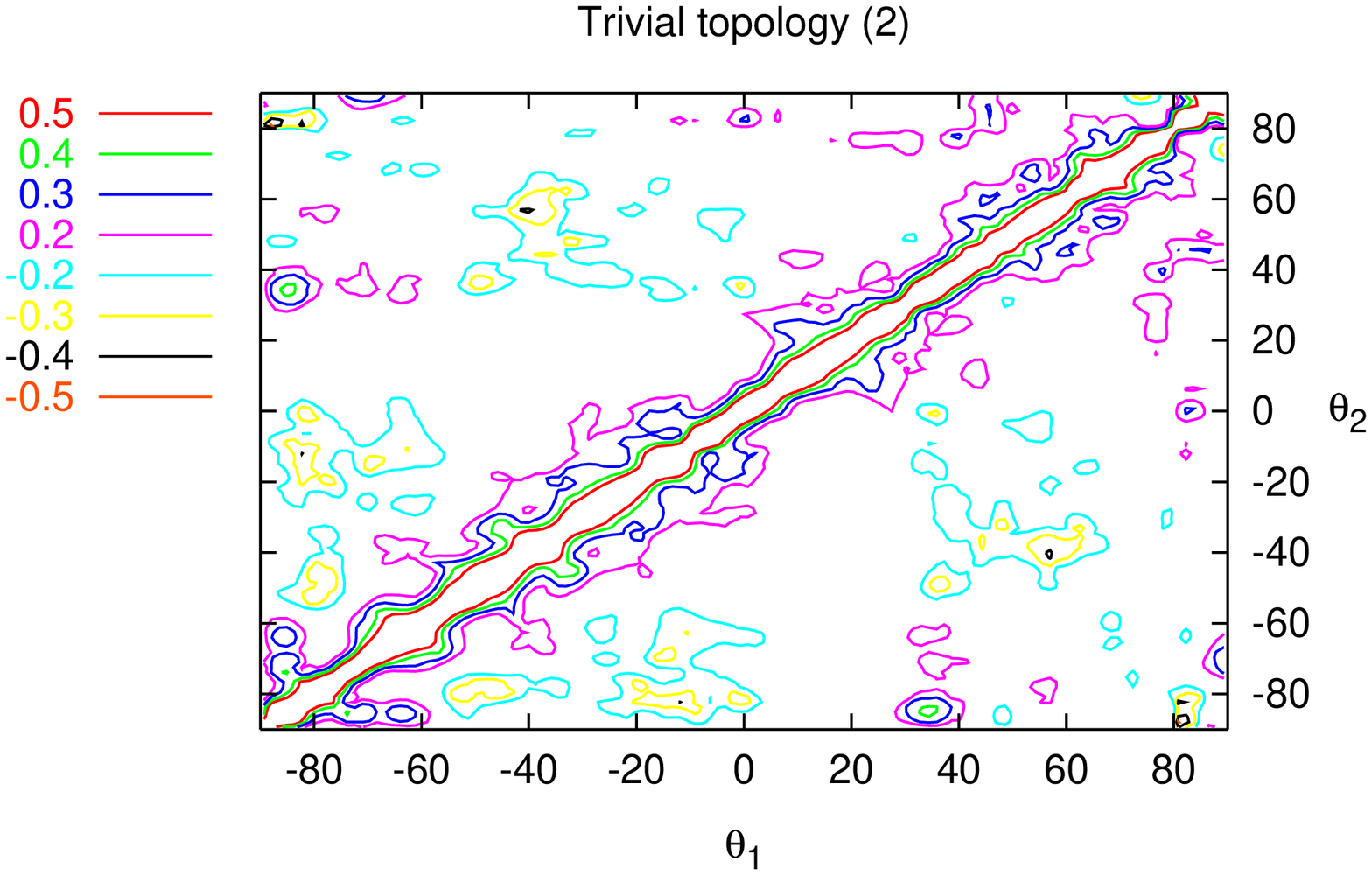,bbllx=135pt,bblly=120pt,
                   bburx=500pt,bbury=660pt,angle=270,width=3.5in}}
\centerline{\psfig{file=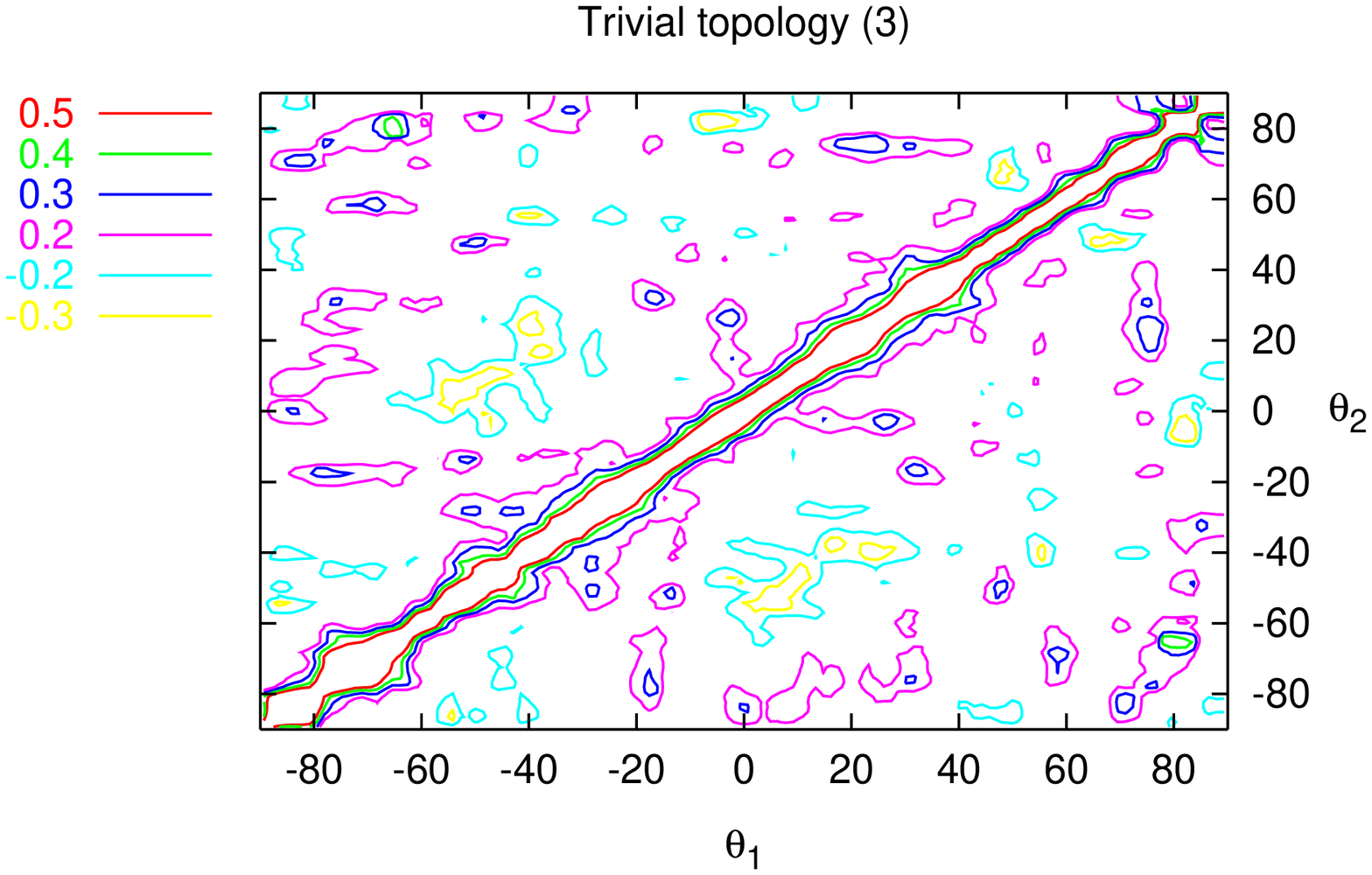,bbllx=135pt,bblly=120pt,
                   bburx=500pt,bbury=660pt,angle=270,width=3.5in}
                   \psfig{file=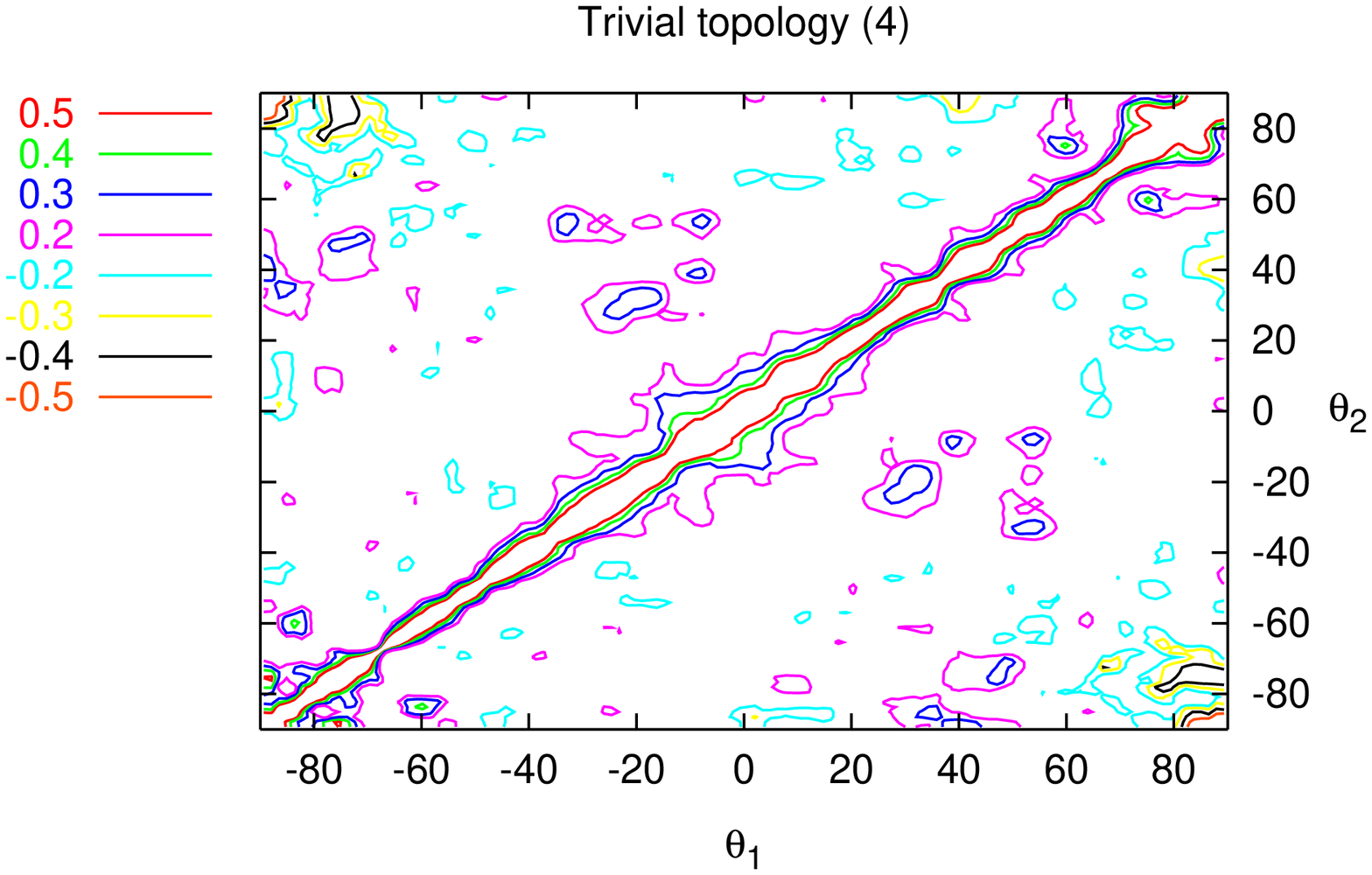,bbllx=135pt,bblly=120pt,
                   bburx=500pt,bbury=660pt,angle=270,width=3.5in}}
\caption{Contour plots of the function $C (\theta_1, \theta_2)$
for four realizations of the temperature anisotropies in a universe
with a simply connected topology. When $|\theta_1 - \theta_2|$ is
small, the correlation between circles exists. When $|\theta_1 -
\theta_2|$ is larger, the apparent correlation between circles
comes from statistical fluctuations, which would be reduced for
higher resolution maps.} \label{fig10}
\end{figure}

We first show in Fig.~\ref{fig10} a few examples of the observed
function $C (\theta_1, \theta_2)$ for a simply connected universe. As
expected, the correlation is quite large when $\theta_1 \sim \theta_2$
because the circles are near each other and the real space correlation
function $C (\theta)$ (the Legendre transform of the $C_\ell$) is not
0 when $\theta \to 0$. With our normalization of $C(\theta_1,
\theta_2)$, one has
\begin{equation}
\lim_{\theta_2 \to \theta_1} C (\theta_1, \theta_2) = 1 ,
\end{equation}
which will of course remain valid when the topology is
multi-connected.  When the separation between $\theta_1$ and
$\theta_2$ is large, one can neglect the correlation between the two
circles, and the main contribution to $C$ comes from statistical
fluctuations: it is always possible that two circles exhibit similar
temperature patterns by chance. The variance of these statistical
fluctuations is probably given by the number of independent pixels on
the map and therefore by a combination of the scale at which the power
spectrum is large and of the resolution of the map (here, $\ell_\MAX =
30$). In any case, the amplitude of the largest statistical
fluctuations of $C(\theta_1, \theta_2)$ gives an idea of the amplitude
of the signal needed to detect a multiconnected topology\footnote{The
signal threshold could therefore be reduced by performing the same
analysis on a higher resolution map, but because the search for
circles is in general a six-parameter problem, it might be necessary
to search low resolution maps first to find likely candidates, and
then search higher resolution maps to confirm them.}. For the maps we
have generated, the correlation reaches $30\%$ for a few pairs of
falsely matched circles.

\begin{figure}
\centerline{\psfig{file=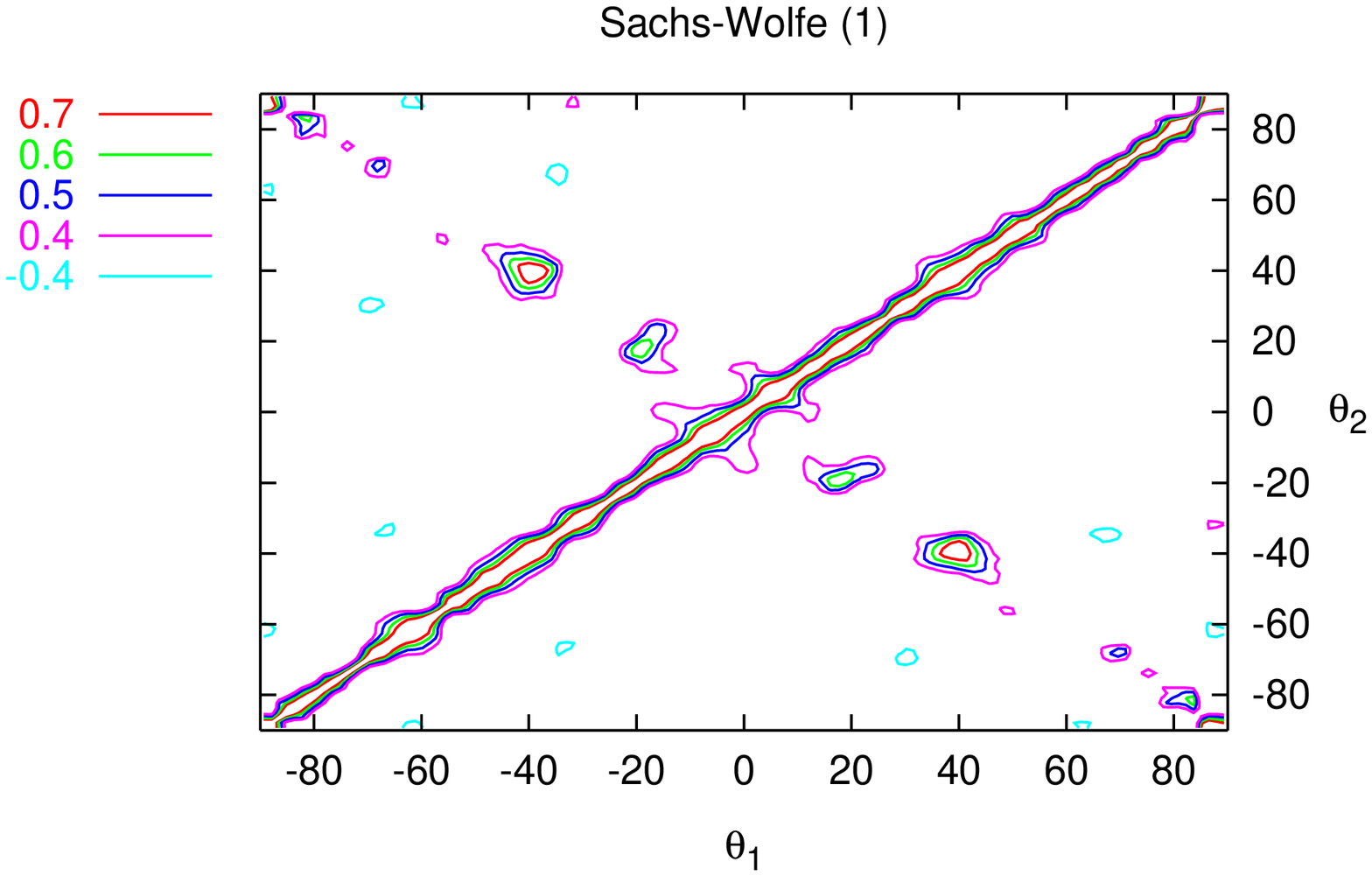,bbllx=135pt,bblly=120pt,
                   bburx=500pt,bbury=660pt,angle=270,width=3.5in}
            \psfig{file=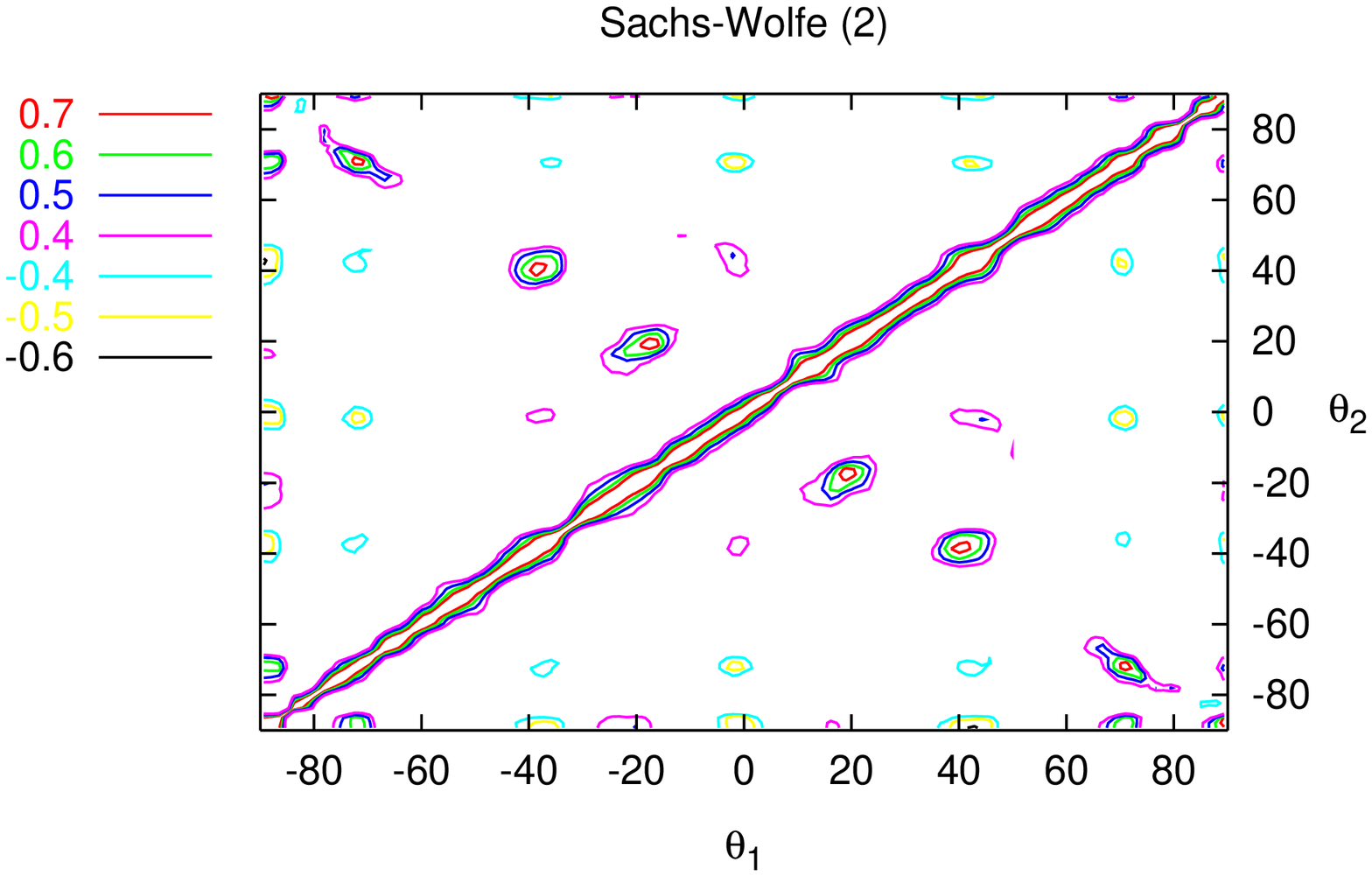,bbllx=135pt,bblly=120pt,
                   bburx=500pt,bbury=660pt,angle=270,width=3.5in}}
\centerline{\psfig{file=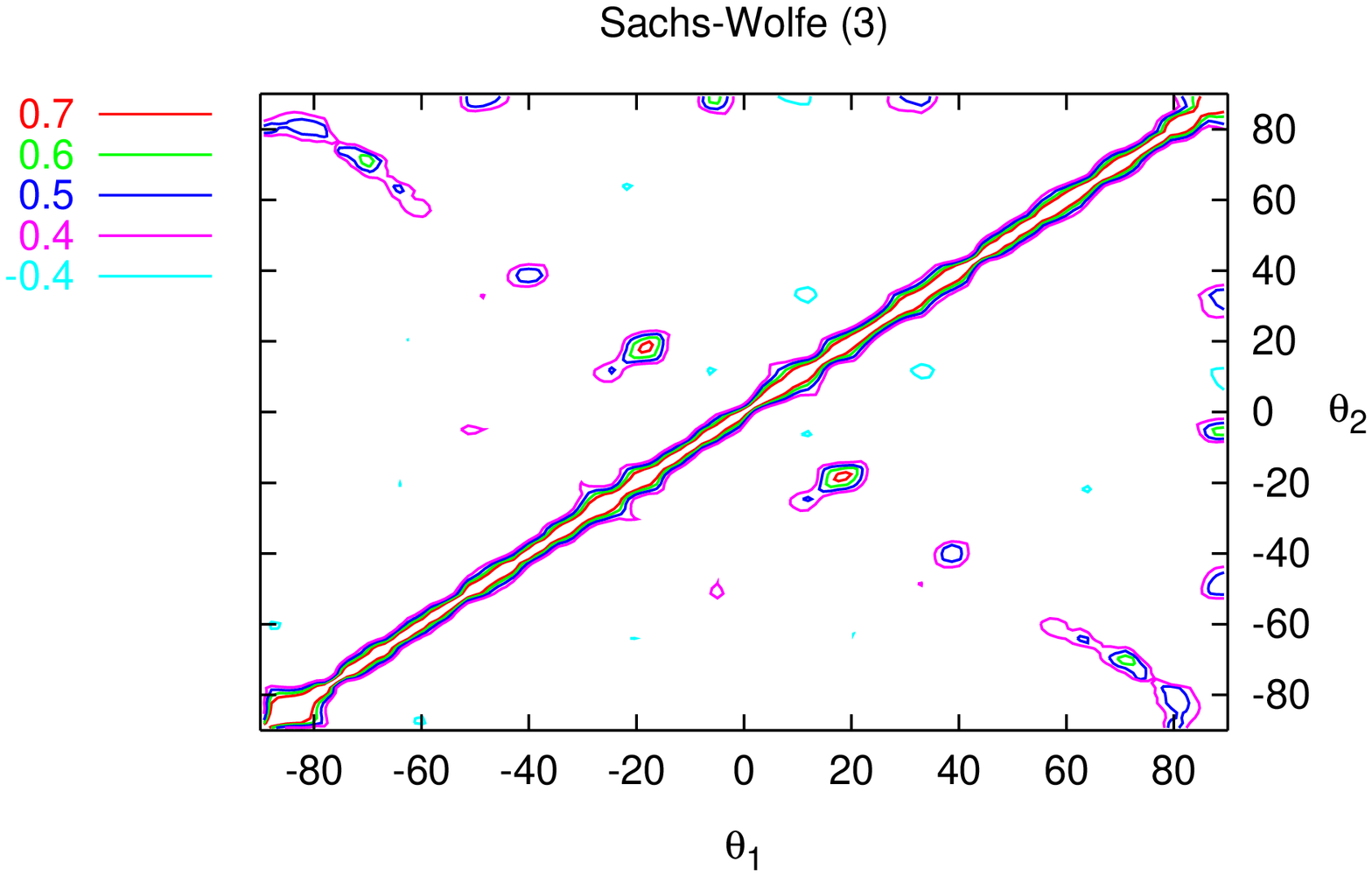,bbllx=135pt,bblly=120pt,
                   bburx=500pt,bbury=660pt,angle=270,width=3.5in}
            \psfig{file=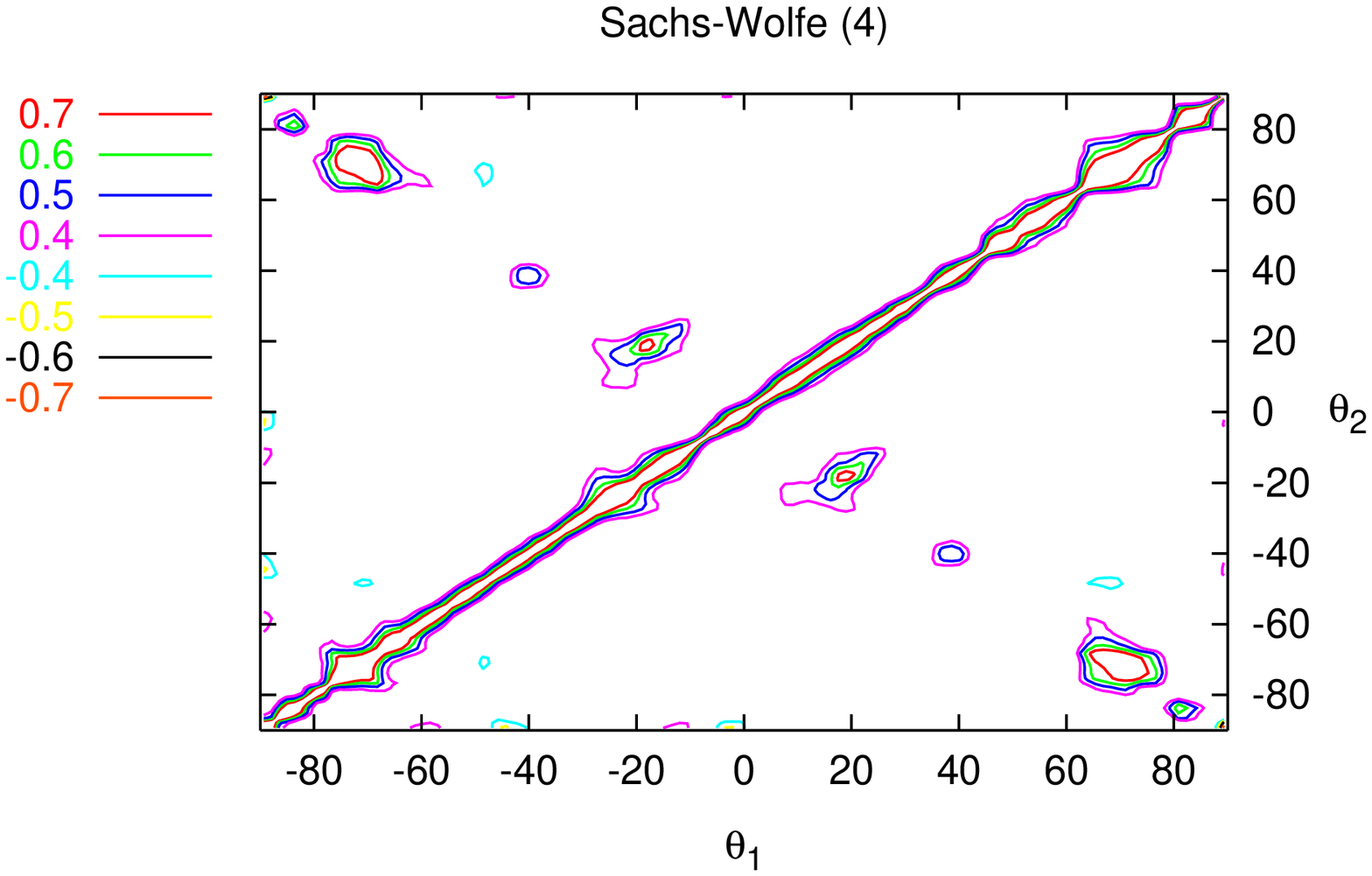,bbllx=135pt,bblly=120pt,
                   bburx=500pt,bbury=660pt,angle=270,width=3.5in}}
\caption{Contour plots of the function $C (\theta_1, \theta_2)$
for four realizations of the Sachs-Wolfe part of the temperature
anisotropies in a toroidal universe. No large correlations are found
except when $|\theta_1 - \theta_2|$ is small and for the three pairs
of matching circles.} \label{fig5}
\end{figure}

\begin{figure}
\centerline{\psfig{file=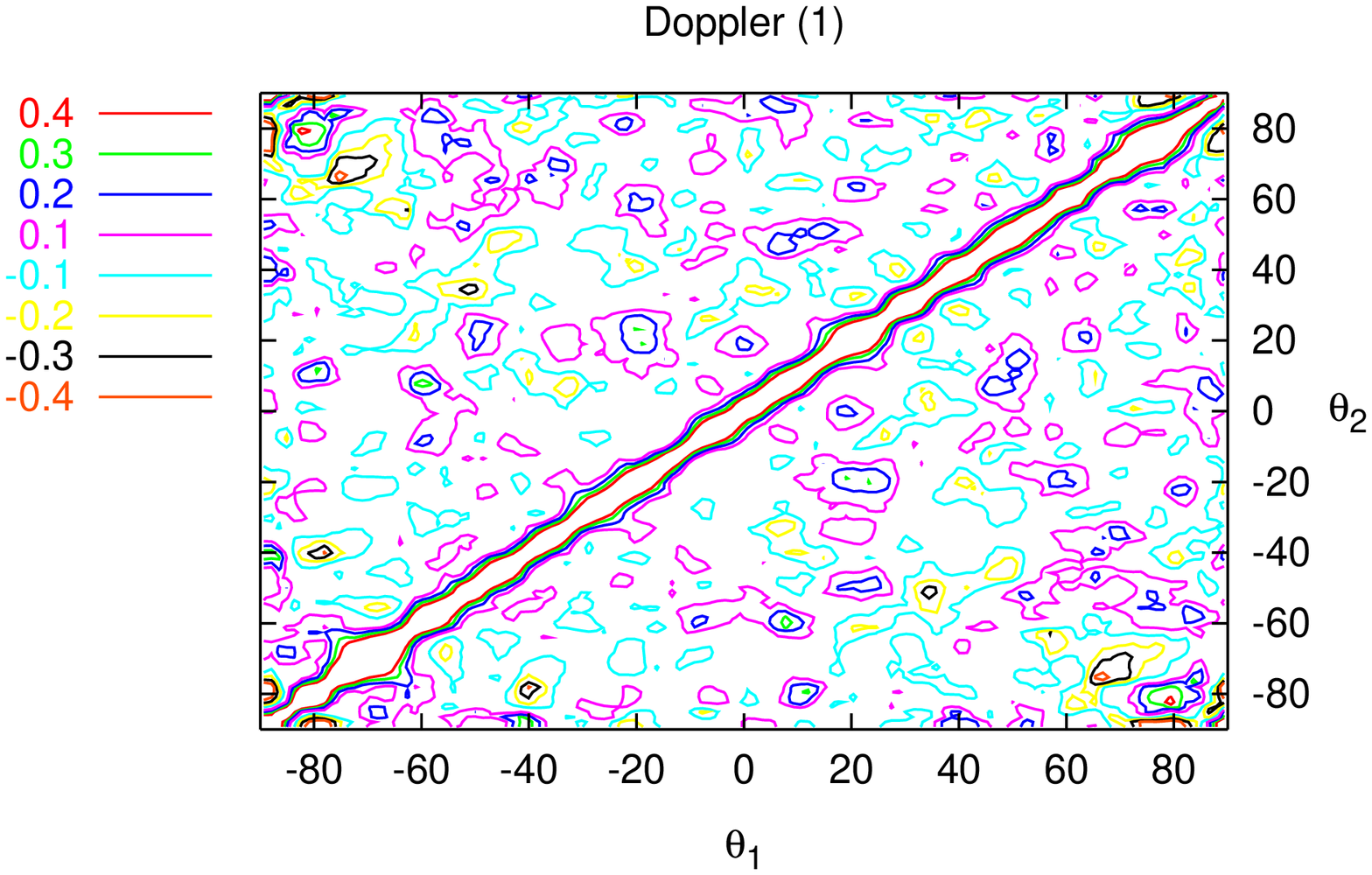,bbllx=135pt,bblly=120pt,
                   bburx=500pt,bbury=660pt,angle=270,width=3.5in}
            \psfig{file=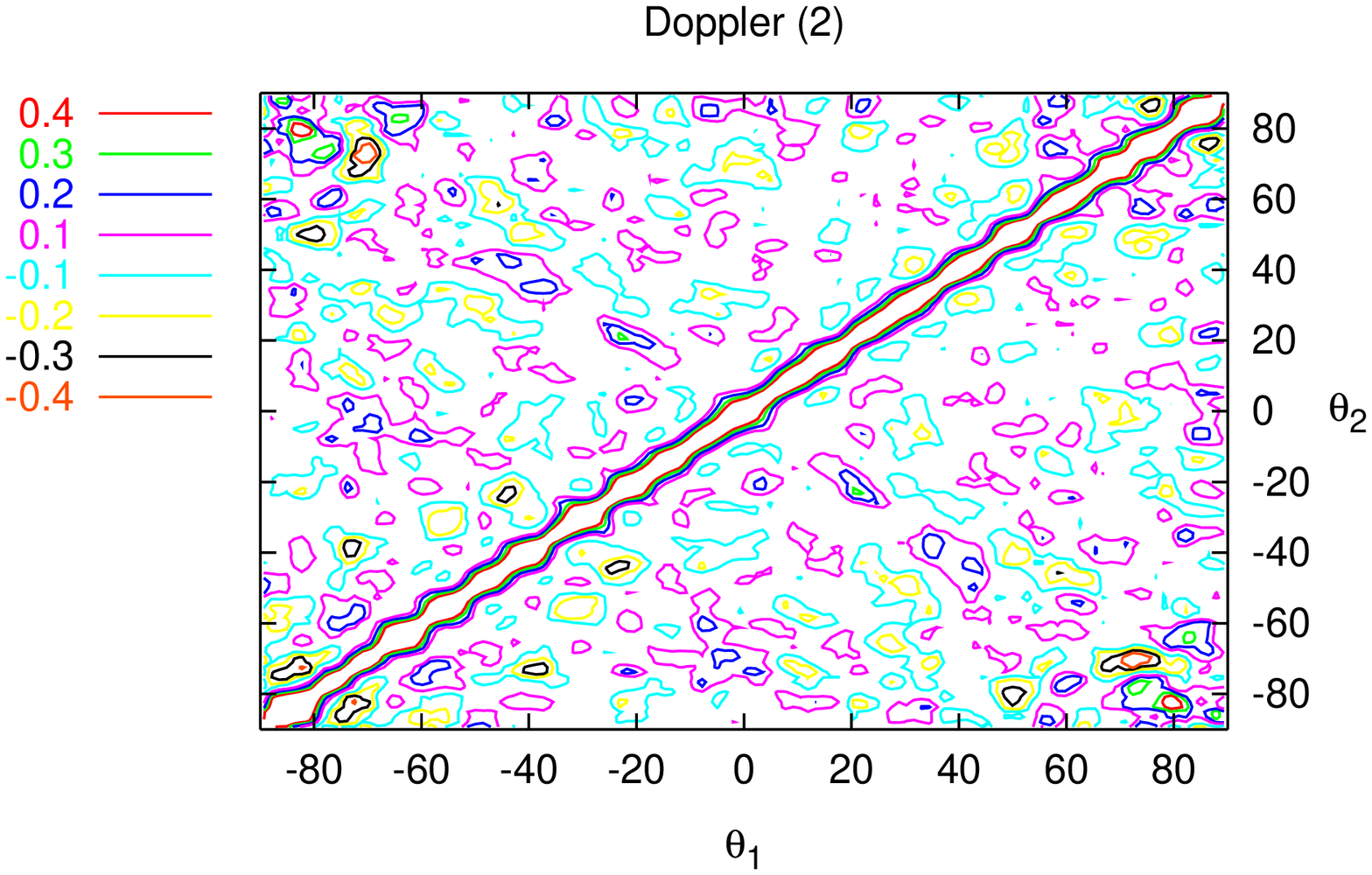,bbllx=135pt,bblly=120pt,
                   bburx=500pt,bbury=660pt,angle=270,width=3.5in}}
\centerline{\psfig{file=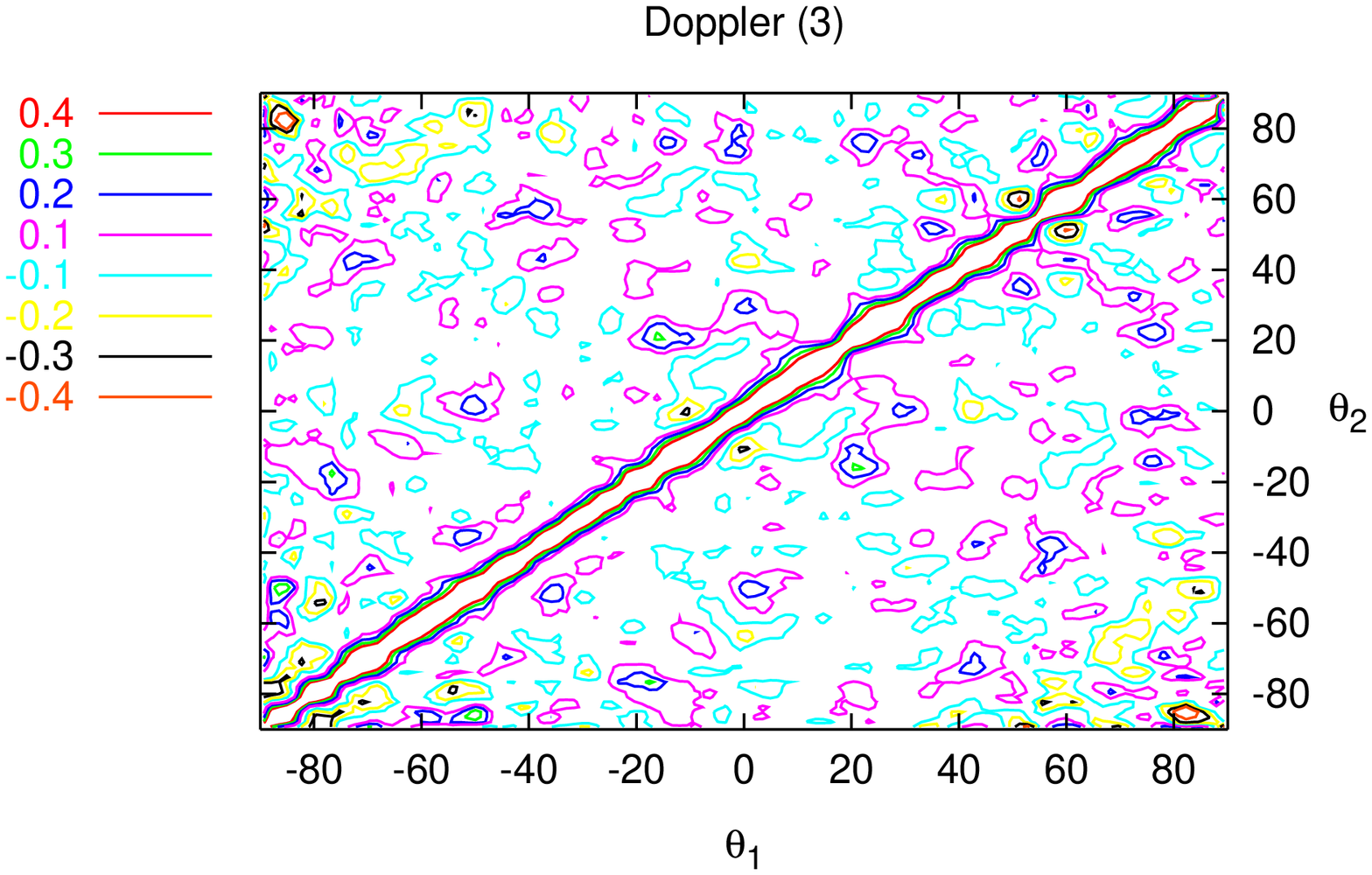,bbllx=135pt,bblly=120pt,
                   bburx=500pt,bbury=660pt,angle=270,width=3.5in}
            \psfig{file=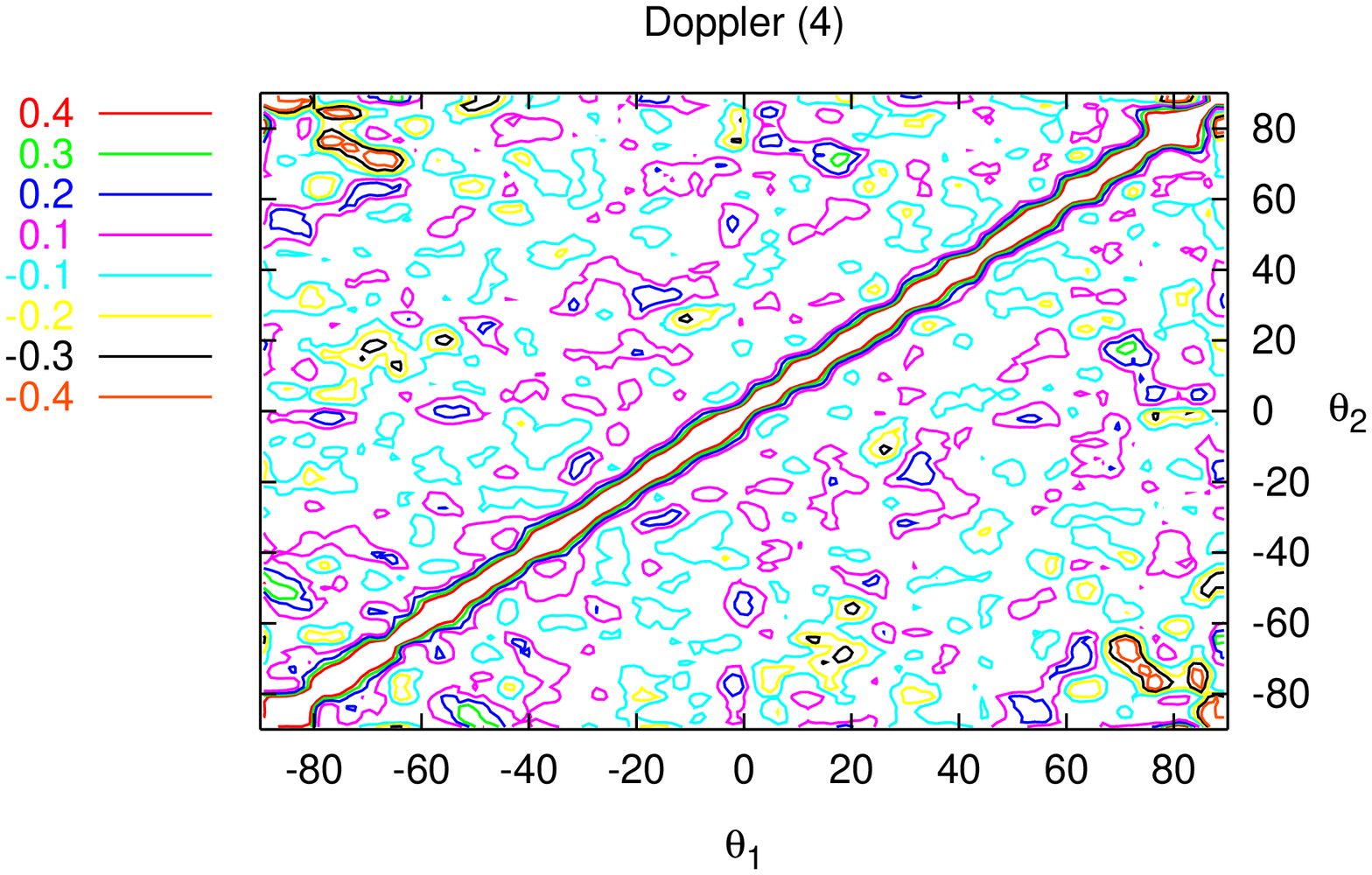,bbllx=135pt,bblly=120pt,
                   bburx=500pt,bbury=660pt,angle=270,width=3.5in}}
\caption{Contour plots of the function $C (\theta_1, \theta_2)$
for four realizations of the Doppler part of the temperature
anisotropies in a toroidal universe. Some correlation or anti
correlation is found for two of the circles, but the signal is not
very large compared to the statistical fluctuations. Note that the
correlation would have been slightly larger if we would have simulated
these maps from a higher resolution correlation matrix.}
\label{fig7}
\end{figure}

\begin{figure}
\centerline{\psfig{file=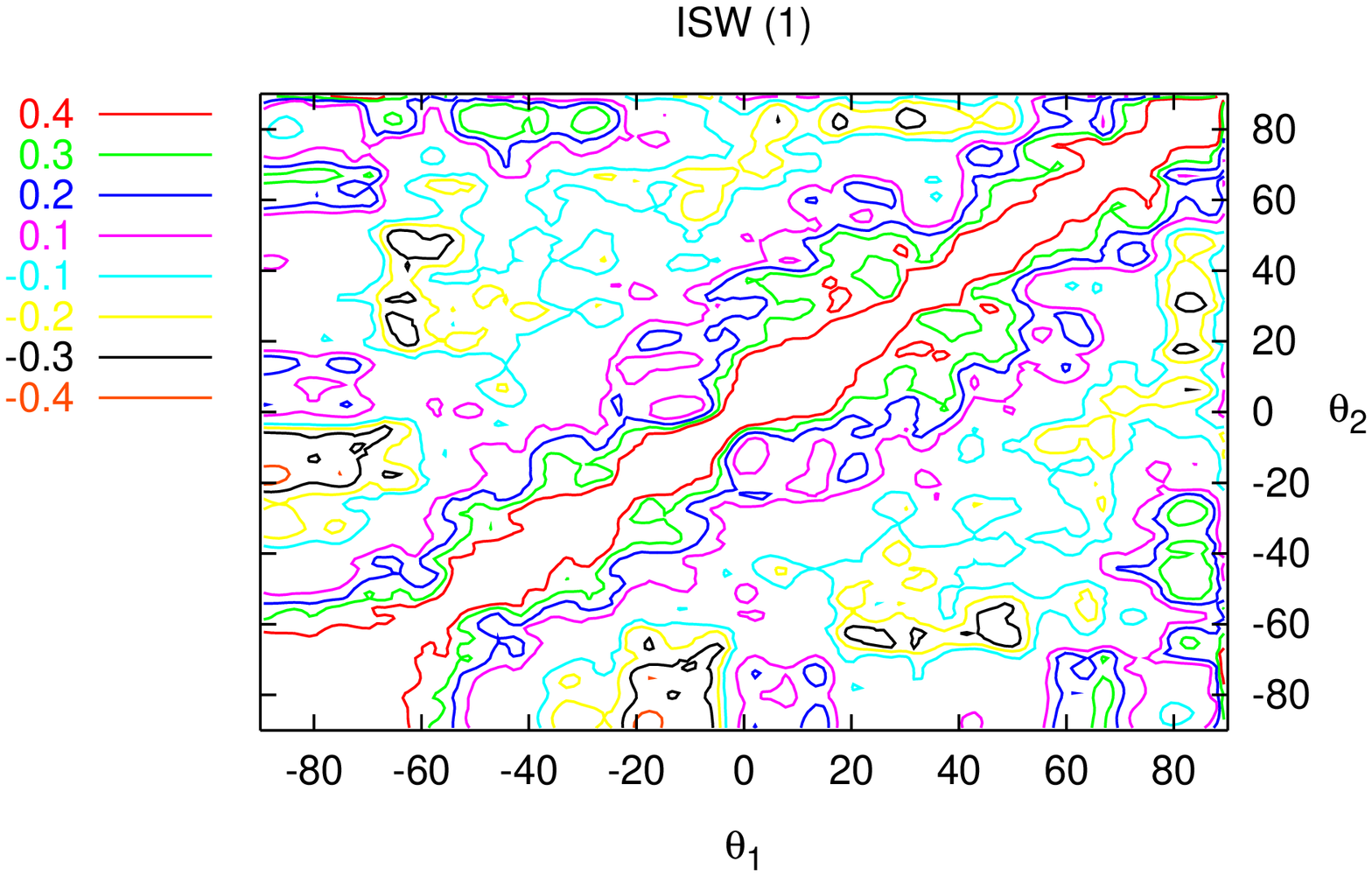,bbllx=135pt,bblly=120pt,
                   bburx=500pt,bbury=660pt,angle=270,width=3.5in}
            \psfig{file=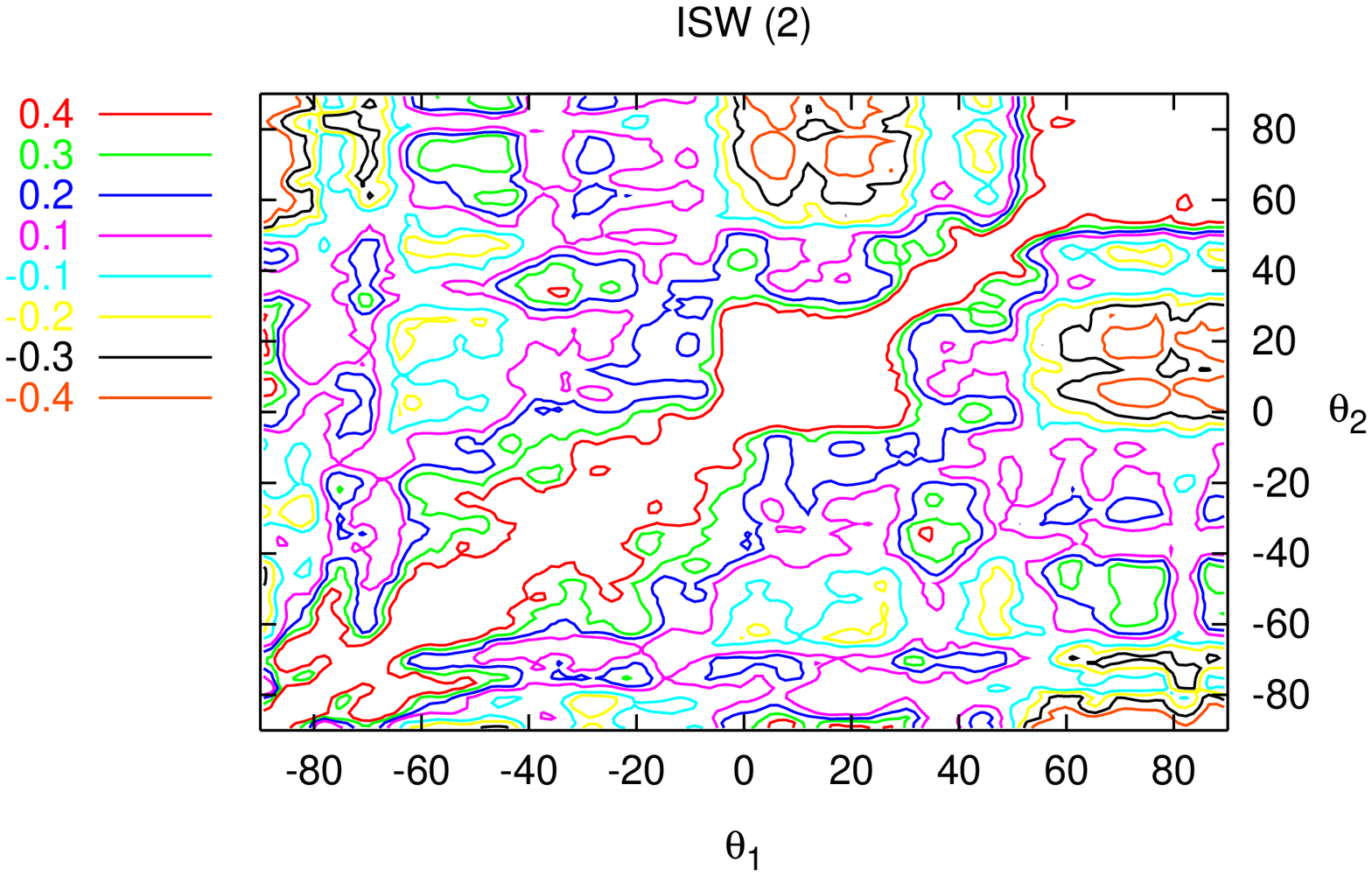,bbllx=135pt,bblly=120pt,
                   bburx=500pt,bbury=660pt,angle=270,width=3.5in}}
\centerline{\psfig{file=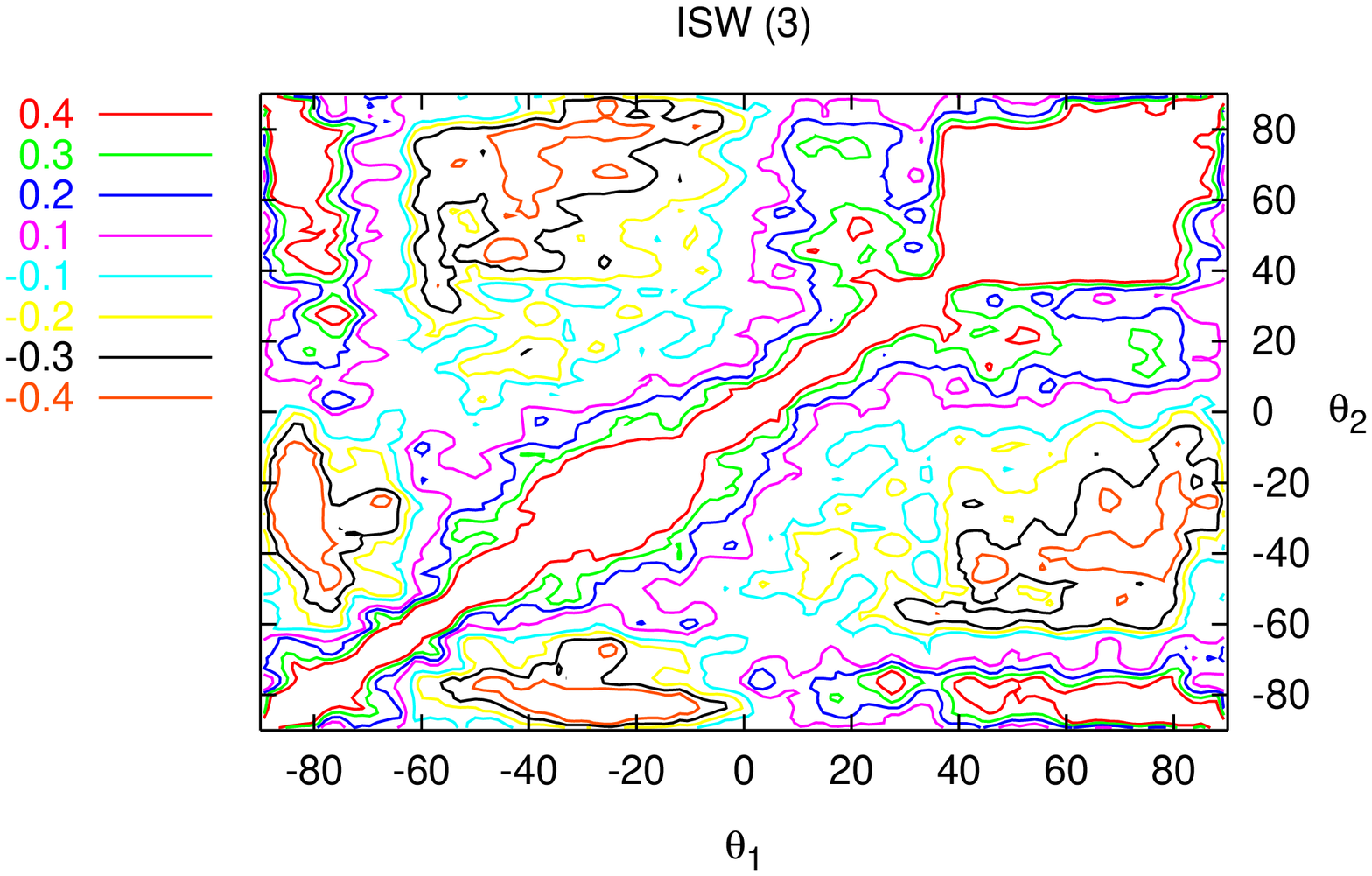,bbllx=135pt,bblly=120pt,
                   bburx=500pt,bbury=660pt,angle=270,width=3.5in}
            \psfig{file=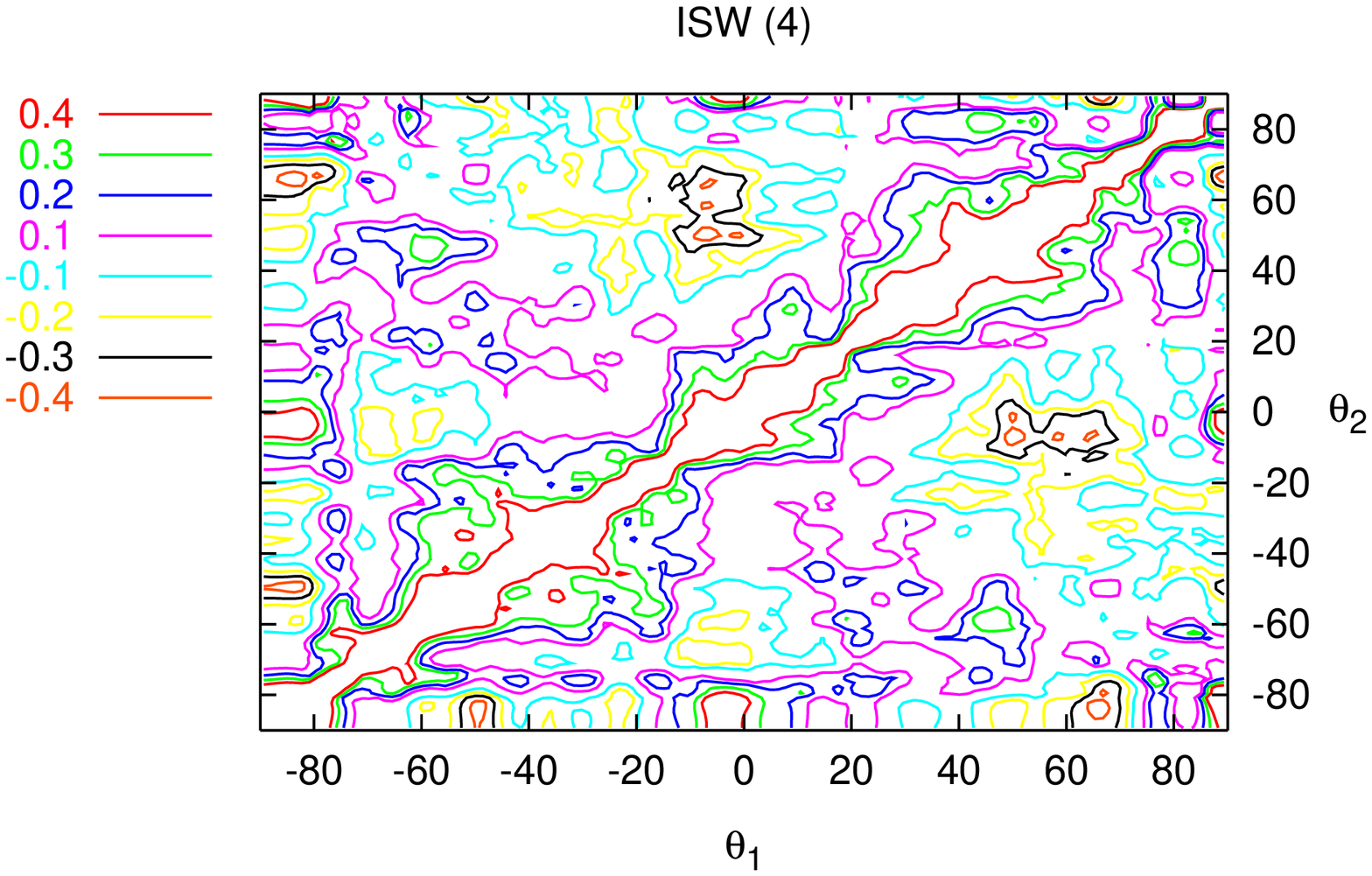,bbllx=135pt,bblly=120pt,
                   bburx=500pt,bbury=660pt,angle=270,width=3.5in}}
\caption{Contour plots of the function $C (\theta_1, \theta_2)$
for four realizations of the integrated Sachs-Wolfe part of the
temperature anisotropies in a toroidal universe. No significant
correlation is seen on the matching circles, whereas large values are
found for $|C (\theta_1, \theta_2)|$ extending across broad regions as
a consequence of the fact that the integrated Sachs-Wolfe effect
appears on large angular scales.} \label{fig8}
\end{figure}

\begin{figure}
\centerline{\psfig{file=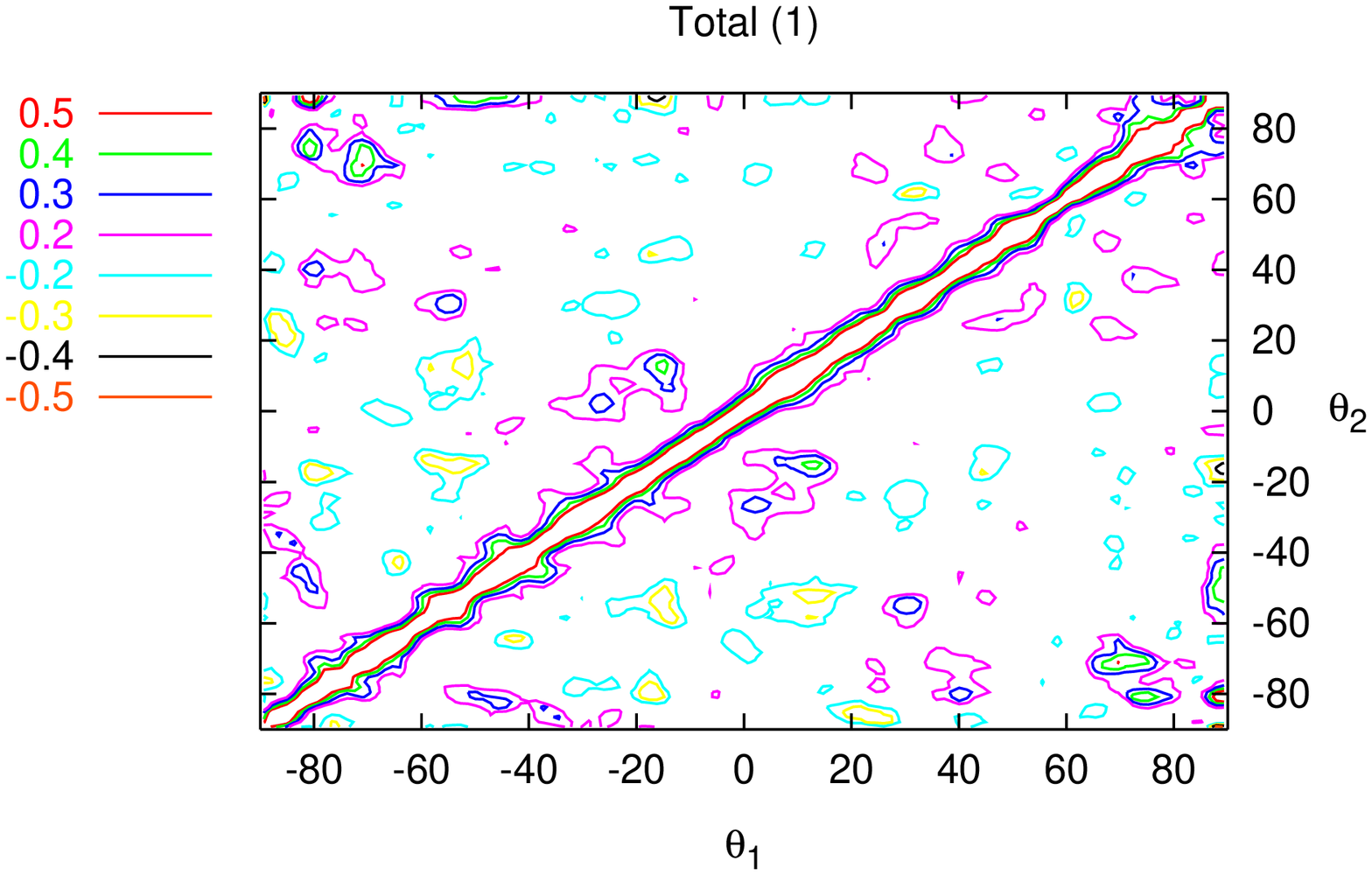,bbllx=135pt,bblly=120pt,
                   bburx=500pt,bbury=660pt,angle=270,width=3.5in}
                   \psfig{file=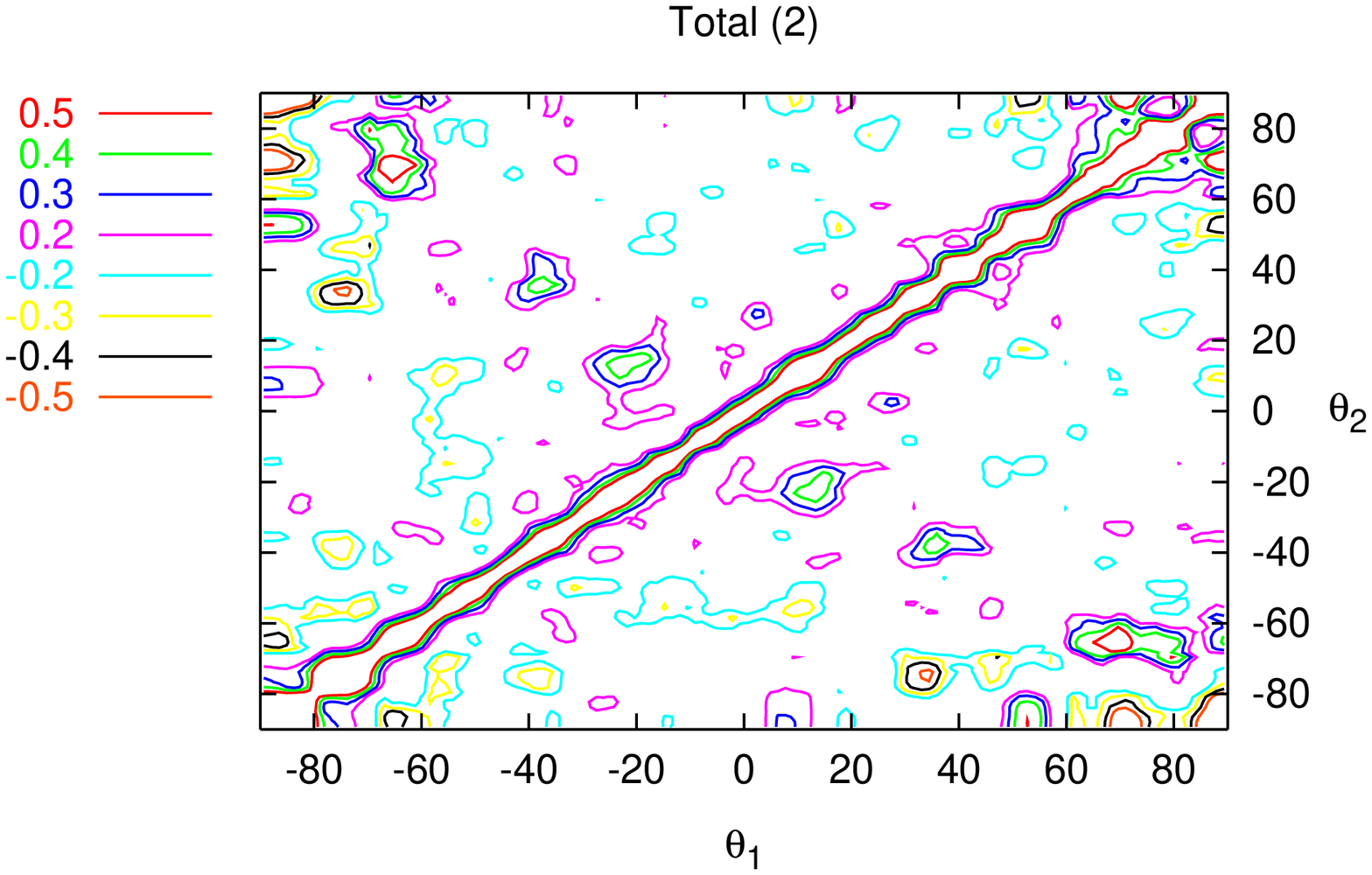,bbllx=135pt,bblly=120pt,
                   bburx=500pt,bbury=660pt,angle=270,width=3.5in}}
\centerline{\psfig{file=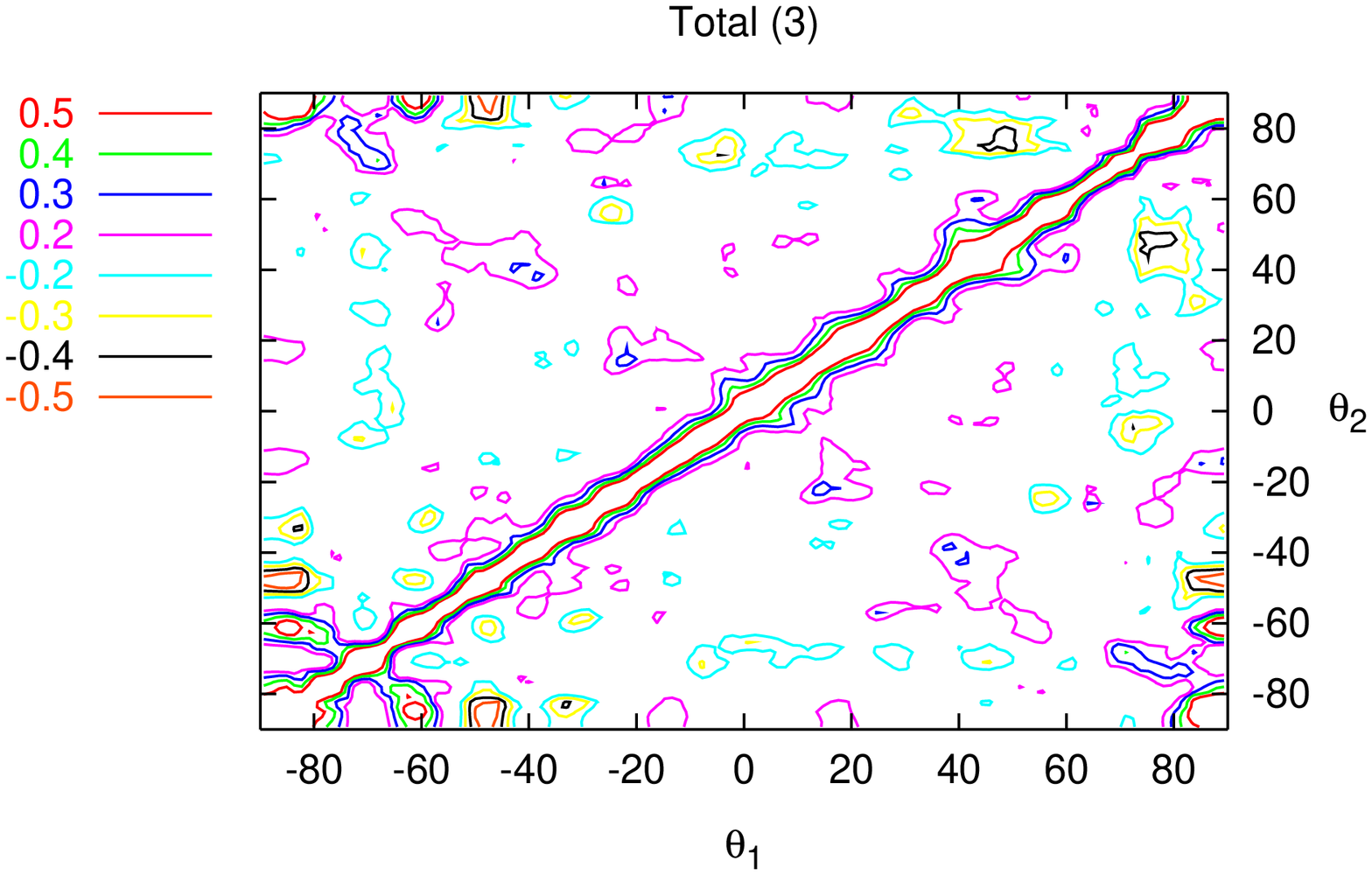,bbllx=135pt,bblly=120pt,
                   bburx=500pt,bbury=660pt,angle=270,width=3.5in}
            \psfig{file=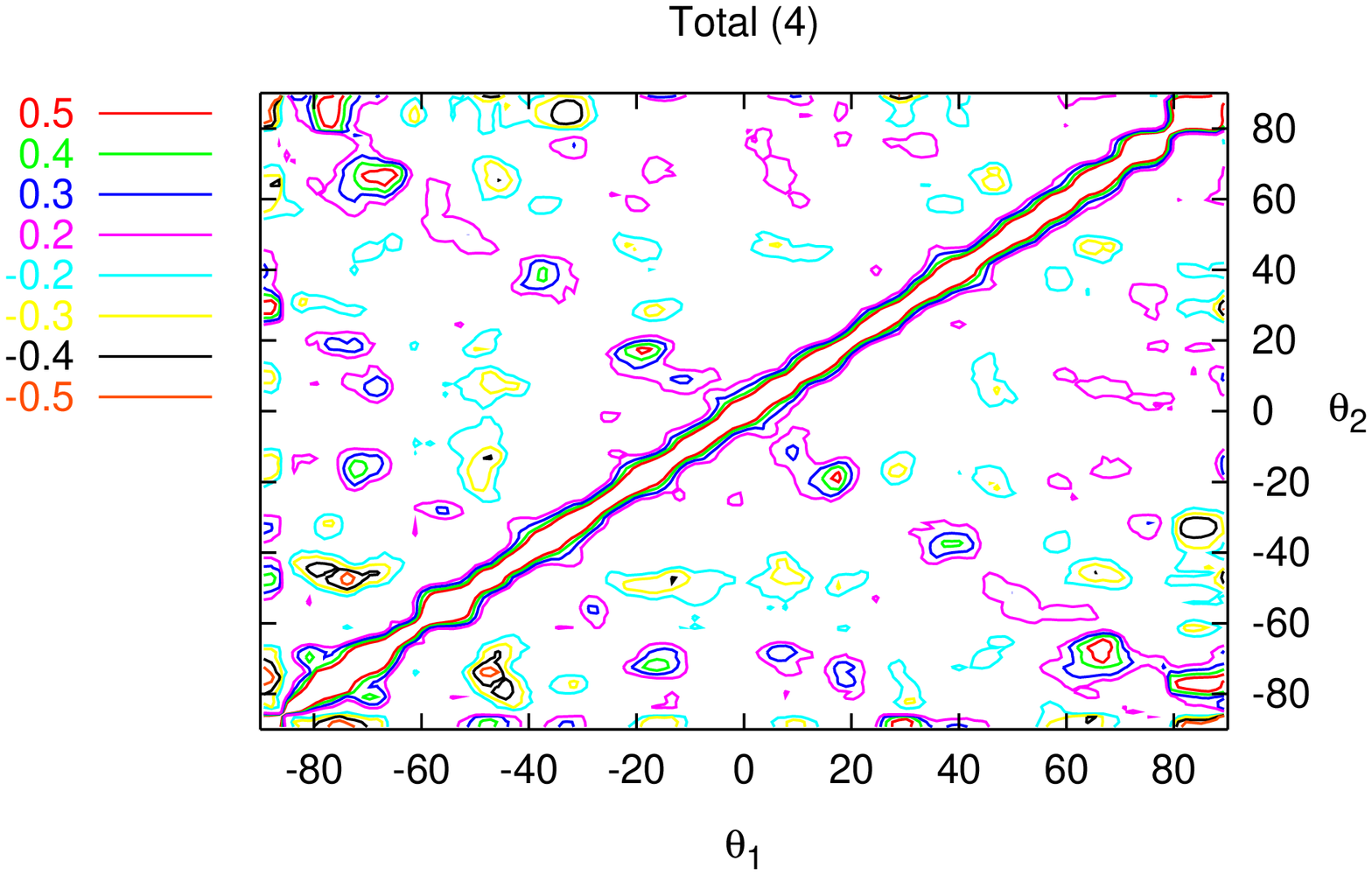,bbllx=135pt,bblly=120pt,
                   bburx=500pt,bbury=660pt,angle=270,width=3.5in}}
\caption{Contour plots of the function $C (\theta_1, \theta_2)$
for four realizations of the total temperature anisotropy in a
toroidal universe (Sachs-Wolfe, Doppler, and ISW effects
combined). The three pairs of matching circles at $\theta_1 = -
\theta_2 = \pm 18^\circ, \pm 39^\circ, \pm 71^\circ$ are still
visible despite the presence of the Doppler and integrated
Sachs-Wolfe contributions.} \label{fig9}
\end{figure}

Figs.~\ref{fig5}, \ref{fig7}, \ref{fig8} and \ref{fig9} show contour
plots for several realizations of the correlation matrix in a torus
universe. The torus aligns naturally with the coordinate system, so
one expects correlated pairs of circles at
\begin{equation}
\theta_1 = - \theta_2 = \pm \arcsin \left(\frac{n L}{2 R_\LSS}
\right) ,
\end{equation}
for each positive integer $n$ such that the arcsine exists. For our
choice of cosmological parameters we have $R_\LSS = 3.17$ and $L = 2$
(in units of the Hubble radius), giving
\begin{equation}
\label{valtheta12} \theta_1 = - \theta_2 = \pm 18^\circ, \pm
39^\circ, \pm 71^\circ.
\end{equation}
For these values of $\theta_1$ and $\theta_2$, one expects a
perfect correlation for the Sachs-Wolfe contribution:
\begin{equation}
C_\SW = 1 .
\end{equation}
This formula holds both when one considers $C_\SW$ as an ensemble
average and when one considers a given realization of the density
field since in both cases it follows from the fact that one sees the
same region from different directions. These correlations appear
clearly in Fig.~\ref{fig5} which considers only the Sachs-Wolfe
contribution. In this case one would have even expected perfect
correlations for the values of $\theta_1$ and $\theta_2$ given in
(\ref{valtheta12}). This is not what we have, but the reason for this
is easy to understand: imposing $\Theta (\theta_1, \varphi) = \Theta
(- \theta_1, \varphi)$ in real space induces in Legendre space
correlations at arbitrary large multipoles $\ell$. Here, for
computational reasons, we were forced to truncate the correlation
matrix at a rather low value of $\ell$, so the matching is significant
but not perfect. It would presumably increase in higher resolution
maps.

If one considers the Doppler contribution to the CMB anisotropies, the
situation is somewhat different. As announced above, the correlation
between two circles depends on their relative angle.  More precisely,
it is given by
\begin{equation}
 C_\DOP = \left< \frac{2 \; (\bhn_1 \cdot \bhn) \; (\bhn_2 \cdot \bhn)}
  {  (\bhn_1 \cdot \bhn)^2 + (\bhn_2 \cdot \bhn)^2 } \right>_\bhn ,
\end{equation}
where $\bhn_1$ and $\bhn_2$ are two constant unit vectors spanning an
angle $|\theta_1 - \theta_2|$, and where the brackets denote an
average over all the directions of the unit vector $\bhn$. After some
manipulations, one obtains
\begin{equation}\label{eq80}
C_\DOP = \tan \left(\frac{\pi}{4} - \frac{|\theta_1 - \theta_2|}{2} \right) .
\end{equation}
Again, for the same reason as for the Sachs-Wolfe contribution, this
formula holds both if one considers $C_\DOP$ as an ensemble average or
if we consider a given realization of the density field.  One recovers
as expected that the correlation is $1$, $0$, $-1$ for $|\theta_1 -
\theta_2| = 0^\circ$, $90^\circ$, $180^\circ$, respectively.  For the
values of the angle given in Eq.~(\ref{valtheta12}), one obtains $C =
0.51$, $0.07$, $-0.49$, respectively.  These are the results that we
obtain qualitatively in Fig.~\ref{fig7}, where no correlation at all
is seen for the circles at $\pm 39^\circ$, and positive (resp.\
negative) correlation is seen for the circles at $\pm 71^\circ$
(resp. $\pm 18^\circ$).

Finally, the correlations due to the integrated Sachs-Wolfe effect are
shown in Fig.~\ref{fig8}. As expected, no particular correlation is
seen for the values of $\theta_1$ and $\theta_2$ of
Eq.~(\ref{valtheta12}). The contour plots are however quite different
from those of the Sachs-Wolfe and Doppler contributions.  The reason
is twofold. First, most of the power lies at the smallest
multipoles. This translates into the fact that the contours are broad
in the sense that they do not vary a lot on small intervals of
$\theta_1$ and $\theta_2$. Second, the fact that most of the power is
at large scales implies that a very small number of modes contributes
to it (since we see a finite region of the universe), so there is a
large cosmic variance that makes the statistical uncertainty very
large (thus serendipitously similar temperature patterns on two
unrelated circles are easily achieved here).

Combining all the contributions to the CMB anisotropies allows one to
simulate realizations of the exact $C (\theta_1, \theta_2)$ as shown
in Fig.~\ref{fig9}. Since the Sachs-Wolfe contribution is dominant,
the spikes are still clearly visible at their expected positions, but
appear less prominent than in Fig.~\ref{fig5}. As expected, it seems
that the circles at $\theta_{1, 2} = \pm 39^\circ$ are slightly less
correlated than the other two pairs because their Doppler contribution
is not correlated, but this deserves a more careful analysis.

\subsection{Spherical case: lens spaces}

Among the spherical spaces, the procedure presented above can be
applied most easily to lens and prism spaces, because their
eigenmodes are known explicitly.  The eigenmodes are known
analytically in toroidal coordinates (see
Section~\ref{subsec_spherical}), and Appendix~\ref{app_B} shows
how to convert them to spherical coordinates. In this section, we
present some sample maps exhibiting the matching circles to
demonstrate that the whole computational chain (computation of the
modes and implementation in a CMB code) is working. A complete and
detailed study, along the same lines as the study done for the
cubic torus in the previous section, will be presented in a
follow-up article.

As explained in Ref.~\cite{wlu02}, because our universe is almost
flat, observational methods such as the circles method will
typically detect only a cyclic subgroup of the holonomy group, so
the universe ``looks like a lens space'' no matter what its true
topology is. It follows that lens spaces are particularly
interesting to capture the observational properties of
multi-connected spherical spaces. In particular, we
showed~\cite{wlu02} that a cyclic factor $Z_p$ creates matching
circles in the CMB only when $\Omega-1>1/p^2$ and that the second
factor, if it exists, is in general undetectable.

Let us emphasize some differences with the torus case. First,
concerning the eigenmodes, let us take the example of a lens space
$L(p, 1)$ of order $p$. For $p = 1$ it reduces to $\STR$ and for
$p = 2$ it reduces to projective space;  more generally the index
plays a role analogous to the size, $L$, of the torus in Euclidean
space. The first non-zero eigenvalue is always $\nu = 2$ and has a
multiplicity $3$ for $p > 2$ and $9$ otherwise. This constancy of
the first eigenvalue contrasts sharply with the case of a cubic
torus, for which the smallest eigenvalue scales as $L^{-1}$. It
can be understood by realizing that when $p$ increases the space
is becoming smaller only in one direction and remains large in
perpendicular directions.

The lens spaces $L(p, 1)$ are globally homogeneous (like the
torus) so that the coefficients $\XITKSLM{\Gamma}{k}{s}{\ell}{m}$
do not depend on the observer's position (i.e., they are the same
no matter where in the space you choose the basepoint).   Thus
neither the correlation matrix $\CLMLPMP{\ell}{m}{\ell'}{m'}$ nor
the positions of the matching circles depend on the basepoint.
Unfortunately, this is not the case for a general lens space $L(p,
q)$. For a general lens space, the coefficients
$\XITKSLM{\Gamma}{k}{s}{\ell}{m}$, the correlation matrix
$\CLMLPMP{\ell}{m}{\ell'}{m'}$ and the positions of the matching
circles all depend on the observer's position.  For instance, the
``canonical'' choice of coordinates used in
Sec.~\ref{subsubsec_lens} for the toroidal coordinate system puts
the preferred symmetry axes in the $(xy)$ and $(zw)$ directions
(i.e., the axis are the intersection of $\STR$ with the $(x,
y)$-plane and the $(z, w)$-plane, respectively, in
four-dimensional Euclidean space). From a cosmological point of
view, this is a poor choice, because the observer's translated
images are ``atypically close''. For example in $L(12, 5)$, which
has cyclic factors $Z_3$ and $Z_4$, a generic observer will see
three lines of four images each, but a nongeneric observer sitting
on a symmetry axis will see a single line of $12$ images.

For a globally homogeneous space $L(p, 1)$, the closest
topological image is located at a distance $\CHN = 2 \pi / p$, so
the topology is detectable just so $p$ isn't too small.  More
precisely, the topology is potentially detectable if and only if
$\pi / p < \CHN_\LSS$.  For example, this implies that the
topology is detectable for all $p > 10$ if $\Omega_\TOT - 1 \sim
10^{-2}$.

Circles match differently in a homogeneous lens space than in a
tours.  In a torus the circles match straight across because the
holonomies are all pure translations.  In a homogeneous lens
space, by contrast, the holonomies are Clifford translations, and
so the matching circles are still diametrically opposite but match
with a twist that's a multiple of $2 \pi / p$, because Clifford
translations twist and translate of the same amount.

\begin{figure}
\centerline{\psfig{file=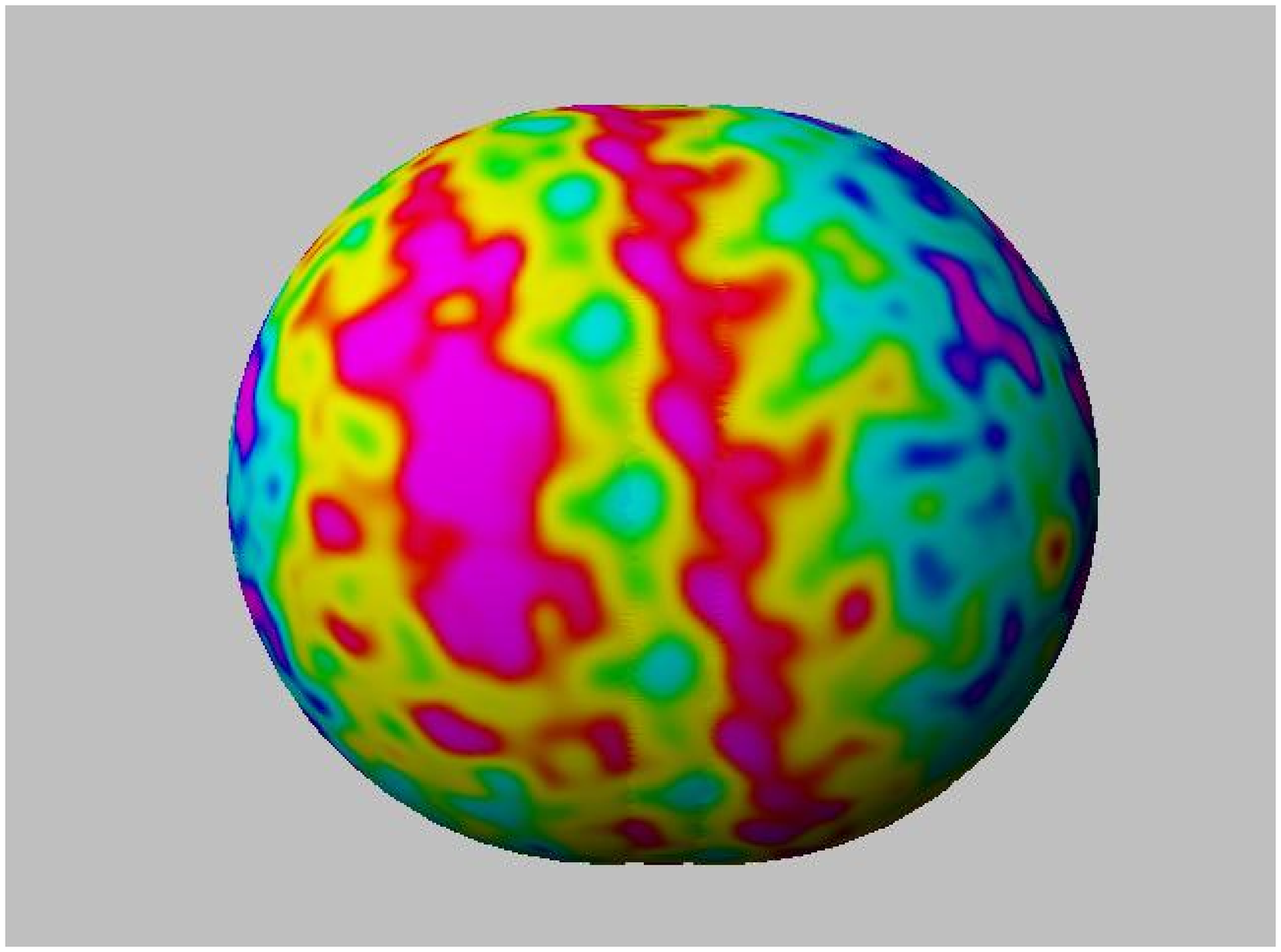,angle=0,width=3.5in}
            \psfig{file=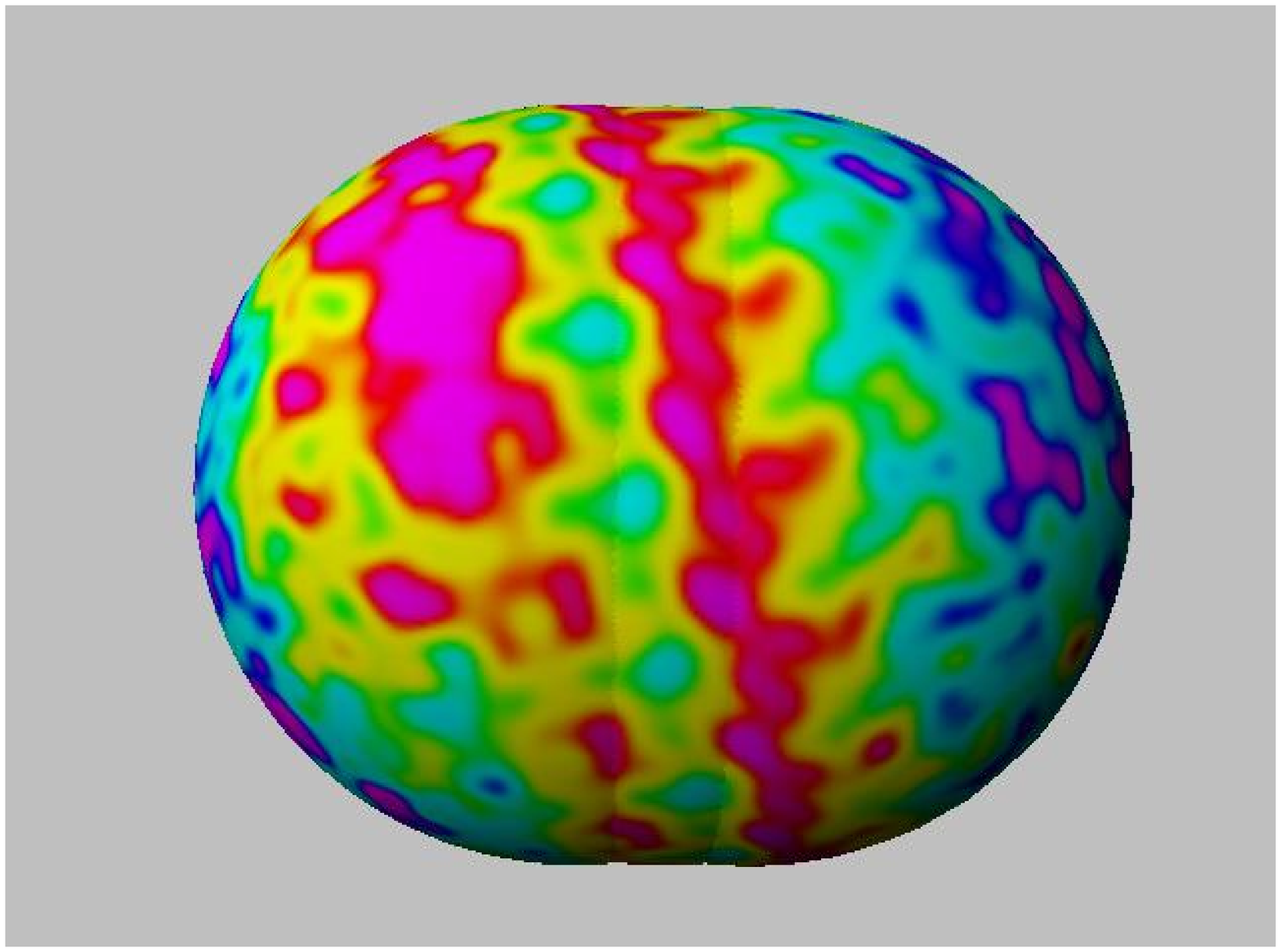,angle=0,width=3.5in}}
\centerline{\psfig{file=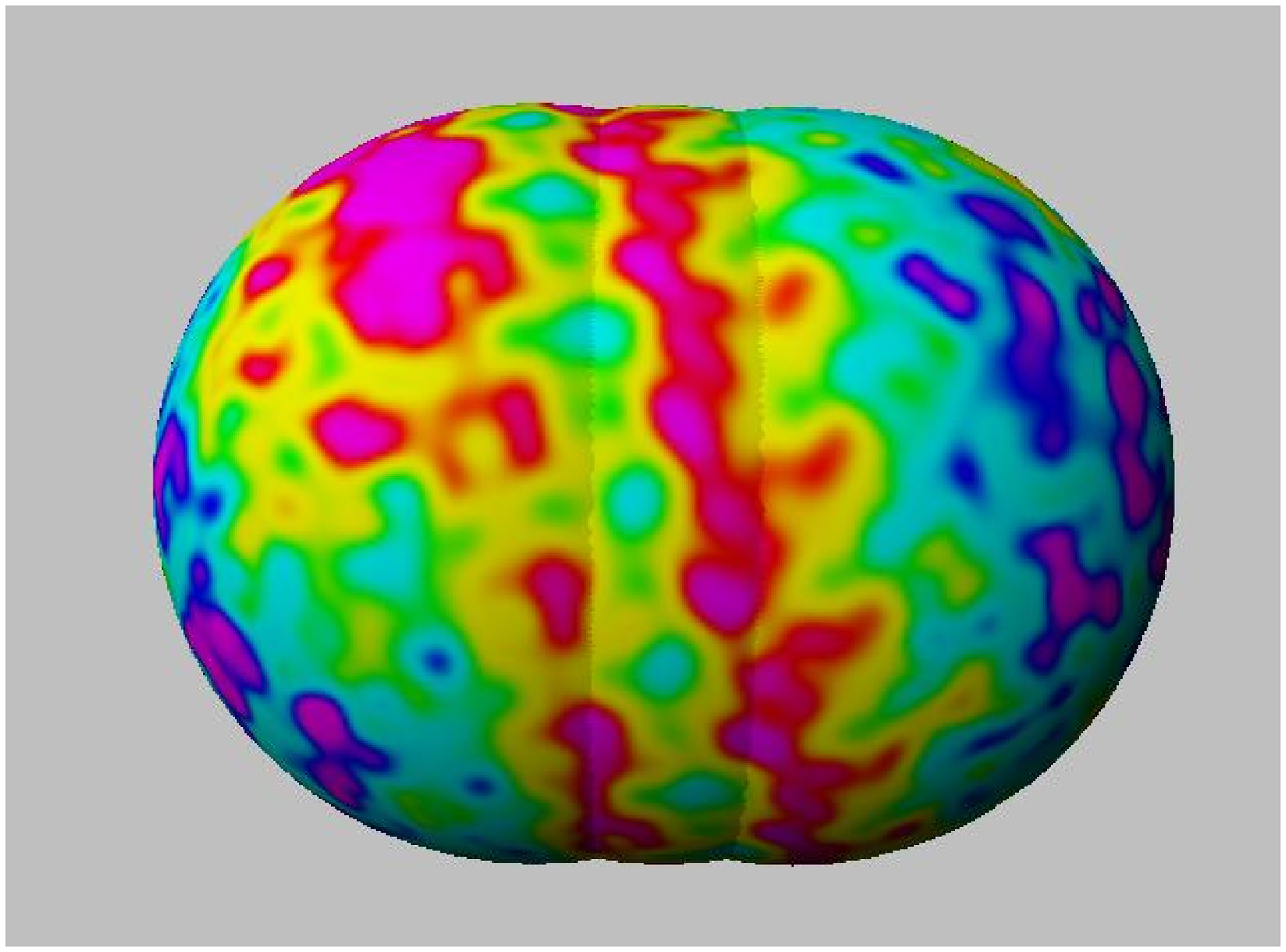,angle=0,width=3.5in}
            \psfig{file=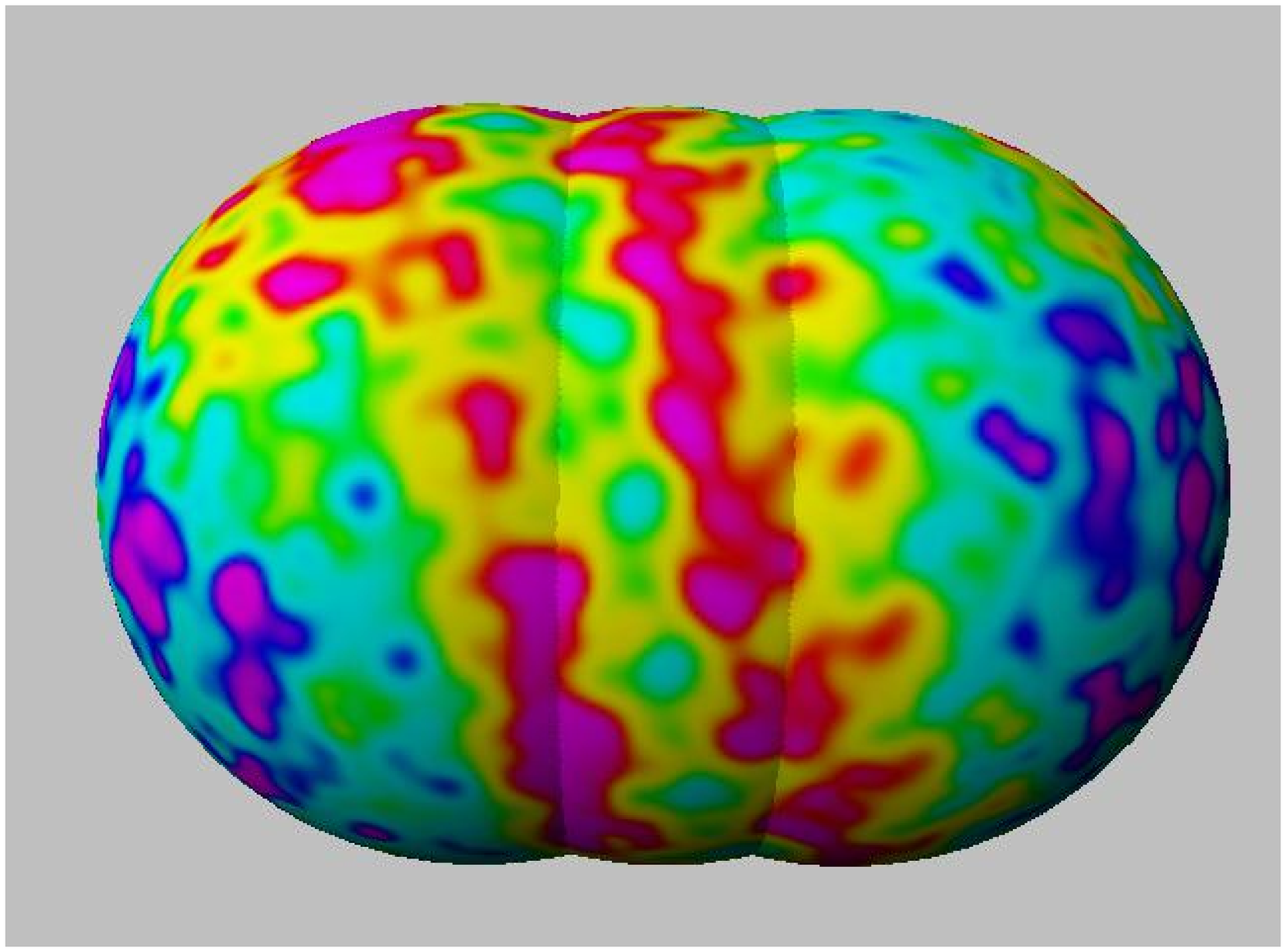,angle=0,width=3.5in}}
\centerline{\psfig{file=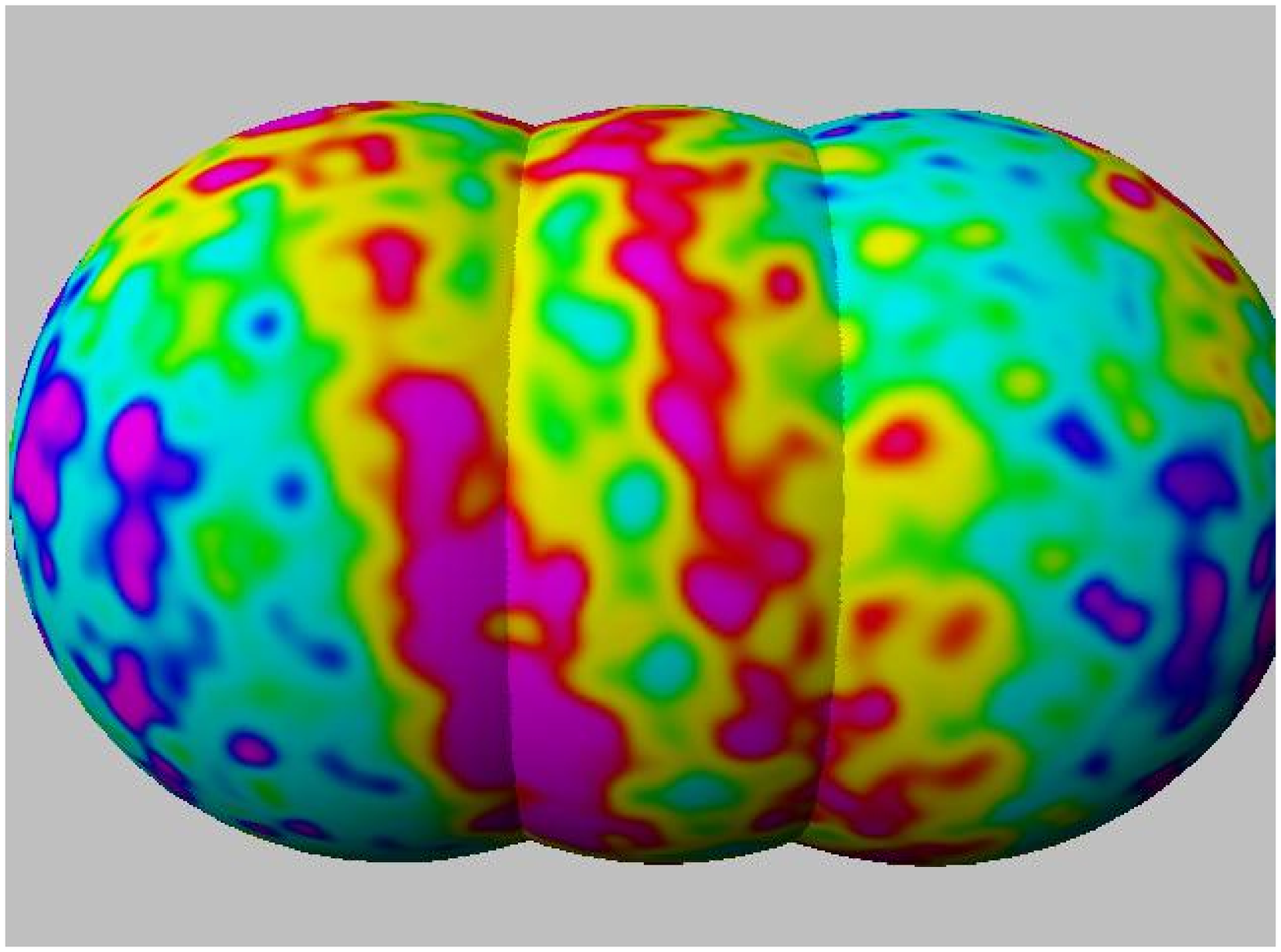,angle=0,width=3.5in}
            \psfig{file=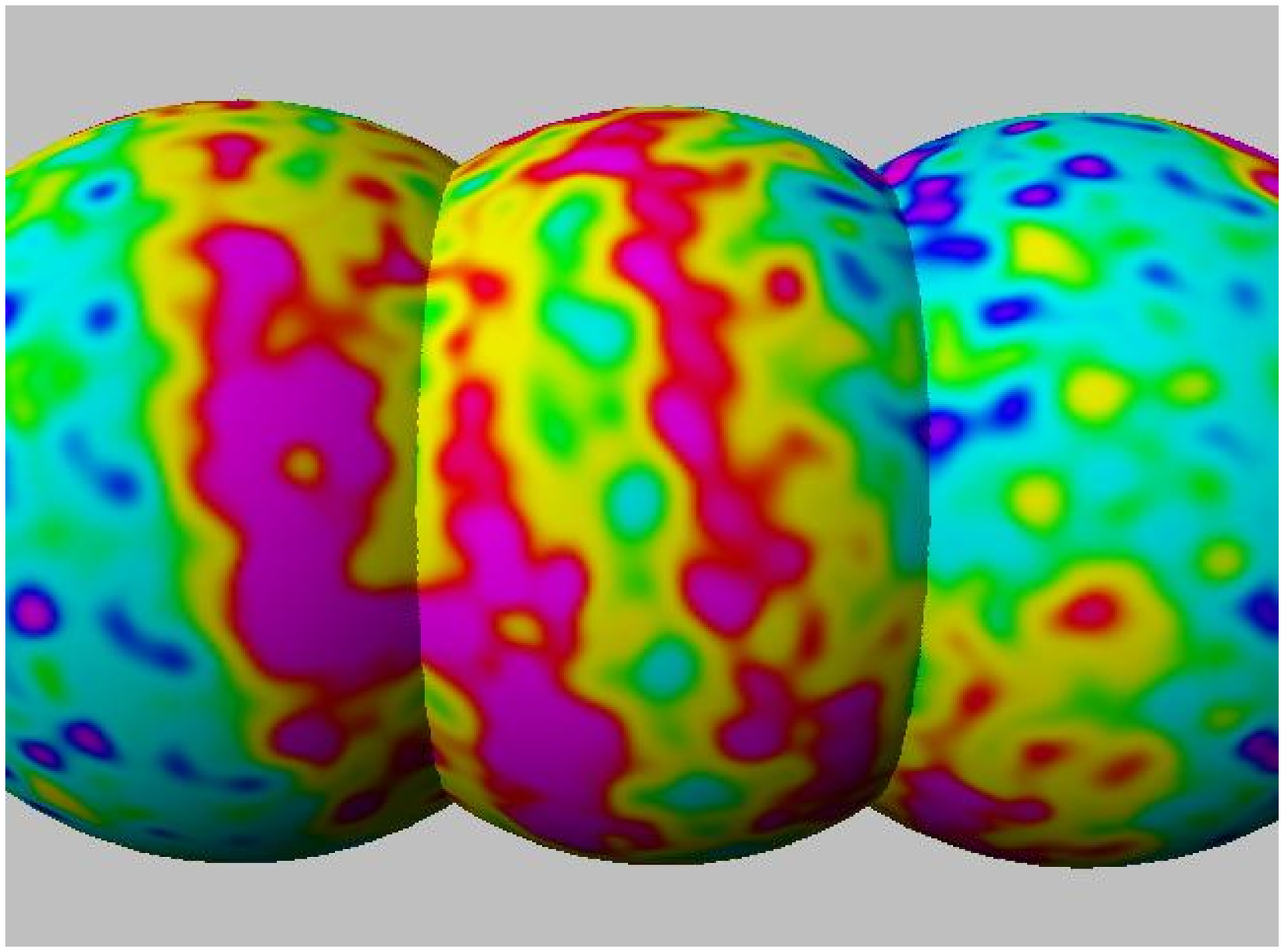,angle=0,width=3.5in}}
\caption{A realization of a CMB map in the case of the lens space
$L(18, 1)$. For purposes of illustration we chose a closed
universe ($\Omega_\TOT = 1.3$) with a large last scattering
surface (the radius $\chi_{LSS}$ is 0.88 times the curvature
radius of the universal covering space $\STR$), in order to
produce higher resolution images with a large number of correlated
circles. In each panel, the matching is obtained by performing a
translation of
$n\, R_c 2 \pi / 18$ and a twist of $n\, 2 \pi / 18$ between the
last scattering surface and its two copies.  The six panels
correspond to $|n| = 2,\dots,7$, with each panel showing both the
positive ($n > 0$) and negative ($n < 0$) translates.}
\label{figgrrr}
\end{figure}

Fig.~\ref{figgrrr} shows a CMB map with resolution $\ell_\MAX =
30$ for the lens space $L(21, 1)$ considering the Sachs-Wolfe term
only. A far more detailed discussion about CMB anisotropies in
lens spaces will appear elsewhere~\cite{lens2}.

\section{Discussion and conclusions}
\label{sec_conclusion}

This articles describes the implementation of topology in CMB codes
and gives explicitly the required tools to perform such an
implementation in flat and spherical spaces.  As emphasized in the
Introduction, these two cases are observationally the most relevant
for an almost flat universe.

Examples of simulated maps were given in the two cases. Here we
presented only low resolution maps due to the computational time
limitation but higher resolution maps will be presented elsewhere.  It
was checked that the expected topological correlations (the matched
circles) were present, confirming the quality of our simulations.

Our method relies on the computation of the correlation matrix of the
coefficients of the decomposition of the temperature fluctuation in
spherical harmonics. This matrix encodes all the topological
information. We emphasize that, due to the breakdown of global
isotropy, this matrix is not purely diagonal. This also offers a
working example to construct tests for the detection of deviation from
global isotropy.

We have illustrated the influence of different effects that will tend
to blur these patterns and affect the perfect circle matching, namely
the Doppler effect and the integrated Sachs-Wolfe effect.  We also
considered the effect of the thickness of the last scattering surface,
but found it to be negligible on the scales considered here. A more
detailed quantitative analysis of these effects on the detectability
of the topological signal is left for future studies~\cite{future}.

A complete investigation of the detectability of the topology in
coming CMB data requires the construction of reliable simulation
tools. Besides the quantification of the amplitude of the effects
cited above, one would also need to include all other observational
effects such as instrumental noise, foreground contamination, etc. The
present work paves the way to all these essential studies.

\section*{Acknowledgments}

We want to thank Nabila Aghanim, Francis Bernardeau, Francois Bouchet,
Gilles Esposito-Far\`ese, Simon Prunet for providing computational
support, Dorian Goldfeld for his help in evaluating some sums of
Appendix~\ref{app_B}, and Jean-Pierre Luminet for very careful reading
of this manuscript and numerous discussions.  J.W.\ thanks the
MacArthur Foundation for its support. J.-P.U.\ thanks the University
of Barcelona for hospitality while a part of this work was
performed. Part of this work was achieved while A.R.\ was a CMBnet
postdoc in the D\'epartement de Physique Th\'eorique of Geneva
University.

\appendix
\section{Eigenmodes of constant curvature three-dimensional spaces}
\label{app_A}

This appendix follows the work by Abbott and Schaeffer~\cite{abott86}
and Harrison~\cite{harrison67} and borrows heavily from Appendix~A of
Ref.~\cite{lwugl}. It summarizes, without proof, the explicit forms of
the scalar harmonic functions solutions of the Helmholtz
equation~(\ref{Helmotz1}).

It is convenient to factor the eigenfunctions into radial and angular
functions as
\begin{equation}
\label{eq:VariablesSeparation}
\YXKLM{X}{k}{\ell}{m} (\CHU, \theta, \varphi)
 = \RXKL{X}{k}{\ell} (\CHU) \YLM{\ell}{m} (\theta, \varphi) ,
\end{equation}
with $\YLM{\ell}{m} (\theta, \varphi)$ being the spherical
harmonics. The associated eigenvalues are $\kappa_k^2 = k^2 - K$, with
\begin{eqnarray}
K < 0 & \Rightarrow & k \in [0, \infty [
                      \quad \mbox{or} \quad
                      i k \in [0, \sqrt{|K|} ] , \\
K = 0 & \Rightarrow & k \in [0, \infty [ , \\
K > 0 & \Rightarrow & k = (\nu + 1) \sqrt{K}
                      \quad,\qquad
                      \nu \in {\bb N}.
\end{eqnarray}
With the normalization
\begin{equation}
\label{eq:Normalization2}
\int \YXKLM{X}{k}{\ell}{m} (\CHU, \theta, \varphi)
     {\YXKLM{X}{k'}{\ell'}{m'}}^{*} (\CHU, \theta, \varphi)
     \; \SK{K}^2(\CHU) \; \ddd \CHU \, \ddd \Omega
 = \frac{1}{k^2} \DIR{k - k'}
   \KRON{\ell}{\ell'} \KRON{m}{m'} ,
\end{equation}
where $\SK{K} (\CHU)$ is defined in Eq.~(\ref{def_sk}), the
normalized radial functions take the form
\begin{eqnarray}
\RXKL{\HTR}{k}{\ell} (\CHU)
 & = & \left(\frac{N_{k \ell}}{k \SK{K} (\CHU)}
       \right)^{1 / 2}
       \PLM{- 1 / 2 + i \omega}{- 1 / 2 - \ell}
         \left(\cosh (\sqrt{|K|} \CHU) \right) , \\
\RXKL{\RTR}{k}{\ell} (\CHU)
 & = & \left(\frac{2}{\pi} \right)^{1 / 2} j_\ell (k \CHU) , \\
\RXKL{\STR}{k}{\ell} (\CHU)
 & = & \left(\frac{M_{k \ell}}{k \SK{K} (\CHU)}
       \right)^{1 / 2}
       \PLM{ 1 / 2 + \nu}{- 1 / 2 - \ell}
         \left(\cos (\sqrt{K} \CHU) \right)
\end{eqnarray}
with
\begin{equation}
\omega = k / \sqrt{|K|}, \qquad \nu = k / \sqrt{K} - 1 .
\end{equation}
In the case of spatially hyperbolic spaces, this normalization is
valid only for sub-curvature modes, and for the super-curvature modes
($i k \in [0, \sqrt{|K|} ]$) the radial function is obtained by
analytic continuation (see Ref.~\cite{lyth} for details).  The two
numerical coefficients are given by
\begin{eqnarray}
\label{eq:Definitions}
N_{k \ell}
 & \equiv & \prod_{n = 0}^\ell \left(\omega^2 + n^2 \right) , \\
M_{k \ell}
 & \equiv & \prod_{n = 0}^\ell \left( (\nu + 1)^2 - n^2 \right)
,\quad\quad M_{k \ell} = 0 \quad\mbox{if}\quad \ell > \nu .
\end{eqnarray}
For any function, we can perform the mode decomposition
\begin{equation}
\label{eq:Decomposiotion}
f (\bx)
 = \sum_{\ell, m}
   \left\lbrace \begin{array}{l}
                 \int k^2 \ddd k\\
                 K^{3 / 2} \sum_{\nu = 2}^\infty (\nu + 1)^2
                \end{array}
   \right\rbrace f_{k \ell m} \YXKLM{X}{k}{\ell}{m}
\, \Longleftrightarrow \,
f_{k \ell m} = \int f(\bx) \YXKLM{X}{k}{\ell}{m}{}^* \sqrt{\gamma} \ddd^3 \bx ,
\end{equation}
choosing the sum or the integral according to whether the universal
covering space is compact or not.  The symbol $\sqrt{\gamma}$ stands
for the square root of the determinant of the spatial metric. In the
case of spatially hyperbolic spaces, the super-curvature modes add a
term to this mode expansion, namely $\int_0^1 |k^2| \ddd(ik)
\sum_{\ell, m} f_{k \ell m} \YXKLM{\HTR}{k}{\ell}{m}$; see
Ref.~\cite{lyth} for details.

In the spherical case, one can however find a solution of the
Helmholtz equation~(\ref{Helmotz1}) which does not involve Legendre
functions. The radial part of the Helmholtz equation reduces, after
setting $\YXKLM{\STR}{k}{\ell}{m} =
\RXKL{\STR}{k}{\ell} (\CHU) \YLM{\ell}{m} (\theta, \varphi)$, to
\begin{equation}
   \frac{1}{\SK{K}^2 (\CHU)} \frac{\ddd}{\ddd\CHU}
   \left(\SK{K}^2 (\CHU) \frac{\ddd}{\ddd\CHU} \RXKL{\STR}{k}{\ell} \right)
 + \left[(k^2 - K)  - \frac{\ell (\ell + 1)}{\SK{K}^2 (\CHU)} \right]
   \RXKL{\STR}{k}{\ell} = 0 .
\end{equation}
It is obviously much more convenient to work in a coordinate system
where the curvature $K$ reduces to $1$. In terms of the dimensionless
radial variable $\CHN$ defined in Eq.~(\ref{def_chn}), the Helmholtz
equation then reduces to
\begin{equation}
   \frac{1}{\sin^2 \CHN} \frac{\ddd}{\ddd \CHN}
   \left(\sin^2 \CHN \frac{\ddd}{\ddd \CHN} \RXKL{\STR}{k}{\ell} \right)
 + \left[\nu (\nu + 2)  - \frac{\ell (\ell + 1)}{\sin^2 \CHN} \right]
   \RXKL{\STR}{k}{\ell} = 0 .
\end{equation}
Note that this is a second order equation and that only one of the two
independent solutions is well behaved at the origin, so the radial
functions are completely determined once the normalization has been
chosen.  After setting $\RXKL{\STR}{k}{\ell} = (\sin
\CHN)^\ell f_{\nu \ell}$, it can be checked that it reduces to
Eq.~(\ref{C12}), the solution of which is simply given in terms of
ultraspherical Gegenbauer polynomials as $f_{\nu \ell} = A_{\nu
\ell} \GNA{\nu - \ell}{\ell + 1}$. The normalization
condition~(\ref{eq:Normalization2}) implies, using the integral
relation~(\ref{C12bis}), that
\begin{equation}
A_{\nu \ell}
 = \frac{2^{\ell + 1 / 2}}{\nu+1}\,{\ell !}\,
   \sqrt{\frac{\nu+1}{\pi}}
   \sqrt{\frac{(\nu - \ell) !}{(\nu + \ell+1) !}} .
\end{equation}
Expressing the spherical harmonics in terms of Gegenbauer polynomials
by means of Eq.~(\ref{C11}), one ends up with an expression of the
eigenmodes in terms of Gegenbauer polynomials only as
\begin{equation}
\YXKLM{\STR}{k}{\ell}{m}
 = A_{\nu \ell} \widetilde A_{\ell m}
   \;\;\sin^{|m|}\theta \;\sin^\ell \CHN
   \;\;\GNA{\nu - \ell}{\ell + 1} (\cos \CHN)
   \;\;\GNA{\ell - |m|}{|m| + 1 / 2} (\cos \theta)
   \;\;\ee^{i m \varphi} ,
\end{equation}
with
\begin{equation}
\widetilde A_{\ell m}
 = \zeta_m (2|m|-1)!!\,
   \sqrt{\frac{(2 \ell + 1)}{4\pi} \frac{(\ell - |m|) !}{(\ell + |m|) !}}
\end{equation}
with $\zeta_m$ given by Eq.~(\ref{C11bis}) and we used the notation
$(2|m| - 1)!! = (2 |m| - 1) (2 |m| - 3) \ldots 1$.

\section{Change of basis between toroidal and spherical coordinates}
\label{app_B}

Sec.~\ref{subsubsec_lens} found the eigenmodes of lens and prism
spaces in toroidal coordinates and converted them to spherical
coordinates. In this appendix, we give the expression for the matrix
$\ALKLMLTMT{\nu}{\ell}{m}{\ell_\ST}{m_\ST}$ necessary to perform the
change of basis.

The spherical coordinate system, as used in Eq.~(\ref{fl_metric}), is
related to the embedding of the $3$-sphere $\STR$ in four-dimensional
Euclidean space by
\begin{eqnarray}
\label{klein}
x & = & \cos{\CHN} , \\
y & = & \sin{\CHN} \cos{\theta} , \\
z & = & \sin{\CHN} \sin{\theta}\cos{\varphi} , \\
w & = & \sin{\CHN} \sin{\theta} \sin{\varphi} ,
\end{eqnarray}
with
\begin{eqnarray}
0 \leq \CHN \leq \pi , \\
0 \leq \theta \leq \pi , \\
0 \leq \varphi \leq 2 \pi .
\end{eqnarray}
The (complex) coefficients $\ALKLMLTMT{\nu}{\ell}{m}{\ell_\ST}{m_\ST}$
characterizing the change of basis are defined by
\begin{equation}
\label{3bis} \ALKLMLTMT{\nu}{\ell}{m}{\ell_\ST}{m_\ST}
 = (\nu+1)^2\int \QKLM{\nu}{\ell_\ST}{m_\ST} (\chi_\ST, \theta_\ST, \varphi_\ST)
        \YXKLM{\STR}{k}{\ell}{m}{}^* (\CHN, \theta, \varphi)
        (\sin \CHN)^2 \ddd\CHN \, \sin \theta \ddd \theta \, \ddd\varphi .
\end{equation}
In this expression the integer $\ell$ ranges from $0$ to $\nu$ and
$m$ ranges from $- \ell$ to $\ell$, while $\ell_\ST $ and $m_\ST$
range from $- \nu$ to $\nu$. To compute this integral, one needs:
\begin{enumerate}

\item to express the eigenmodes $\QKLM{\nu}{\ell_\ST}{m_\ST}$ as
polynomials in $x$, $y$, $z$ and $w$ (see Ref.~\cite{luw02}), and
replace these rectangular coordinates by their
expressions~(\ref{klein}),

\item use the relations
\begin{equation}
\label{p}
\cos^{|\ell_\ST|} \chi_\ST
 \left\lbrace \begin{array}{l}
  \cos (|\ell_\ST| \theta)\\
  \sin(|\ell_\ST| \theta)
\end{array}
   \right.
 = \left\lbrace \begin{array}{ll}
                 \Re \left[(\cos \CHN + i \sin \CHN \cos \theta)^{|\ell_\ST|}
                     \right] & \ell_\ST \geq 0 , \\
                 \Im \left[(\cos \CHN + i \sin \CHN \cos \theta)^{|\ell_\ST|}
                     \right] & \ell_\ST < 0 ,
                \end{array}
   \right.
\end{equation}
and
\begin{equation}
\label{p2}
\sin^{|m_\ST|}\chi_\ST
 \left\lbrace \begin{array}{l}
 \sin (|m_\ST| \varphi)\\
 \cos(|m_\ST| \varphi)
\end{array}
   \right.
 = (\sin \CHN \sin \theta)^{|m_\ST|}
   \left\lbrace \begin{array}{ll}
                 \cos (|m_\ST| \varphi) & m_\ST \geq 0 , \\
                 \sin (|m_\ST| \varphi) & m_\ST < 0 ,
                \end{array}
   \right.
\end{equation}

\item develop the Jacobi polynomial appearing in
Eq.~(\ref{PsiSolution}) by using Eq.~(\ref{C12}) and with
\begin{equation}
\cos 2 \chi_\ST - 1 = - 2 \sin^2 \CHN \sin^2 \theta .
\end{equation}

\end{enumerate}
This leads, after an easy integration on $\varphi$, to the somewhat
heavy expressions involving two sums [arising from the development of
the Jacobi polynomials and the power in Eq.~(\ref{p})] and an integral
over $\theta$ and $\CHN$,
\begin{equation}
\label{alpha2} \ALKLMLTMT{\nu}{\ell}{m}{\ell_\ST}{m_\ST}
 = (\nu+1)^2 \pi A_{\nu\ell} \widetilde A_{\ell m}
   \BKLM{\nu}{\ell_\ST}{m_\ST}
   \CKLM{\nu}{\ell_\ST}{m_\ST}
   \left\lbrace\begin{array}{l} \Re \\ \Im \end{array}
   \right.\left[{\cal J}(\nu; \ell, m; \ell_\ST, m_\ST)\right]
   \times \left\lbrace \begin{array}{l}
                        \left(  \KRON{m}{|m_\ST|}
                              + \KRON{m}{-|m_\ST|} \right) \\
                        - i\left(  \KRON{m}{|m_\ST|}
                                 - \KRON{m}{- |m_\ST|} \right) \\
                       \end{array}
          \right. .
\end{equation}
Here, the first and second line of the first brace are for $\ell_\ST
\geq 0$ and $\ell_\ST < 0$, respectively [see Eq.~(\ref{p})], the first
and second line of the second brace are for $m_\ST \geq 0$ and $m_\ST
< 0$, respectively [see Eq.~(\ref{p2})], and the numerical coefficient
$\CKLM{\nu}{\ell_\ST}{m_\ST}$ is given by
\begin{equation}
\CKLM{\nu}{\ell_\ST}{m_\ST}
 = \frac{\Gamma (|m_\ST| + d + 1)}
        {d !\, \Gamma(|m_\ST| + |\ell_\ST| + d + 1)} .
\end{equation}
The function ${\cal J}$ is explicitly given by
\begin{eqnarray}
{\cal J}(\nu; \ell, m; \ell_\ST, m_\ST)
 & = & \sum_{q = 0}^d \CNP{d}{q} (- 1)^q
                      \frac{ \Gamma (|m_\ST| + |\ell_\ST | + d + q + 1)}
                           { \Gamma (|m_\ST| + q + 1)}
\nonumber \\ & & \times
   \sum_{j = 0}^{|\ell_\ST|} i^j \CNP{|\ell_\ST|}{j}
    I \left(j, 2 q + |m| + |m_\ST |; \ell - |m|, |m| + \frac{1}{2} \right)
\nonumber \\ & & \label{alpha3} \qquad \qquad \qquad \qquad \times
    I \left(|\ell_\ST| - j, 2 q + j + \ell + |m_\ST| + 1 ;
            \nu - \ell, \ell + 1 \right) .
\end{eqnarray}
Note that the index $j$ is even when $\ell_\ST \geq 0$ and odd when
$\ell_\ST <0$.  This quantity involves only two sums, once the
quantity $I(p, q; n, \alpha)$, defined by
\begin{equation}
I(p, q; n, \alpha)
 \equiv \int_{-1}^{1} x^p (1 - x^2)^{q / 2} \GNA{n}{\alpha} (x) \ddd x ,
\end{equation}
is known. Using the expression~(\ref{C15}) for the Gegenbauer
polynomials in term of hypergeometric functions and the
integral~(\ref{C16}), it can be shown that
\begin{eqnarray}
I(p, q; 2 m, \alpha)
 & = & \frac{(- 1)^m}{2 (\alpha + m)}
       \frac{B (q / 2 + 1, (p + 1) / 2)}{B (\alpha , m + 1)}
       \left[1 + (- 1)^p \right] \,
       {}_3F_2 \left(- m, m + \alpha, \frac{p + 1}{2};
                     \frac{1}{2}, \frac{q + p + 3}{2}; 1 \right) , \\
I(p, q; 2 m + 1, \alpha)
 & = & (- 1)^m \frac{B (q / 2 + 1, (p + 2) / 2)}{B (\alpha, m + 1)}
       \left[1 - (- 1)^{p} \right] \,
       {}_3F_2 \left(- m, m + \alpha + 1, \frac{p + 2}{2};
                     \frac{3}{2}, \frac{q + p + 4}{2}; 1 \right) ,
\end{eqnarray}
where $B$ is the Euler Beta function. It follows directly from these
expressions that ${\cal J} (\nu; \ell, m; \ell_\ST, m_\ST) = 0$ when
$\nu - |m| + |\ell_\ST|$ is odd.

\section{Some properties of some special functions}
\label{app_C}

This appendix gathers some useful relations used in the article,
to make the article more self-contained.\\

The spherical harmonics $\YLM{\ell}{m}$ are related to the associated
Legendre polynomials $\PLM{\ell}{m}$ by (see Eq.~(5.2.1) of
Ref.~\cite{vmk})
\begin{equation}
\label{C1}
\YLM{\ell}{m} (\theta, \varphi)
 = \sqrt{\frac{2 \ell + 1}{4 \pi} \frac{(\ell - m)!}{(\ell + m)!}}
   \;\PLM{\ell}{m} (\cos \theta)\; \ee^{i m \varphi} .
\end{equation}
They satisfy the conjugation relation (Eq.~(5.4.1) of Ref.~\cite{vmk})
\begin{equation}
\label{C2}
\YLM{\ell}{m}{}^* (\theta, \varphi)
 = (- 1)^m \YLM{\ell}{- m} (\theta, \varphi)
 = \YLM{\ell}{m} (\theta, - \varphi),
\end{equation}
the normalization (Eq.~(5.6.1) of Ref.~\cite{vmk})
\begin{equation}
\label{C3}
\int_0^{2\pi} \ddd \varphi \int_0^{\pi} \sin \theta \, \ddd \theta \;
     \YLM{\ell}{m}{}^* (\theta, \varphi) \YLM{\ell'}{m'} (\theta, \varphi)
 = \KRON{\ell}{\ell'} \KRON{m}{m'},
\end{equation}
the closure relation (Eq.~(5.2.2) of Ref.~\cite{vmk})
\begin{equation}
\label{C4}
\sum_{\ell=0}^\infty\sum_{m=-\ell}^\ell \YLM{\ell}{m} (\theta, \varphi)
               \YLM{\ell}{m}{}^* (\theta', \varphi')
 = \DIR{\cos \theta - \cos \theta'} \DIR{\varphi - \varphi'} ,
\end{equation}
and the addition theorem (Eq.~(5.17.2.9) of Ref.~\cite{vmk})
\begin{equation}
\label{C5}
\sum_{m = - \ell}^\ell \YLM{\ell}{m} (\theta, \varphi)
                       \YLM{\ell}{m}{}^* (\theta', \varphi')
 = \frac{2 \ell + 1}{4 \pi} P_\ell(\cos \alpha) ,
\end{equation}
where $\alpha$ is the angle between the two directions $(\theta,
\varphi)$ and $(\theta', \varphi')$, and $P_\ell$ is the Legendre
polynomial. The Fourier transform of the spherical harmonics is given
by (Eq.~(5.9.2.6) of Ref.~\cite{vmk})
\begin{equation}
\label{C6}
\int_0^\infty r^2 \ddd r
\int_0^{2 \pi} \ddd \varphi \int_0^\pi \sin \theta \, \ddd \theta
     \frac{\ee^{i \bk . \br}}{(2 \pi)^{3 / 2}}j_\ell (k' r)
     \YLM{\ell}{m} (\theta, \varphi)
 = \sqrt{\frac{2}{\pi}} i^\ell \frac{\DIR{k' - k}}{k^2}
   \YLM{\ell}{m} (\theta_\bk, \varphi_\bk) ,
\end{equation}
where $j_\ell$ is a spherical Bessel function, from which it follows
that (Eq.~(5.17.3.14) of Ref.~\cite{vmk})
\begin{equation}
\label{C7}
\ee^{i \bk . \br}
 = \sum_{\ell = 0}^\infty (2 \ell + 1) i^\ell
        j_\ell(kr) P_\ell (\cos \theta_{\bk, \br}) ,
\end{equation}
and (Eq.~(5.17.4.18) of Ref.~\cite{vmk})
\begin{equation}
\label{C8}
 \DIR{\br_1 - \br_2}
 = \frac{\DIR{r_1 - r_2}}{r_1^2}
   \sum_{\ell = 0}^\infty \frac{2 \ell + 1}{4 \pi} P_\ell(\cos \theta_{1 2}) .
\end{equation}

The spherical harmonics can also be expressed in terms of Gegenbauer
polynomials as (Eq.~(5.2.6.39c) of Ref.~\cite{vmk})
\begin{equation}
\label{C11}
\YLM{\ell}{m} (\theta, \varphi)
 = \zeta_m \ee^{i m \varphi}
   \sqrt{\frac{2 \ell + 1}{4 \pi}}
   \sqrt{\frac{(\ell - |m|)!}{(\ell + |m|)!}}
   (2 |m| - 1)!! (\sin \theta)^{|m|}
   \GNA{\ell - |m|}{|m| + 1 / 2} (\cos\theta)
\end{equation}
with $\zeta_m$ defined by
\begin{equation}\label{C11bis}
\zeta_m
 = \left \lbrace \begin{array}{ll}
                 (-1)^m & m > 0 \\
                 1      & m \leq 0
                 \end{array}\right. .
\end{equation}
The ultraspherical (or Gegenbauer) polynomials $\GNA{n}{\alpha}$ are
solutions of the differential equation (Eq.~(22.6.5) of
Ref.~\cite{AbSt})
\begin{equation}
\label{C12}
(1 - x^2) y'' - (2 \alpha + 1) y'+ n (n + 2 \alpha) y = 0 ,
\end{equation}
and they satisfy the normalization condition (Eq.~(7.313) of
Ref.~\cite{gr})
\begin{equation}
\label{C12bis}
\int_{- 1}^{1} (1 - x^2)^{\alpha - 1 / 2}
               \left[\GNA{n}{\alpha} (x)\right]^2 \; \ddd x
 = \frac{\pi 2^{1 - 2 \alpha} \Gamma(2 \alpha + n)}
        {n ! (n + \alpha) \left[\Gamma (\alpha) \right]^2} ,
\end{equation}
if $\Re (\alpha) > -1 / 2$.

The Jacobi polynomials are given by (Eq.~(22.3.2) of Ref.~\cite{AbSt})
\begin{equation}
\label{C13}
 \JAC{\alpha}{\beta}{n} (x)
 = \frac{\Gamma (\alpha + n + 1)}{n ! \Gamma(\alpha + \beta + n + 1)}
   \sum_{j = 0}^n \CNP{n}{j}
   \frac{\Gamma (\alpha + \beta + n + j + 1)}
        {2^j \, \Gamma (\alpha + j + 1)}
   (x - 1)^j ,
\end{equation}
under the conditions $\alpha > - 1$ and $\beta > -1$.  Interestingly,
the Gegenbauer polynomials can be expressed in terms of hypergeometric
functions as (Eqs.~(8.932.2,8.932.3) of Ref.~\cite{gr})
\begin{eqnarray}
\label{C15}
\GNA{2 n}{\lambda} (x)
 & = & \frac{(- 1)^n}{(\lambda + n) B (\lambda, n + 1)}\,
       F \left(- n, \lambda + n; 1 / 2; x^2 \right) , \\
\GNA{2 n + 1}{\lambda} (x)
 & = & \frac{(- 1)^n}{B (\lambda, n + 1)} (2x) \,
       F \left(- n, \lambda + n + 1; 3 / 2; x^2 \right) ,
\end{eqnarray}
which satisfies the integral property (Eqs.~(7.513) of Ref.~\cite{gr})
\begin{equation}
\label{C16}
\int_0^1 x^{s - 1}(1 - x^2)^\nu F \left(- n, a; b; x^2 \right) \; \ddd x
 = \frac{1}{2} B(\nu + 1, s / 2) \;
   {}_3 F_2 \left(- n, a, s / 2; b, \nu + 1 + s / 2; 1 \right)
\end{equation}
if $\Re(s)>0$ and $\Re(\nu)>-1$.


\end{document}